\documentclass[12pt]{amsart}

\advance\textheight\topskip
\usepackage{upgreek}
\newcommand{\psim}{\uppsi_-}
\newcommand{\psip}{\uppsi_+}
\newcommand{\xiz}{\upxi}

\usepackage{amsfonts,amsthm,amssymb}
\allowdisplaybreaks
\usepackage{afterpage}

\usepackage[mathscr]{eucal}
\usepackage{stmaryrd}
\usepackage{graphicx}

\usepackage[dvips,bookmarks=false,hyperfigures=false,a4paper,
  colorlinks=true, linkcolor=black, citecolor=black]{hyperref}

\usepackage[OT2,T1]{fontenc}
\newcommand\cyr{%
  \renewcommand\rmdefault{wncyr}%
  \renewcommand\sfdefault{wncyss}%
  \renewcommand\encodingdefault{OT2}%
  \normalfont
  \selectfont}
\newcommand{\Zh}{{\cyr Zh}}
\usepackage{times}
\usepackage{mathptmx}

\usepackage[curve,ps,line,matrix,arrow,tips]{xy} 
\UsePSspecials{dvips}

\newcommand{\Sact}{\mathsf{s}}
\newcommand{\Tact}{\mathsf{t}}
\newcommand{\utauii}{\underline{\tau^2}}
\newcommand{\utau}{\underline{\tau}}
\newcommand{\uiitau}{\underline{2\tau}}
\newcommand{\uone}{\underline{1}}

\newcommand{\minor}{\omega}
\newcommand{\minorplus}{\pi}
\newcommand{\skewminor}{\varpi}
\newcommand{\tauskewminor}{\varsigma}
\newcommand{\skewmajor}{\varphi}

\newcommand{\SLiiZ}{SL(2,\oZ)}

\newcommand{\Rmin}{\mathscr{R}_{\mathrm{int}}(p)}
\newcommand{\Rpi}{\mathscr{R}_{p+1}}

\newcommand{\soc}{\mathop{\mathrm{soc}}\nolimits}

\newcommand{\kerker}{\ker\Qminus\cap\ker\Qplus}

\newcommand{\smatrix}[4]{\mbox{\scriptsize%
    $\displaystyle\begin{pmatrix}#1&#2\\
      #3&#4\end{pmatrix}$}}

\newcommand{\upperH}{\mathfrak{h}}

\newcommand{\one}{\boldsymbol{1}}

\def\Charge#1{*{\scriptstyle[#1]}\ar@{}[0,0];[0,0]+<0pt,3pt>;}
\def\Bharge#1{*{\scriptstyle\boldsymbol{[#1]}}\ar@{}[0,0];[0,0]+<0pt,3pt>;}

\def\verma{\bullet\rule[-2pt]{0pt}{9pt}
  \ar@{-}@*{[|(2.5)]}[0,0]+<0pt,0pt>;[0,0]-<10pt,0pt>
  \ar@{-}@*{[|(2.5)]}[0,0]-<1pt,-1pt>;[0,0]+<7pt,-7pt>
  \ar@{}[0,0]+<0pt,0pt>;}

\def\Cverma{\circ\rule[-2pt]{0pt}{9pt}
  \ar@{-}@*{[|(2.5)]}[0,0]+<-3pt,0pt>;[0,0]-<10pt,0pt>
  \ar@{-}@*{[|(2.5)]}[0,0]+<2pt,-2pt>;[0,0]+<6pt,-7pt>
  \ar@{}[0,0]+<0pt,0pt>;}

\def\iverma{\bullet\rule[-2pt]{0pt}{9pt}
  \ar@{-}@*{[|(2.5)]}[]+<0pt,0pt>;[0,0]+<10pt,0pt>
  \ar@{-}@*{[|(2.5)]}[]+<1pt,1pt>;[0,0]+<-7pt,-7pt>
  \ar@{}[0,0]+<0pt,0pt>;}

\def\Civerma{\circ\rule[-2pt]{0pt}{9pt}
  \ar@{-}@*{[|(2.5)]}[]+<3pt,0pt>;[0,0]+<10pt,0pt>
  \ar@{-}@*{[|(2.5)]}[]-<2pt,2pt>;[0,0]+<-6pt,-7pt>
  \ar@{}[0,0]+<0pt,0pt>;}

\newcommand{\q}{\mathfrak{q}}

\def\nothing{*{\mbox{}}\ar@{};[0,0];}

\def\Bullet{*{\rule[-1pt]{0pt}{7pt}\bullet}\ar@{};[0,0];}
\def\Circ{*{\rule[-1pt]{0pt}{7pt}\circ}\ar@{};[0,0];}

\newcommand{\Wright}{W^-}

\newcommand{\Wleft}{W^+}

\newcommand{\Qminus}{Q_-}
\newcommand{\Qplus}{Q_+}

\newcommand{\varphim}{\varphi_{-}}
\newcommand{\varphip}{\varphi_{+}}
\newcommand{\varphiz}{\varphi_{0}}

\newcommand{\MFFplus}[1]{\ket{\textsc{mff}^+(#1)}}
\newcommand{\MFFminus}[1]{\ket{\textsc{mff}^-(#1)}}
\newcommand{\jplus}{\lambda^+}
\newcommand{\jminus}{\lambda^-}

\newcommand{\thetap}{\theta_+}
\newcommand{\thetam}{\theta_-}

\newcommand{\plus}{+}
\newcommand{\minus}{-}

\newcommand{\repY}{\mathscr{Y}}

\newcommand{\thevert}[1]{e_{\phantom{h}}^{\left(#1\right)\ldot\varphi}}
\newcommand{\Thevert}[2]{e_{\phantom{h}}^{\left(#1\right)\ldot\varphi(#2)}}
\newcommand{\tensor}{\mathbin{\otimes}}
\newcommand{\Uop}[2]{U_{#2}[#1]}
\newcommand{\Uwak}[2]{\rep{U}_{#2}[#1]}
\newcommand{\Rop}[2]{R_{#2}[#1]}
\newcommand{\Rwak}[2]{\rep{R}_{#2}[#1]}
\newcommand{\Lop}[2]{L_{#2}[#1]}
\newcommand{\Vop}[2]{V_{#2}[#1]}
\newcommand{\Lwak}[2]{\rep{L}_{#2}[#1]}

\newcommand{\pUop}[2]{U'_{#2}[#1]}
\newcommand{\pUwak}[2]{\rep{U}'_{#2}[#1]}
\newcommand{\pRop}[2]{R'_{#2}[#1]}
\newcommand{\pRwak}[2]{\rep{R}'_{#2}[#1]}
\newcommand{\pLop}[2]{L'_{#2}[#1]}
\newcommand{\pVop}[2]{V'_{#2}[#1]}
\newcommand{\pLwak}[2]{\rep{L}'_{#2}[#1]}

\newcommand{\repK}{\rep{K}}

\newcommand{\rep}{\mathscr}

\newcommand{\MFFpm}[1]{\ket{\textsc{mff}^{\pm}(#1)}}

\newcommand{\dd}{\partial}

\newcommand{\bref}[1]{\textbf{\textup{\ref{#1}}}}

\newcommand{\ldot}{\mathbin{\boldsymbol{.}}}

\newcommand{\WW}[1]{\mathscr{W}^{(2)}_{#1}}

\newcommand{\polP}{\mathscr{P}}

\renewcommand{\geq}{\geqslant}
\renewcommand{\leq}{\leqslant}

\newcommand{\Jplus}{E}
\newcommand{\Jminus}{F}
\newcommand{\Jnaught}{H}

\newcommand{\SSL}[2]{s\ell(#1|#2)}

\newcommand{\hSSL}[2]{\widehat{s\ell}(#1|#2)}
\newcommand{\hSL}[1]{\widehat{s\ell}(#1)}
\newcommand{\SL}[1]{s\ell(#1)}

\newcommand{\hD}{\widehat{D}(2|1;\alpha)}

\newcommand{\Univ}{\mathscr{U}}

\newcommand{\oC}{\mathbb{C}}
\newcommand{\oN}{\mathbb{N}}

\newcommand{\oZ}{\mathbb{Z}}

\newcommand{\res}{\mathop{\mathrm{res}}\limits}
\newcommand{\Tr}{\mathop{\mathrm{Tr}}^{\phantom{y}}\nolimits}
\newcommand{\spFlow}[1]{\mathop{{\mathscr{U}_{#1}}}}

\newcommand{\actedby}{\raisebox{-4.5pt}{\mbox{\LARGE${\cdot}$}}}

\newcommand{\simtimesr}{%
  \mathrel{{\times}\kern-2.6pt\raisebox{1.2pt}{\mbox{\tiny $|$}}}}
\newcommand{\simtimesl}{%
  \mathrel{\raisebox{1.2pt}{\mbox{\tiny $|$}}\kern-2.6pt{\times}}}

\newcommand{\Verma}{\rep{M}}  
\newcommand{\mC}{\rep{C}}  
\newcommand{\mI}{\rep{I}}  
\newcommand{\mN}{\rep{N}}

\newcommand{\charSL}[2]{\chi_{%
    {\phantom{h}\kern-3pt #2}}^{\phantom{y}\kern-3pt #1}}
\newcommand{\charW}[2]{\chi_{%
    {\phantom{h}\kern-3pt #2}}^{\phantom{y}\kern-3pt #1}}

\newcommand{\ket}[1]{\mathchoice{%
    {\left|\smash[t]{#1}\right\rangle}}{|{#1}\rangle}{|{#1}\rangle}{|{#1}\rangle}}

\newcommand{\ketV}[1]{\mathchoice{%
    {\left|\smash[t]{#1}\right\rangle\!\rangle}}{|{#1}\rangle\!\rangle}{|{#1}\rangle\!\rangle}{|{#1}\rangle\!\rangle}}

\newcommand{\mfrac}[2]{\mbox{\small$\displaystyle\frac{#1}{#2}$}}
\newcommand{\ffrac}[2]{\raisebox{.5pt}{\mbox{\footnotesize$\displaystyle\frac{#1}{#2}$}}}
\newcommand{\half}{%
  \mathchoice{\ffrac{1}{2}}{\frac{1}{2}}{\frac{1}{2}}{\frac{1}{2}}}

\numberwithin{equation}{section}
\makeatletter
\@addtoreset{equation}{section}
\@addtoreset{subsubsection}{section}

\@addtoreset{figure}{section}
\renewcommand{\thefigure}{\thesection.\arabic{figure}}

\def\@secnumfont{\bfseries}
\def\subsubsection{\@startsection{subsubsection}{3}%
  \z@{.5\linespacing\@plus.7\linespacing}{-.5em}%
  {\normalfont\bfseries}}
\def\paragraph{\@startsection{paragraph}{4}%
  \z@\z@{-\fontdimen2\font}%
  \normalfont\bfseries}
\def\subparagraph{\@startsection{subparagraph}{5}%
  \z@\z@{-\fontdimen2\font}%
  \normalfont\bfseries}

\makeatother

\swapnumbers
\newtheorem{Thm}[subsection]{Theorem}

\newtheorem{lemma}[subsubsection]{Lemma}

\theoremstyle{definition}

\newtheorem{rem}[subsubsection]{Remark}

\begin{document}

\title[Logarithmic $\hSL2$]{%
  \vspace*{-4\baselineskip}
  \mbox{}\hfill\texttt{\small\lowercase{hep-th}/0701279}
  \\[\baselineskip]
  Toward logarithmic extensions of $\smash{\hSL2_k}$\\
  conformal field models}

\author[Semikhatov]{A.M.~Semikhatov}%

\address{\mbox{}\kern-\parindent Lebedev Physics Institute
  \hfill\mbox{}\linebreak \texttt{ams@sci.lebedev.ru}}

\begin{abstract}
  For positive integer $p\!=\!k\!+\!2$, we construct a logarithmic
  extension of the $\hSL2_k$ conformal field theory of integrable
  representations by taking the kernel of two fermionic screening
  operators in a three-boson realization of $\hSL2_k$.  The currents
  $\Wright(z)$ and $\Wleft(z)$ of a $W$-algebra acting in the kernel
  are determined by a highest-weight state of dimension $4p\,{-}\,2$
  and charge $2p\,{-}\,1$, and a $(\theta\!=\!1)$-twisted
  highest-weight state of the same dimension $4p\,{-}\,2$ and charge
  $-2p\!+\!1$.  We construct $2p$ $W$-algebra representations,
  evaluate their characters, and show that together with the
  $p\,{-}\,1$ integrable representation characters they generate a
  modular group representation whose structure is described as a
  deformation of the $(9p\,{-}\,3)$-dimensional representation
  $\Rpi{}\oplus{}\,\oC^2{\tensor}\,\Rpi{}\oplus{}
  \mathscr{R}_{p-1}{}\oplus{}\,\oC^2\!\tensor\mathscr{R}_{p-1}
  {}\oplus{}\,\oC^3\!\tensor\mathscr{R}_{p-1}$, where
  $\mathscr{R}_{p-1}$ is the $\SLiiZ$ representation on integrable
  representation characters and $\Rpi$ is a $(p\!+\!1)$-dimensional
  $\SLiiZ$ representation known from the logarithmic $(p,1)$ model.
  The dimension $9p\,{-}\,3$ is conjecturally the dimension of the
  space of torus amplitudes, and the $\oC^n$ with $n=2$ and~$3$
  suggest the Jordan cell sizes in indecomposable $W$-algebra modules.
  Under Hamiltonian reduction, the $W$-algebra currents map into the
  currents of the triplet~$W$-algebra of the logarithmic $(p,1)$
  model.
\end{abstract}

\maketitle
\thispagestyle{empty}
\enlargethispage{\baselineskip}

\vspace*{-\baselineskip}

\begin{footnotesize}%
  \addtolength{\parskip}{-6pt}%
  \renewcommand{\textbf}[1]{\mbox{}\kern20pt#1}%
  \tableofcontents
\end{footnotesize}

\vspace*{-1.5\baselineskip}

\section{Introduction}
Logarithmic conformal field theories in two
dimensions~\cite{[Gurarie],[GK2]} are attracting some attention from
different standpoints: in the context closest to the subject of this
paper, in~\cite{[G-alg],[N-SU2],[LMRS],[R-j]} and, as regards the
extended ($W$-)algebras,
in~\cite{[K-first],[GK3],[N-red],[FHST],[FGST],[FGST3]}; in the
context of boundary conformal field theories,
in~\cite{[GR],[CQS],[GT]}; mathematically, various aspects of
logarithmic conformal models and related structures in vertex-operator
algebras were considered in~\cite{[My],[CF],[Fuchs],[HLZ-2006]};
relations to statistical-mechanics models have been studied
in~\cite{[PRZ],[JPR],[RS]}; various aspects of logarithmic models were
elaborated
in~\cite{[GK3],[N-vac],[FFHST],[FHST],[R-sle],[GL],[FG],[SS],[EF],[FGK]};
relations to quantum groups and a ``nonsemisimple'' extension of the
Kazhdan--Lusztig correspondence~\cite{[KL]} were investigated
in~\cite{[FGST],[FGST2],[FGST-q],[GT]}.  Logarithmic conformal field
theories can be viewed as an extension of rational conformal field
theories~\cite{[MS],[FbZ],[BK],[fuRs]} to the case involving
indecomposable representations of the chiral algebra.  Identification
of the chiral algebra itself requires some care in logarithmic models:
in the known examples, starting with the pioneering
works~\cite{[K-first],[GK2],[GK3]}, the chiral algebra is not the
``naive,'' manifest symmetry algebra (e.g., Virasoro) but its
nonlinear extension, i.e., some $W$-algebra
(cf.~\cite{[N-SU2],[FHST],[FGST],[FGST3]}).

A systematic way to define a logarithmic conformal field theory model
is to take the kernel of the differential in a complex associated with
screening operators acting in appropriate free-field spaces.
Constructed this way, logarithmic models are a natural generalization
of rational ones (which are just the cohomology of the same
differential, cf.~\cite{[F],[BMP]}), but can also be defined in the
case where the cohomology is trivial and therefore the rational model
is empty~\cite{[FHST]}. \ Furthermore, defining logarithmic models in
terms of a kernel of screenings suggests chiral $W$-algebras of these
models; in the known $(p,1)$ and $(p,q)$ cases, the $W$-algebra that
is the symmetry of the model is the maximum local algebra acting in
the kernel.

In this paper, the ``screening-based'' approach is used to
logarithmically extend the well-known $\hSL2_k$ minimal models of
integrable representations.  Part of the motivation is in the general
popularity of WZW-related models and the possibility of constructing
coset models in particular.  But success is not guaranteed a priori.

Two \textit{related} difficulties can be perceived in carrying the
previously developed methods over to theories where the ``naive''
symmetry algebra (the one that is manifest before identifying the
$W$-algebra) is an affine Lie algebra.  First, the characters acquire
a dependence on $\nu\in\oC$, in addition to the modular parameter
$\tau\in\upperH$; then, whenever the $W$-algebra characters involve
derivatives of theta-functions (which is a typical feature of
logarithmic conformal field theories), predictable complications with
modular transformation properties occur.  Second, representations
multiply under the spectral flow action, to infinity in general; this
seems to take us even farther away from the rational class than the
generous setting of logarithmic conformal field theory may allow.  But
the situation with infinitely many inequivalent representations
produced by the spectral flow is in fact already encountered in the
more familiar setting of admissible representations~\cite{[KW]} of
affine Lie algebras, $\hSL2$ in particular.  If the characters are
understood in appropriate analytic-continuation terms, the number of
the resulting \textit{character functions} is finite
(cf.~\cite{[FSST],[STT],[LMRS1]}) and, moreover, a finite-dimensional
modular group representation is realized on them.  An extra
complication occurring in the logarithmic$/$nonsemisimple case is that
the space of torus amplitudes is not exhausted by the characters, and
therefore some other functions, which are not characters, come into
play.  In the ``$\nu$-free'' cases studied previously, these
\textit{generalized characters} typically had the form of characters
times polynomials in~$\tau$, with the degree of the polynomials
determined by the Jordan cell size~\cite{[FGST3]}; we have to see how
this behavior is affected by the appearance of a $\nu$ variable.
Continuing with challenges encountered in the study of logarithmic
conformal field theories, we mention that their $W$-symmetries are
rather complicated \hbox{algebras whose representation theory is
  poorly understood in general}.

Our aim is to report that despite these complications, it is
nevertheless possible to achieve certain consistency in constructing
logarithmic extensions of the minimal $\hSL2$ models following the
strategy
``\textsl{screenings${}\longrightarrow{}$kernel${}\longrightarrow{}$%
  {}$W$-algebra${}\longrightarrow{}$characters
  $\longrightarrow{}$generalized characters and modular
  transformations}.''  \textit{Consistency} here refers to modular
transformations, whose closure is a very strong consistency check for
various structures in conformal field theory.  It has been observed in
somewhat different situations in~\cite{[STT],[FSST]} (and maybe
elsewhere) that the closure of a set of character functions under the
spectral flow tends to imply their closure under modular
transformations.  To a certain extent, this is also the case with the
proposed logarithmic $\hSL2$ theory, where, as in other logarithmic
models, generalized characters occur in addition, but where also
``absorbing'' the explicit $\nu$ dependence requires introducing a
matrix automorphy factor, i.e., changing 
a (right) $\SLiiZ$-action $\gamma: f(\tau,\nu)\mapsto
f(\gamma\tau,\gamma\nu)$, $\gamma\in\SLiiZ$, into $\gamma:
f(\tau,\nu)\mapsto j(\gamma;\tau,\nu) f(\gamma\tau,\gamma\nu)$, where
$j$ is a function on $\SLiiZ\times\upperH\times\oC$ (matrix-valued if
$f$ is a vector)
satisfying the cocycle condition
\begin{equation*}
  j(\gamma\gamma';\tau,\nu)=
  j(\gamma';\tau,\nu)\, j(\gamma;\gamma'\tau,\gamma'\nu),
  \quad
  j(\one;\tau,\nu) = 1.
\end{equation*}

We start with screenings that single out the $\hSL2$ algebra as their
centralizer in a three-boson realization.  There are two a priori
inequivalent possibilities for this, involving either one bosonic and
one fermionic or two fermionic screenings.  The option chosen in this
paper is the one with \textit{two fermionic screenings}.\pagebreak[3]
Two fermionic screenings $\Qminus$ and $\Qplus$ give rise to complexes
of a somewhat unusual \Zh-shape
\begin{equation} 
  \label{eq:bfly2}
  \includegraphics[bb=2in 7in 8in 10.4in, clip, scale=.6]{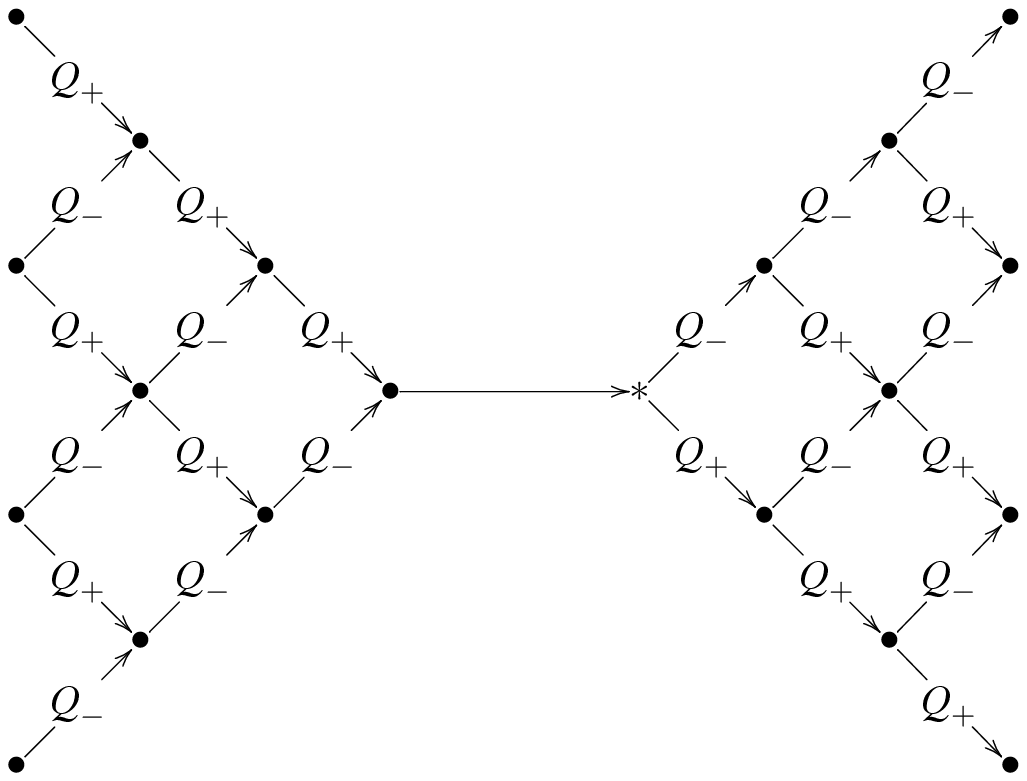}
\end{equation}
which is an $\hSL2$ version of the butterfly resolution
in~\cite{[b-fly]} (also see~\cite{[P]}).  The sites denote twisted
(and typically highly reducible) $\hSL2_k$-modules.\footnote{The
  butterfly resolution differs from Felder-type
  resolutions~\cite{[F],[BMP]} not only in its shape but also in that
  the modules farther away from the horizontal symmetry axis of the
  butterfly have progressively higher twists (parameters of spectral
  flow transformations).  The spectral flow can be visualized to map
  vertically in~\eqref{eq:bfly2}, with the result that the butterfly
  starts ``flying.''}  The two types of arrows, north-east and
south-east, correspond to the two fermionic screenings, whose
composition gives the middle link.  \textit{For positive integer
  $k+2$, the two screenings commute with each other}.  Whenever the
cohomology is nontrivial, it sits at the ``right eye'' ($\ast$) and
gives an integrable representation.  We also construct and use several
acyclic butterfly complexes.

That the resolution involves $\hSL2$-representations of arbitrarily
large twists leads to a number of problems (e.g., reasonably weighted
sums of their characters diverge).  In following the
``screenings${}\longrightarrow{}$kernel${}\longrightarrow{}$%
{}$W$-algebra${}\longrightarrow{}\dots$'' strategy, we therefore take
the kernels not in all the modules constituting the butterfly
resolution but only in the untwisted ones (those at the horizontal
symmetry line).  Accordingly, the $W$-algebra that we identify maps
only horizontally between the $\hSL2_k$-modules associated with the
sites in~\eqref{eq:bfly2}.  The currents $\Wright(z)$ and $\Wleft(z)$
generating the $W$-algebra are determined by singular vectors
representing a highest-weight state of dimension $4p\,{-}\,2$ and
charge $2p\,{-}\,1$, and a ($\theta\!=\!1$)-twisted highest-weight
state of the same dimension $4p\,{-}\,2$ and charge $-2p\!+\!1$, where
$p=k+2$.  More precisely, we let $\Jplus_n$ and $\Jminus_n$ be the
$\hSL2$ generators (with
$[\Jplus_m,\,\Jminus_n]={}k\,m\,\delta_{m+n,0} + 2\Jnaught_{m+n}$) and
write $\ket{\lambda;\theta}$ for a highest-weight vector with
spin~$\lambda$ and twist~$\theta$ (see Sec.~\bref{sec:twisted-verma}
for the details).  Let $\jplus(r,s)=\frac{r-1}{2}-p\frac{s-1}{2}$ and
$\jminus(r,s)=-\frac{r+1}{2}+p\frac{s}{2}$.  The $W$-algebra currents
$\Wright(z)$ and $\Wleft(z)$ are the operators corresponding to the
states
\begin{align}
  \label{Wright-0}
  \ket{\Wright}&=
  (\Jminus_{-1})^{3 p - 1}(\Jplus_{0})^{2 p - 1}(\Jminus_{-1})^{p - 1}
  (\Jplus_{0})^{-1}(\Jminus_{-1})^{-p - 1}
  \ket{\jplus(p\,{-}\,1,3);1},\\
  \label{Wleft-0}
  \ket{\Wleft}&=(\Jplus_{-1})^{3 p - 1}
  \ket{\jminus(3p-1,1);0},
\end{align}
where negative powers are to be understood as explained
in~\cite{[MFF]}.

For this $W$-algebra, we construct $2p$ of its representations,
denoted by $\repY^{\pm}_r$, $1\leq r\leq p$, evaluate their characters
$\charW{\pm}{r}(\tau,\nu)$, and study their spectral-flow and modular
transformation properties.  Because the representation theory of this
$W$-algebra is largely unexplored,\footnote{$W$-algebras in
  logarithmic models can be viewed as extensions of ``minimal'' (e.g.,
  Virasoro or, in our case, $\hSL2$) algebras by (descendants of)
  certain vertex operators.  Vertex-operator extensions have attracted
  some general interest, e.g.,
  in~\cite{[BFST],[FM-ext],[FJM],[JM],[MR]}.} we actually use the
spectral flow to generate a set of character functions on which the
modular group action may be expected to close if an appropriate number
of generalized characters (involving polynomials in~$\tau$) are added.
In a sense, we compensate for the poorly known $W$-algebra
representation theory by seeking a modular group representation
generated from a set of those $W$-algebra characters that we can
explicitly evaluate by honest representation theory.  The precise
result is as follows.

\smallskip

\begin{it}
  \noindent
  \textup{\textbf{Main result.}} \ 
  The modular group representation generated from $W$-algebra
  characters in the logarithmic $\hSL2_k$ model with positive integer
  $p=k+2$ is a deformation, via a matrix automorphy factor, of the
  direct sum
  \begin{equation}\label{eq:main-result}
    \Rpi\oplus\oC^2\!\tensor\Rpi
    \,\oplus\,
      \Rmin\oplus\oC^2\!\tensor\Rmin
      \oplus
      \oC^3\!\tensor\Rmin,
  \end{equation}
  where $\Rmin$ is the $(p\,{-}\,1)$-dimensional $\SLiiZ$
  representation on the integrable $\hSL2_k$ characters, $\Rpi$ is a
  $(p\,{+}\,1)$-dimensional representation, $\oC^2$ is the defining
  two-dimen\-sional representation, and $\oC^3$ is its symmetrized
  square; the matrix automorphy factor becomes equal to the identity
  matrix at $\nu=0$.
\end{it}

Fulfilling the general expectation~\cite{[N-red],[R-j]}, we next show
(Theorem~\bref{thm:HamRed}) that the Hamiltonian reduction indeed
relates the logarithmic $\hSL2_k$ to the logarithmic
$(p\!=\!k\!+\!2,1)$ models: \textit{the $W$-algebra generators
  in~\eqref{Wright-0} and~\textup{\eqref{Wleft-0}} map under the
  Hamiltonian reduction to generators of the triplet
  $W$-algebra~\cite{[K-first],[GK2],[GK3]} of the $(p,1)$ logarithmic
  model}, which were defined in~\cite{[FHST]} in terms of a Virasoro
screening.  We note that the $\SLiiZ$ representation on generalized
characters in the logarithmic $(p,1)$ model was evaluated as
$\Rpi\oplus\oC^2\!\tensor\Rmin$ in~\cite{[FGST]}; the Hamiltonian
reduction argument, in particular, ``explains'' the occurrence of
$\Rmin$ there.\footnote{Hamiltonian reduction at the level of
  conformal blocks (solutions of the Knizhnik--Zamolodchikov
  equations, see~\cite{[GP]} and the references therein) was studied
  in~\cite{[N-red]} to analyze logarithmic extensions of $(p,q)$
  minimal models and, in particular, gave evidence in favor of the
  existence of a $W$-algebra in these models.
}

The total dimension $9p\,{-}\,3$ of the $\SLiiZ$-representation
in~\eqref{eq:main-result} is a likely candidate for the dimension of
the space of torus amplitudes, if such a finite-dimensional space can
be constructed at all following one of the more direct approaches.
The $n=2$ and $3$ in the $\oC^n$ tensor factors
entering~\eqref{eq:main-result} suggest the Jordan cell sizes to be
encountered in indecomposable $W$-algebra modules.

The $\nu$ variable in the argument of the characters can in hindsight
be seen to result in producing deformations of direct sums of $\SLiiZ$
representations.  The $\bigl(2p\,{+}\,(p\,{-}\,1)\bigr)$-dimensional
space spanned by the $W$-algebra characters
$\bigl(\charW{\pm}{r}(\tau,\nu)\bigr)_{1\leq r\leq p}$ and the
integrable $\hSL2_k$-representation characters
$\bigl(\charSL{}{r}(\tau,\nu)\bigr)_{1\leq r\leq p-1}$ is extended to
the $(9p\,{-}\,3)$-dimensional space in~\eqref{eq:main-result} due to
two mechanisms.  First, the spectral flow closes if $2p$ functions
$\minor^{\pm}_r(\tau,\nu)$, $1\leq r\leq p$, are added.  We treat them
on equal footing with characters, although their representation-theory
meaning is not discussed here.  Second, as with the ``$\nu$-free''
characters of logarithmic conformal field theories known
previously~\cite{[FGST],[FGST3]}, certain combinations of the
$\charW{\pm}{r}(\tau,\nu)$, $\charSL{}{r}(\tau,\nu)$,
$\minor^{\pm}_r(\tau,\nu)$ become parts of multiplets, i.e., transform
in representations of the form $\oC^n\tensor\pi$, where $\pi$ is some
$\SLiiZ$ representation and $\oC^n$ is $\oC^2$ or $\oC^3$ realized on
polynomials in $\tau$ of the respective degrees $1$ and $2$.  But it
then turns out that the explicit occurrences of the $\nu$ variable in
modular transformation formulas results not only in ``mixing''
different modular-group representations with each other but also in
proliferating the number of functions involved, as is already clear
from the $\nu\mapsto\nu/(c\tau+d)$ transformations introducing a
fractional-linear factor.  This behavior can be ``absorbed'' into a
matrix automorphy factor, isolating which leaves us just with the
representation in~\eqref{eq:main-result}.\footnote{The need to
  introduce a matrix automorphy factor in order to extract a
  finite-dimensional $\SLiiZ$-representation of course reflects
  certain ``pathologies'' inherent in the adopted setting, where the
  number of free fields ($3$) is larger than the number of
  screenings~($2$).  As can be seen in~\cite{[W2n]}, a more natural
  object from the ``screening$/$quantum-group'' standpoint may be
  given by the coset~$\hSL2/u(1)$, but we leave its
  ``logarithmization'' for the future.}

\subsubsection*{Notation}
We fix the level $k$ as any complex number not equal to $-2$ in
Sec.~\bref{sec:sl2-general} and as $k\in\{0,1,2,\dots\}$ (and
occasionally $k=-1$) starting with~\bref{partA}.  We also use the
notation $p=k+2$, with the apologies for a certain lack of
consistency, in that a formula or two neighboring formulas may contain
both $k$ and $p$.  Similar negligence is shown regarding another
global notation, $j=\frac{r-1}{2}$: both $j$ and $r$ are used
interchangeably.

\subsubsection*{Remark}
The paper contains quite a few pictures of the subquotient structure
of the relevant modules and maps between them.  An alternative way of
delivering the same information would be a comparable abundance of
formal notation, making sense out of which would anyway require some
visualization.  The reader inclined to giving each object a special
name and a defining formula that makes all the parameters explicit
must be able to reconstruct the details from the numerous labels in
the pictures~(e.g., as in Fig.~\ref{fig:Lwak} (p.~\pageref{fig:Lwak})
below).

\medskip

This paper is organized as follows. In Sec.~\bref{sec:sl2-general}, we
fix the notation and conventions and recall standard facts about the
$\hSL2$ algebra, the spectral flow, and singular vectors in $\hSL2$
Verma modules, and then introduce the bosonization and the
corresponding screenings to be used in what follows.  In
Sec.~\bref{sec:resolutions}, we construct the butterfly resolution of
integrable representations and several acyclic butterfly complexes.
This gives enough information for constructing the $W$-algebra
generators $W^{\pm}(z)$ and representations $\repY^{\pm}_r$, $1\leq
r\leq p$, and evaluating the characters of the latter in
Sec.~\bref{sec:W-char}. \ We also study the spectral-flow and modular
transformation properties of the characters (extended by other
functions) in Sec.~\bref{sec:W-char}. \ In Sec.~\bref{sec:HRR}, we
evaluate the Hamiltonian reduction of the $W^{\pm}(z)$ generators,
showing that they map into the generators of the triplet $W$-algebra
of the logarithmic $(p,1)$-models. \ Section~\bref{sec:todo} is a list
of things that have not been done in this paper but are potentially
interesting, even if some of them prove impracticable.

Appendix~\bref{app:Verma} pertains entirely to
Sec.~\bref{sec:resolutions} and serves to recall the embedding
structure of $\hSL2_k$ Verma modules; most readers may ignore it
altogether. \ Appendix~\bref{app:theta} sets the notation and
summarizes some facts about theta-functions, extensively used in
Sec.~\bref{sec:W-char}. \ Appendix~\bref{app:SL2Z} contains a rather
explicit description of extensions among $\SLiiZ$ representations in
their ``functional'' realization, which occur in
Sec.~\bref{sec:W-char}.

\section{The $\smash{\protect\widehat{s\ell}(2)}$
  algebra}\label{sec:sl2-general}
In this section, we set the notation for the $\hSL2$ algebra, its
twisted modules, and singular vectors in Verma modules.  We then
introduce the three-boson realization of $\hSL2$ and the corresponding
screenings.

\subsection{The algebra, spectral flow, and twisted Verma modules}
The level-$k$ affine algebra $\hSL2_k$ is defined by the commutation
relations
\begin{equation}
  \begin{aligned}
    {[}\Jnaught_m,\,\Jplus_n]&=\Jplus_{m+n},\quad
    {[}\Jnaught_m,\,\Jminus_n]=-\Jminus_{m+n},\quad
    [\Jnaught_m,\,\Jnaught_n]=\ffrac{k}{2}\,m\,\delta_{m+n,0},
    \\
    {[}\Jplus_m,\,\Jminus_n]&={}k\,m\,\delta_{m+n,0} + 2\Jnaught_{m+n},
  \end{aligned}
  \label{sl2modes}
\end{equation}
with $m,n\in\oZ$.  In terms of the currents
$X(z)=\sum_{n\in\oZ}X_n\,z^{-n-1}$, $X=\Jplus$, $\Jnaught$, $\Jminus$,
the above commutation relations are reformulated as the OPEs
\begin{equation}\label{sl2-OPE}
  \begin{alignedat}{2}
    \Jnaught(z) \Jplus(w) &= \ffrac{\Jplus(w)}{z\,{-}\,w},&\quad
    \Jnaught(z) \Jminus(w) &= \ffrac{-\Jminus(w)}{z\,{-}\,w},\\
    \Jplus(z) \Jminus(w) &= \ffrac{k}{(z\,{-}\,w)^2} + \ffrac{2
      \Jnaught(w)}{z\,{-}\,w},& \quad \Jnaught(z) \Jnaught(w) &=
    \ffrac{k/2}{(z\,{-}\,w)^2}.
  \end{alignedat}
\end{equation}
and the Sugawara energy-momentum tensor is given by the standard
expression
\begin{gather*}
  T_{\mathrm{Sug}}(z)
  =\ffrac{1}{k\!+\!2}\Bigl(\half\,\Jplus(z)\Jminus(z) +
  \half\,\Jminus(z)\Jplus(z) 
  +\Jnaught(z)\Jnaught(z)\!\Bigr)
\end{gather*}
(here and below, normal ordering is understood; for brevity, we
sometimes write $AB(z)$ instead of the normal-ordered product
$A(z)B(z)$).  Generators of the Virasoro algebra with central charge
$c=\ffrac{3k}{k+2}$ are then introduced via
$T_{\mathrm{Sug}}(z)=\sum_{n\in\oZ}L_n\,z^{-n-2}$.

For each $\theta\in\oZ$, there is an $\hSL2_k$ automorphism given by
the so-called spectral flow transformation (see~\cite{[Kac]})
\begin{equation}
  \spFlow\theta:\quad
  \Jplus_n\mapsto \Jplus_{n+\theta},\qquad
  \Jminus_n\mapsto \Jminus_{n-\theta},\qquad
  \Jnaught_n\mapsto \Jnaught_n+\ffrac{k}{2}\,\theta\delta_{n,0}
  \label{spectral-sl2}
\end{equation}
Spectral-flow transforming any $\hSL2$-module~$\mC$ gives
\textit{twisted} modules $\spFlow\theta\mC=\mC_{;\theta}$.\footnote{An
  automorphism $\alpha$ of an algebra $\mathfrak{a}$ maps an
  $\mathfrak{a}$-module $M$ into a module $\alpha M$ on which the
  algebra acts as $a.(\alpha\,m)=\alpha(\alpha^{-1}(a).m)$,
  $a\in\mathfrak{a}$, $m\in M$.  The $\mathfrak{a}$-representations on
  $M$ and $\alpha M$ are \textit{not} equivalent in general.}

For any $\hSL2_k$-module $\mC$, its character is
\begin{equation}\label{lyuboiverma}
  \charSL{\mC}{}(q,z)=\Tr_{\mC}\bigl(
    q^{L_0-\frac{c}{24}}\,z^{\Jnaught_0}
  \bigr).
\end{equation}
We let $\charSL{\mC}{;\theta}(q,z)$ denote the character
$\charSL{\mC_{;\theta}}{}(q,z)$ of twisted modules.  In what follows,
we frequently use the following elementary result.

\begin{lemma}[\cite{[FSST]}] Let $\mC$ be an $\hSL2_k$-module.  Then
  \begin{equation}\label{spectral-sl2-general}
    \charSL{\mC}{;\theta}(q,z)=
    q^{\frac{k}{4}\theta^2}\,z^{-\frac{k}{2}\theta}\,
    \charSL{\mC}{}(q,z\,q^{-\theta}).
  \end{equation}
\end{lemma}

\subsubsection{Twisted Verma modules}\label{sec:twisted-verma}
We next fix our conventions regarding twisted Verma modules.  For
$\lambda\in\oC$ and $\theta\in\oZ$, the twisted Verma module
$\rep{M}_{\lambda;\theta}$ is freely generated by
$\Jplus_{\leq\theta-1}$, $\Jminus_{\leq-\theta}$, and
$\Jnaught_{\leq-1}$ from a twisted highest-weight vector
$\ket{\lambda;\theta}$ defined by the conditions
\begin{equation}\label{twistedhw}
  \begin{split}
    \Jplus_{\geq\theta}\,\ket{\lambda;\theta}&=
    \Jnaught_{\geq1}\,\ket{\lambda;\theta}=
    \Jminus_{\geq-\theta+1}\,\ket{\lambda;\theta}=0,\\
    \bigl(\Jnaught_{0}+\ffrac{k}{2}\,\theta\bigr)\,\ket{\lambda;\theta}&=
    \lambda\,\ket{\lambda;\theta}.
  \end{split}
\end{equation}
It follows that
\begin{equation*}
  L_0\,\ket{\lambda;\theta}
  = \Delta_{\lambda;\theta}\,\ket{\lambda;\theta},\quad
  \Delta_{\lambda;\theta}=
  \ffrac{\lambda^2+\lambda}{{k\!+\!2}}-\theta \lambda +
  \ffrac{k}{4}\theta^2.
\end{equation*}
For $k\theta\neq0$, we must therefore distinguish between the
eigenvalue of $\Jnaught_0$ on a twisted highest-weight state and the
spectral-flow-independent parameter $\lambda$ (which, e.g., determines
the existence of singular vectors in $\rep{M}_{\lambda;\theta}$).
\emph{We say that the eigenvalue of $\Jnaught_0$ is the \emph{charge}
  and $\lambda$ is the \emph{spin} of $\ket{\lambda;\theta}$.}
Setting $\theta=0$ gives the usual (``untwisted'') Verma modules.  We
write $\ket{\lambda}=\ket{\lambda;0}$ and, similarly,
$\Verma_{\lambda}=\rep{M}_{\lambda;0}$.

We write $\ket{\alpha}\doteq\ket{\lambda;\theta}$ whenever
conditions~\eqref{twistedhw} are satisfied for a state~$\ket{\alpha}$.

The character of a twisted Verma module $\Verma_{\lambda;\theta}$ can
be conveniently written in terms of $h=\lambda-\frac{k}{2}\,\theta$,
the eigenvalue of $\Jnaught_0$ in \eqref{twistedhw}, as
\begin{equation}\label{Verma-twisted}
  \charSL{\Verma}{\lambda;\theta}(q,z)
  =(-1)^{\theta}\,
  \frac{
    q^{\frac{(h-\theta+\half)^2}{k+2} -\frac{1}{8}}
    \,z^{h-\theta}}{
    \vartheta_{1,1}(q,z)}
\end{equation}
(see~\eqref{vartheta} for $\vartheta_{1,1}$).

\smallskip

Twists, although producing nonequivalent modules, do not alter the
submodule grid structure, and we can therefore reformulate a classic
result as follows.

\begin{Thm}[\cite{[KK],[MFF]}]\label{MFFthm}
  \addcontentsline{toc}{subsection}{\thesubsection. \ \ Singular
    vectors} A singular vector exists in a twisted $\hSL2$ Verma
  module $\Verma_{\lambda;\theta}$ if and only if $\lambda$ can be
  written as $\lambda=\jplus(r,s)$ or $\lambda=\jminus(r,s)$ with $r,s
  \in\oN$, where\pagebreak[3]
  \begin{align*}
    \jplus(r,s)&=\ffrac{r\!-\!1}{2}-(k\!+\!2)\ffrac{s-1}{2},
    \\
    \jminus(r,s)&=-\ffrac{r\!+\!1}{2}+(k\!+\!2)\ffrac{s}{2}.
  \end{align*}
  Whenever $\lambda=\jplus(r,s)$, the singular vector is given by
  \begin{multline}\label{mffplus}
    \MFFplus{r,s;\theta|\lambda}=
    (\Jminus_{-\theta})^{r+(s-1)(k+2)}
    (\Jplus_{\theta-1})^{r+(s-2)(k+2)}(\Jminus_{-\theta})^{r+(s-3)(k+2)}
    \ldots\\
    \cdot(\Jplus_{\theta-1})^{r-(s-2)(k+2)}
    (\Jminus_{-\theta})^{r-(s-1)(k+2)}\ket{\lambda;\theta}.
  \end{multline}
  Whenever $\lambda=\jminus(r,s)$, the singular vector is given by
  \begin{multline}\label{mffminus}
    \MFFminus{r,s;\theta|\lambda}=
    (\Jplus_{\theta-1})^{r+(s-1)(k+2)}(\Jminus_{-\theta})^{r+(s-2)(k+2)}
    (\Jplus_{\theta-1})^{r+(s-3)(k+2)}\ldots\\
    {}\cdot(\Jminus_{-\theta})^{r-(s-2)(k+2)}
    (\Jplus_{\theta-1})^{r-(s-1)(k+2)}\ket{\lambda;\theta}.
  \end{multline}
\end{Thm}
We recall that these formulas yield polynomial expressions in the
currents via repeated application of (the spectral-flow transform of)
the formulas
\begin{align}
  (\Jminus_0)^\alpha\,\Jplus_m &= -\alpha
  (\alpha - 1) \Jminus_m(\Jminus_0)^{\alpha-2} - 2 \alpha
  \Jnaught_m\,(\Jminus_0)^{\alpha-1} + \Jplus_m\,(\Jminus_0)^{\alpha}
  ,\notag\\
  (\Jminus_0)^{\alpha}\,\Jnaught_m &= \alpha
  \Jminus_m(\Jminus_0)^{\alpha-1} + \Jnaught_m\,(\Jminus_0)^{\alpha}
  ,\notag\\
  \label{properties}
  (\Jplus_{-1})^\alpha\,\Jminus_m &= -\alpha (\alpha - 1)
  \Jplus_{m-2}(\Jplus_{-1})^{\alpha-2} - k\,\alpha\,\delta_{m - 1, 0}
  (\Jplus_{-1})^{\alpha-1}\\*
  &\qquad\qquad\qquad\quad
  {}+ 2 \alpha \Jnaught_{m-1}\,(\Jplus_{-1})^{\alpha-1} +
  \Jminus_m\, (\Jplus_{-1})^{\alpha},\notag\\
  (\Jplus_{-1})^{\alpha}\,\Jnaught_m &= -\alpha
  \Jplus_{m-1}(\Jplus_{-1})^{\alpha-1} +
  \Jnaught_m\,(\Jplus_{-1})^{\alpha},\notag
\end{align}
which can be derived for positive integer $\alpha $ and then continued
to arbitrary complex~$\alpha $.

\subsubsection{}For $s=1$, singular vectors~\eqref{mffplus}
and~\eqref{mffminus} do not require any algebraic rearrangements and
take the simple form
\begin{equation}\label{s=1}
  \MFFplus{r,1;\theta|\lambda}
  =(\Jminus_{-\theta})^{r}\ket{\lambda;\theta},\qquad
  \MFFminus{r,1;\theta|\lambda}
  =(\Jplus_{\theta-1})^{r}\ket{\lambda;\theta}.
\end{equation}

\subsubsection{}\label{degeneration2}Another special case 
to be used in what follows occurs for positive integer $p=k+2$ and
$\lambda=\jplus(p,s)$.  {}From~\eqref{mffplus}, we then have
\begin{equation}\label{plus-factor}
  \MFFplus{p,s}=(\Jminus_0)^{s p}(\Jplus_{-1})^{(s-1) p}\ket{\lambda}.
\end{equation}
If $s=1$, we return to~\eqref{s=1}, but if $s\geq 2$, then the
corresponding state $\ket{\lambda}$ with $\lambda= p - \frac{p s +
  1}{2}$ also admits the singular vector $\MFFminus{p(s - 1), 1}$,
through which $\MFFplus{p,s}$ is actually seen to factor
in~\eqref{plus-factor}.  

Similarly, if $\lambda=\jminus(p,s)$, the corresponding singular
vector becomes
\begin{equation}
  \MFFminus{p,s}=
  (\Jplus_{-1})^{s p}(\Jminus_{0})^{(s-1)p}\ket{\lambda},
\end{equation}
which factors through $\MFFplus{p(s - 1), 1}$ whenever $s\geq2$.

\subsection{Integrable representation characters}
\label{sec:integrable}For positive integer $k+2$, the
characters of the integrable representations $\mI_{r}$,
$r=1,\dots,k+1$, are given by
\begin{equation}\label{integrable-sl2}
  \charSL{}{r}(q,z)\equiv\charSL{\mI}{r}(q,z) =
  \mfrac{\theta_{r,p}(q,z) - \theta_{-r,p}(q,z)}{
    \Omega(q,z)
  },\quad r=1,\dots,p-1
\end{equation}
where
\begin{equation}\label{Omega}
  \Omega(q,z)=q^{\frac{1}{8}}z^{\half}\vartheta_{1,1}(q,z).
\end{equation}
The integrable representation characters are holomorphic in~$z\in\oC$
and transform under the spectral flow as
$\charSL{}{r;1}(q,z)=\charSL{}{p-r}(q,z)$.

\subsection{Bosonization and fermionic screenings}\label{bosonize}
We keep the level $k$ fixed, temporarily at any complex value not
equal to~$-2$.  We introduce the well-known bosonization of the
$\hSL2_k$ algebra associated with two fermionic screenings, following
the conventions in~\cite{[W2n]} (where two $\hSL2$ bosonizations are
discussed from a unified standpoint and a more general case is also
considered; the bosonization chosen here is termed \textit{symmetric}
in~\cite{[W2n]}, for the reasons that become quite obvious when it is
compared with the other, nonsymmetric bosonization also discussed
there).

\subsubsection{``Symmetric'' three-boson realization}
\label{sec:bosonization}
Let $\xiz$, $\psim$, and $\psip$ be three vectors in $\oC^3$ with the
scalar products
\begin{alignat}{3}
  \xiz\ldot\xiz&=0,&\quad \xiz\ldot\psim&=1,&
  \quad \xiz\ldot\psip&=-1,\notag\\
  &&\psim\ldot\psim&=1,&\quad\psim\ldot\psip&=k+1,\label{Gram2}\\
  &&&&\psip\ldot\psip&=1\notag
\end{alignat}
(the determinant of the Gram matrix is equal to $-2(k+2)$, and hence
the vectors are defined uniquely modulo an overall rotation).  We
introduce a triple of scalar fields
$\varphi=(\varphi_1,\varphi_2,\varphi_3)$, in the standard basis, with
the OPEs
\begin{equation*}
  \dd\varphi_i(z)\,\dd\varphi_j(w)=\ffrac{\delta_{i,j}}{(z\!-\!w)^2},
\end{equation*}
where $\dd f(z)=\frac{\dd f(z)}{\dd z}$.  \ For any $a\in\oC^3$, let
$a\ldot\dd\varphi$ (as well as $a\ldot\varphi$) denote the Euclidean
scalar product.

It is easy to verify that the currents
\begin{equation}\label{bosonization}
  \begin{split}
    \Jplus(z)&=\psip\ldot\dd\varphi(z)\,e^{\xiz\ldot\varphi(z)},\\
    \Jnaught(z)&=\half(k \xiz + \psim - \psip)\ldot\dd\varphi(z),\\  
    \Jminus(z)&=\psim\ldot\dd\varphi(z)\,e^{-\xiz\ldot\varphi(z)}
  \end{split}
\end{equation}
satisfy the $\hSL2_k$ OPEs. We refer to these formulas as the
three-boson realization (bosonization) of $\hSL2_k$ (its relation to
the Wakimoto representation~\cite{[W],[FF]} is established by
bosonizing the first-order $\beta\gamma\,$ system involved in the
Wakimoto representation).  The bosonized form of the Sugawara
energy-momentum tensor is given by a standard formula involving the
inverse of the Gram matrix of $\xiz$, $\psim$, and
$\psip$~\cite{[W2n]}.  We keep the notation $L_n$ for the
corresponding Virasoro generators.

\subsubsection{Screenings}\label{sec:screenings} The bosonization
in~\eqref{bosonization} is associated with two fermionic screenings
\begin{equation}\label{the-screenings}
  \Qminus = \oint e^{\psim\ldot\varphi},\qquad
  \Qplus = \oint e^{\psip\ldot\varphi}.
\end{equation}
Simple calculation shows that they indeed commute with the $\hSL2_k$
currents in~\eqref{bosonization}.

In what follows, we no longer use the components
$(\dd\varphi_1,\dd\varphi_2,\dd\varphi_3)$ of $\dd\varphi$, and
instead sometimes use the notation
\begin{equation*}
  \dd\varphim(z)=\psim\ldot\dd\varphi(z),\quad
  \dd\varphip(z)=\psip\ldot\dd\varphi(z),\quad
  \dd\varphiz(z)=\xiz\ldot\dd\varphi(z).
\end{equation*}

\section{Twisted Wakimoto-type modules and butterfly complexes}
\label{sec:resolutions}
The bosonization of $\hSL2_k$ introduced in~\bref{bosonize} gives rise
to Wakimoto-type free-field modules~\cite{[W],[FF]}.  The aim of this
section is to construct complexes of twisted free-field modules using
the two fermionic screenings.  We begin in~\bref{sec:wakimoto} with
defining the relevant vertex operators and give simple formulas for
the action of $\hSL2$ generators on them; these formulas are then used
in evaluating singular vectors in terms of the above bosonization and
screenings.  A foregone conclusion is that ``half'' the singular
vectors in Wakimoto-type modules vanish.  The emerging pattern can
then be rephrased as the existence of the desired complexes.
In~\bref{partA}, we construct the butterfly resolutions of integrable
representations, with the cohomology concentrated in the ``right
eye.''  The complexes in~\bref{partB} and~\bref{sec:steinberg} are
acyclic.

The reader may wish to skip this long and rather technical section and
come back to the results in it when they are actually needed, and
proceed directly to Sec.~\bref{sec:W-char}.

\subsection{Vertex operators and states}
\label{sec:wakimoto} We introduce the family of vertex operators
\begin{equation}\label{family}
  U_{\lambda,h,\thetam,\thetap}(z)=
  e^{(h\xiz + (\frac{\lambda}{k+2}-\thetam)\psim
    + (\frac{\lambda}{k+2}-\thetap)\psip)\ldot\varphi(z)}
\end{equation}
and the corresponding states $\ketV{\lambda,h,\thetam,\thetap}$.  The
parameterization is redundant, the vertex being unchanged under
\begin{equation}\label{redundancy}
  \lambda\mapsto \lambda+p\alpha, \quad
  \theta_\pm\mapsto\theta_\pm+\alpha
\end{equation}
for arbitrary $\alpha$.  When we restrict to positive integer $k+2$ in
what follows, we take $2\lambda+1\in\{1,\dots,2(k+2)\}$,
$h\mp\lambda\in\oZ$, and $\theta_{\pm}\in\oZ$, with $\thetam$ and
$\thetap$ of the same sign (hence the two, not four, wings of the
butterfly).

The $\ketV{\lambda,h,\thetam,\thetap}$ states are Virasoro primaries
and, moreover, $\hSL2_k$ \textit{relaxed highest-weight states}: the
Verma-module highest-weight conditions in~\eqref{twistedhw} are
relaxed to\pagebreak[3]
\begin{align}
  \Jplus_{\geq\theta + 1}
  \ketV{\lambda,h,\thetam,\thetap} &=0,\notag
  \pagebreak[3]\\
  \Jminus_{\geq -\theta + 1}
  \ketV{\lambda,h,\thetam,\thetap}&=0,\label{relaxed-hw}\\
  \Jnaught_{\geq1}
  \ketV{\lambda,h,\thetam,\thetap} &=0,\notag
  \qquad\theta=\thetam - \thetap
\end{align}
(which, strictly speaking, requires $\thetam - \thetap\in\oZ$);
moreover, as is easy to verify,
\begin{equation}\label{eq:factors}
  \begin{aligned}
    \Jplus_{\thetam - \thetap}\ketV{\lambda,h,\thetam,\thetap} &=
    (\lambda - h - \thetam (k\!+\!1) - \thetap)
    \ketV{\lambda, h + 1,\thetam,\thetap},\\
    \Jminus_{\thetap - \thetam}\ketV{\lambda,h,\thetam,\thetap}&=
    (\lambda + h - \thetam - (k\!+\!1)\thetap)
    \ketV{\lambda, h - 1,\thetam,\thetap}.
  \end{aligned}
\end{equation}
It also follows that
\begin{equation*}
  \Jnaught_0 \ketV{\lambda,h,\thetam,\thetap} =
  h \ketV{\lambda,h,\thetam,\thetap}
\end{equation*}
(and therefore $h$ is the charge of the state, in accordance with with
the terminology introduced in~\bref{sec:twisted-verma}).

The reader is invited to appreciate the significance of the factors
$(\lambda-h-\thetam(k+1) - \thetap)$ and $(\lambda+h-\thetam -
(k+1)\thetap)$ in~\eqref{eq:factors} in ``strengthening'' the relaxed
highest-weight conditions in~\eqref{relaxed-hw} to the twisted
highest-weight conditions~\eqref{twistedhw} by appropriately choosing
the parameters (e.g., $h$).  We now elaborate on this and several
other simple technical details (the next subsection may be skipped
until its results are actually used).

\subsubsection{}\label{sec:modes-act}
First, it is obvious that
\begin{equation*}
  \Jplus_{\thetam - \thetap}\ketV{\lambda, h,\thetam,\thetap}=0\
  \Longleftrightarrow \
  \Qplus\,U_{\lambda,h,\thetam,\thetap}(z)=0.
\end{equation*}
Moreover, when this condition is satisfied, the state $\ketV{\lambda,
  h+1,\thetam,\thetap}$ is mapped by $\Qplus$
(see~\bref{sec:screenings}) into a twisted highest-weight state with
the twist $\thetam - \thetap + 1$.  This is an immediate consequence
of the OPE
\begin{equation*}
  e^{\psip\ldot\varphi(z)}\,U_{\lambda,h,\thetam,\thetap}(w)
  =(z-w)^{-h+\lambda-\thetam(k+1)-\thetap}
  e^{\psip\ldot\varphi(z)}U_{\lambda,h,\thetam,\thetap}(w)
\end{equation*}
(with the normal ordered product in the right-hand side).  Properly
developing this observation shows that
\begin{multline}\label{L-type-from-Qminus}
  (\Jplus_{\thetam-\thetap-1})^{N}
  U_{\lambda,\lambda-\thetam(k+1)-\thetap,\thetam,\thetap}(z)=\\
  {}=(-1)\dots(-N)\Qplus
  U_{\lambda,\lambda-\thetam(k+1)-\thetap+N,\thetam,\thetap+1}(z).
\end{multline}
For the $U$ operator in the left-hand side, we have
\begin{equation}\label{L-type}
  U_{\lambda,\lambda-\thetam(k+1)-\thetap,\thetam,\thetap}
  \doteq\ket{\lambda-(k+2)
    \smash{\ffrac{\thetam + \thetap}{2}};\thetam -
    \thetap}.
\end{equation}
It also follows that the ``top'' $\Jminus$-mode acts on it as
\begin{multline}\label{Jminus-on-L-type}
  (\Jminus_{\thetap - \thetam})^N
  U_{\lambda,\lambda-\thetam(k+1)-\thetap,\thetam,\thetap}(z)=\\
  {}=
  \ffrac{\Gamma(2\lambda + 1 - (\thetam + \thetap)(k+2))}{
    \Gamma(2\lambda + 1 - N - (\thetam + \thetap)(k+2))}\,
  U_{\lambda,\lambda-N-\thetam(k+1)-\thetap,\thetam,\thetap}(z).
\end{multline}
The ratio of $\Gamma$-functions here is of course a simple product of
$N$ factors as they follow from~\eqref{eq:factors}.\pagebreak[3]
But~\eqref{Jminus-on-L-type} can be analytically continued to complex
values of $N$, consistently with the continuation underlying the
construction of the MFF singular vectors (see~\eqref{properties}).
Formula~\eqref{L-type-from-Qminus} can be continued similarly, up to
an (inessential) sign, if $N!$ is replaced with the $\Gamma$-function.

Similarly, $\Jminus_{\thetap - \thetam}\ketV{\lambda,
  h,\thetam,\thetap}=0\ \Longleftrightarrow \
\Qminus\,U_{\lambda,h,\thetam,\thetap}(z)=0$
and\enlargethispage{\baselineskip}
\begin{multline}\label{R-type-from-Qplus}
  (\Jminus_{\thetap-\thetam})^M
  U_{\lambda,-\lambda+\thetam-1+(k+1)\thetap,\thetam-1,\thetap}(z)=\\
  =(-1)\dots(-M)\Qminus\,
  U_{\lambda,-\lambda + \thetam - 1 - M + (k+1)\thetap,\thetam,\thetap}(z).
\end{multline}
Up to a sign, this can be continued to complex $M$ by replacing $M!$
with the $\Gamma$-function.  For the operator in the left-hand side
of~\eqref{R-type-from-Qplus}, we have
\begin{equation}\label{R-type}
  U_{\lambda,-\lambda+\thetam-1+(k+1)\thetap,\thetam-1,\thetap}
  {}\doteq
  \ket{-\lambda-1+(k+2)\smash{\ffrac{\thetam + \thetap}{2}};\thetam -
    \thetap}.
\end{equation}
It also follows that the ``top'' $\Jplus$-mode acts on this state as
\begin{multline}\label{Jplus-on-R-type}
  (\Jplus_{\thetam - \thetap - 1})^N
  U_{\lambda,-\lambda+\thetam-1+(k+1)\thetap,\thetam-1,\thetap}=\\
  {}=
  \ffrac{\Gamma(2\lambda + 1 - (\thetam + \thetap - 1)(k+2))}{
    \Gamma(2\lambda + 1 - N - (\thetam + \thetap - 1)(k+2))}
  U_{\lambda,-\lambda + N + \thetam - 1 + (k+1)\thetap, \thetam-1,\thetap}.
\end{multline}

\subsubsection{}\label{wakimoto}
The above formulas provide a bridge between the MFF singular vectors
and the screenings, in accordance with the well-known fact that
``half'' the singular vectors in Wakimoto modules vanish.  For
example, if $\lambda$ is chosen as $\lambda=\frac{r-1}{2}$ with a
positive integer~$r$, the $\MFFplus{r,\thetam + \thetap + 1;\thetam -
  \thetap}$ singular vector can be constructed on the state
in~\eqref{L-type}.  Recalling~\eqref{mffplus}, we then evaluate the
action of the first (rightmost) operator factor $(\Jminus_{-\thetam +
  \thetap})^{r-(\thetam + \thetap)(k+2)}$
using~\eqref{Jminus-on-L-type}.  As noted above,
Eq.~\eqref{Jminus-on-L-type} is applicable for any complex $k\neq-2$
and hence a complex exponent $r-(\thetam + \thetap)(k+2)$.  The result
vanishes when evaluated at $N=r-(\thetam + \thetap)(k+2)$.


\specialsection*{\textbf{Butterfly complexes}}

{}From now on, we assume that $p=k+2\in\{1,2,\dots\}$ and consider
$\hSL2$ modules whose elements are given by the
vertices~\eqref{family} and their descendants with
\begin{equation}\label{range}
  2 \lambda+1\in\{1,\dots,2p\},\qquad
    h-\lambda\in\oZ
\end{equation}
and integer $\theta_{\pm}$.

The first consequence of restricting to integer $k\geq-1$ is that the
two fermionic screenings $\Qminus$ and $\Qplus$ become local with
respect to each other and, moreover, (super)commute.  Indeed, we have
the regular operator product
$e^{\psim\ldot\varphi(z)}\,e^{\psip\ldot\varphi(w)}\propto
(z\!-\!w)^{k+1}$.

The range of $2p$ different values of~$\lambda$ in~\eqref{range} is
covered in what follows by considering $\lambda=\frac{r-1}{2}$ with
$r=1,\dots,p-1$ in~\bref{partA} (with the spins of integrable
representations), $\lambda=\frac{p}{2}+\frac{r-1}{2}$ in~\bref{partB},
and two remaining values in~\bref{sec:steinberg}.

\subsection{Butterfly resolutions of integrable
  representations}\label{partA} The aim of this subsection is to
construct resolutions of integrable representations of
form~\eqref{eq:bfly2}, with the map in the center given by the
composition $\Qminus\circ\Qplus$.\pagebreak[3]

In each wing, the modules are labeled by two integers of the same
sign.  With a ``global'' numbering of all of them (e.g., with
$\thetam$ and $\thetap$ ranging from minus to plus infinity), those in
one of the wings would be labeled with negative integers; but
analyzing the structure of modules dependent on expressions like
$-\thetam-\thetap-1$ with negative $\thetam$ and $\thetap$ is somewhat
counter-intuitive, and we therefore choose a ``local'' numbering in
each wing, with positive integers in either case:
\begin{equation}\label{leftmn}
  \begin{split}
    m,n&\geq1\quad \text{in the left wing},
    \\
    m,n&\geq0\quad \text{in the right wing},
  \end{split}
\end{equation}
but with the notation for right-wing objects acquiring a prime.

\subsubsection{Left wing}\label{left-wing-A}For compactness of the
formulas, we use the notation
\begin{equation}\label{eq:j}
  j=\ffrac{r-1}{2}.
\end{equation}

For positive integer $m$ and $n$, we define the operator
\begin{equation}\label{Uop}
  \Uop{r}{m,n}(z)=\Thevert{[n-1+m(k+1)-j]\xiz
    +\frac{j}{k+2}(\psim + \psip) - n \psim - m \psip}{z}
\end{equation}
(which is $U_{j,n-1+m(k+1)-j,n,m}(z)$ in the nomenclature
of~\bref{sec:wakimoto}) and let $\Uwak{r}{m,n}$ denote the
corresponding Wakimoto-type module, i.e., the $\hSL2_k$-mod\-ule on
the free-field space generated from $\Uop{r}{m,n}$ (abusing the
terminology, we sometimes say for brevity that $\Uwak{r}{m,n}$ is
``generated'' from $\Uop{r}{m,n}$).  The extremal diagram of the
module (tilted in accordance with the twist $n-m$) can be
represented~as\enlargethispage{1.5\baselineskip}
\begin{equation}\label{eq:Utop}
  \includegraphics[bb=2in 8.45in 6.5in 10.26in, clip, scale=.8]{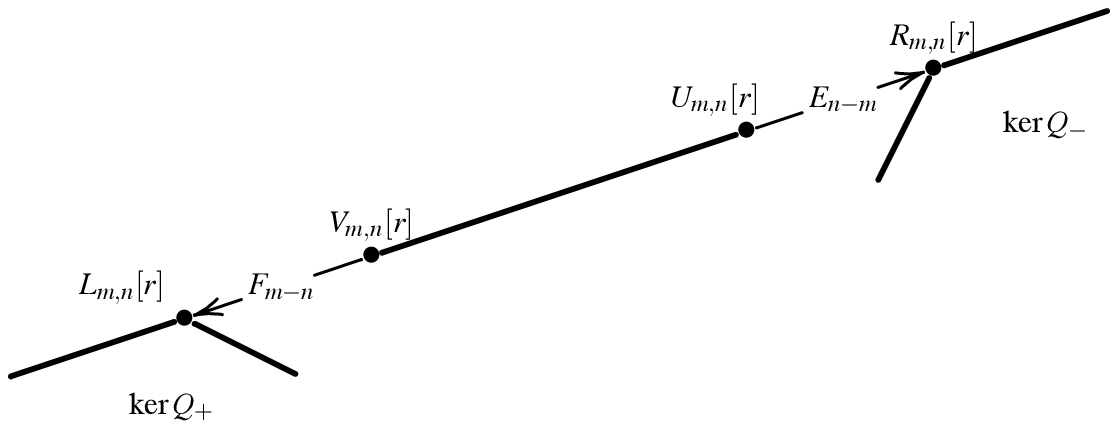}
\end{equation}
As follows from~\bref{sec:wakimoto}, $\Uop{r}{m,n}$ defines a relaxed
highest-weight state,\footnote{For notational simplicity, we no longer
  distinguish between operators and the corresponding states.  We also
  omit nonzero factors in the normalization of states.}
\begin{equation*}
  \Jplus_{n-m+1}\Uop{r}{m,n}=\Jminus_{m-n+1}\Uop{r}{m,n}=0,
\end{equation*}
and acting on $\Uop{r}{m,n}$ with
$(\Jminus_{m-n})^{(m+n)(k+2)-(2j+1)}$ gives the operator$/$state
\begin{align}\label{Lop}
  \Lop{r}{m,n}(z)&=\Thevert{[j-n(k+1)-m]\xiz
    +\frac{j}{k+2}(\psim + \psip) - n\psim - m
    \psip}{z}\\*
  &\doteq\ket{
    \ffrac{r\!-\!1}{2}-(k\!+\!2)\ffrac{m\!+\!n}{2};n\!-\!m},\notag
\end{align}
a twisted highest-weight state with the spin
$\jplus(r,n\,{+}\,m\,{+}\,1)=\jminus((m+n+1)(k+2)-r,$\linebreak[0]$1)$.
\ The $\Jplus_{n-m}$ and $\Jminus_{m-n}$ arrows in~\eqref{eq:Utop} map
into twisted highest-weight states (``charged'' singular vectors): for
example, as is easy to verify,
\begin{align}\label{Rop}
  \Jplus_{n-m}\Uop{r}{m,n}\equiv
  \Rop{r}{m,n}(z)&=\Thevert{[n+m(k+1)-j]\xiz
    +\frac{j}{k+2}(\psim + \psip) - n \psim - m
    \psip}{z}\\*
  &\doteq\ket{
    \jplus((m\,{+}\,n\,{+}\,1)p\,{-}\,r,1);
    n\!-\!m\!+\!1}.
  \notag  
\end{align}
We also note the OPE $e^{\psip\ldot\varphi(u)}
\Lop{r}{m+1,n}(z)=\text{reg}$, whence $\Qplus \Lop{r}{m,n}(z)=0$; it
follows similarly that $\Qminus\Rop{r}{m,n}(z)=0$.

Sugawara dimensions of the operators
$\Lop{r}{m,n}$ 
and $\Rop{r}{m,n}$ are
\begin{align*}
  \Delta_{\Lop{r}{m,n}}&=\ffrac{(j - n(k+2))(j - n(k+2) + 1)}{k+2}
  -\half (m - n)(m - n + 1),\\
  \Delta_{\Rop{r}{m,n}}&=\ffrac{(j - m(k+2))(j - m(k+2) + 1)}{k+2}
  -\half(m - n - 1)(m - n).
\end{align*}

\begin{figure}[tb]
  \centering
  \includegraphics[bb=1.3in 5.9in 7.4in 10.2in, clip,
  scale=.85]{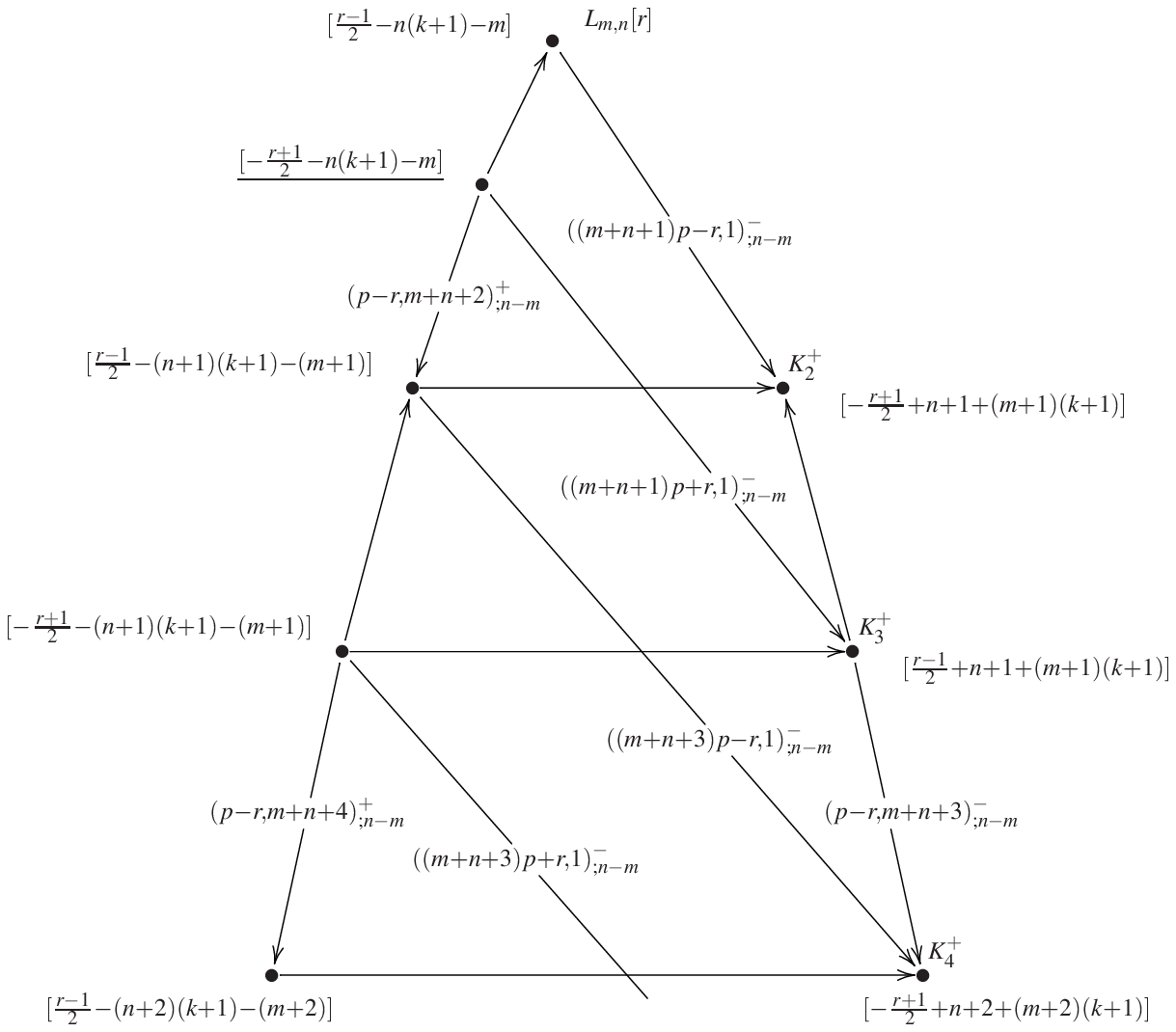}
  \caption{\small Subquotient structure of the left-wing twisted
    Wakimoto module $\Lwak{r}{m,n}$.  For visual clarity, the view is
    ``rotated back'' by the spectral flow with $\theta=m-n$ (in the
    original view, the horizontal arrows are tilted by the angle
    $\alpha$ such that $\tan\alpha=n-m$).  The values of $h$ for
    operators~\eqref{gen-vert} are shown in square brackets (one of
    these is underlined for later reference).  The
    $(a,1)^{\minus}_{;n-m}$ arrows represent nonvanishing singular
    vectors $\MFFminus{a,1;n-m}$, given by simple
    formula~\eqref{s=1}.}\label{fig:Lwak}
\end{figure}

The subquotient structure of the module $\Lwak{r}{m,n}$ ``generated''
from $\Lop{r}{m,n}$ is shown in the well-known picture in
Fig.~\ref{fig:Lwak}.
We use the same convention as in Appendix~\bref{app:Verma} to direct
arrows toward submodules.  The embedding structure of $\Lwak{r}{m,n}$
can be considered a result of the vanishing of ``half'' the singular
vectors in the free-field realization~\eqref{bosonization}.  The
filled dots in the figure represent operators of the form
\begin{equation}\label{gen-vert}
  \polP[\dd\varphi(z)]\,
  \Thevert{h\xiz +\frac{j}{k+2}(\psim + \psip) - n \psim - m
    \psip}{z},
\end{equation}
where $\polP$ is a differential polynomial (in the three currents
$\dd\varphiz(z)$, $\dd\varphim(z)$, and $\dd\varphip(z)$) and
\textit{the values of $h$ are shown in square brackets at the
  corresponding nodes}.  In particular, for the operators$/$states
labeled $K^+_b$, $b=2,3,\dots$, in Fig.~\ref{fig:Lwak}, we have
\begin{equation*}
  K^+_b(z)=\polP_b[\dd\varphi(z)]\,
  \Thevert{h_b\xiz +\frac{j}{k+2}(\psim + \psip) - n \psim - m
    \psip}{z},
\end{equation*}
where
\begin{equation}\label{the-h}
  h_b=
  \begin{cases}
    -\frac{r\!+\!1}{2}+n+\frac{b}{2}
    +\bigl(m+\frac{b}{2}\bigr)(k+1),&
    b\text{ even},\\[6pt]
    \ffrac{r\!-\!1}{2}+n+\frac{b\!-\!1}{2}
    +\bigl(m+\frac{b\!-\!1}{2}\bigr)(k+1),&
    b\text{ odd}.
  \end{cases}
\end{equation}

The character of the irreducible subquotient $\repK_b$ corresponding
to $K^+_b$ follows from \eqref{Verma-char-even}, \eqref{Verma-char-odd},
and~\eqref{spectral-sl2-general}: for $b\geq1$, we have (with the
dependence on $m$ and $n$ indicated as a subscript)
\begin{multline}\label{V-even}
  \charSL{\repK_{2b}}{m,n}(q,z)
  {}=
  \mfrac{(-1)^{n-m}}{q^{\frac{1}{8}
  }\,
  \vartheta_{1,1}(q,z)}
  \Bigl(\sum_{a\geq0} + \!\!\sum_{a\leq-n-m-2b}\Bigr)
  \Bigl(
  q^{p\left(\frac{r}{2p}-(m+b+a)\right)^2}
  z^{-\frac{r+1}{2}+(m+b+a)p}\\*
  {}-
  q^{p\left(\frac{r}{2p}-(n+b+a)\right)^2}
  z^{\frac{r-1}{2}-(n+b+a)p}
  \Bigr),
\end{multline}
\mbox{}\vspace*{-1.2\baselineskip}
\begin{multline}\label{V-odd}
  \charSL{\repK_{2b+1}}{m,n}(q,z)
  {}=
  \mfrac{(-1)^{n-m}}{q^{\frac{1}{8}
  }\,
  \vartheta_{1,1}(q,z)}
  \Bigl(\sum_{a\geq0} + \!\!\sum_{a\leq-n-m-2b-1}\Bigr)
  \Bigl(
  q^{p\left(\frac{r}{2p}+m+b+a\right)^2}
  z^{\frac{r-1}{2}+(m+b+a)p}\\*
  {}-
  q^{p\left(\frac{r}{2p}+n+b+a\right)^2}
  z^{-\frac{r+1}{2}-(n+b+a)p}
  \Bigr).
\end{multline}

\begin{figure}[tbp]
  \centering
  \includegraphics[bb=1.3in 6.1in 7.4in 10.2in, clip, scale=.8]{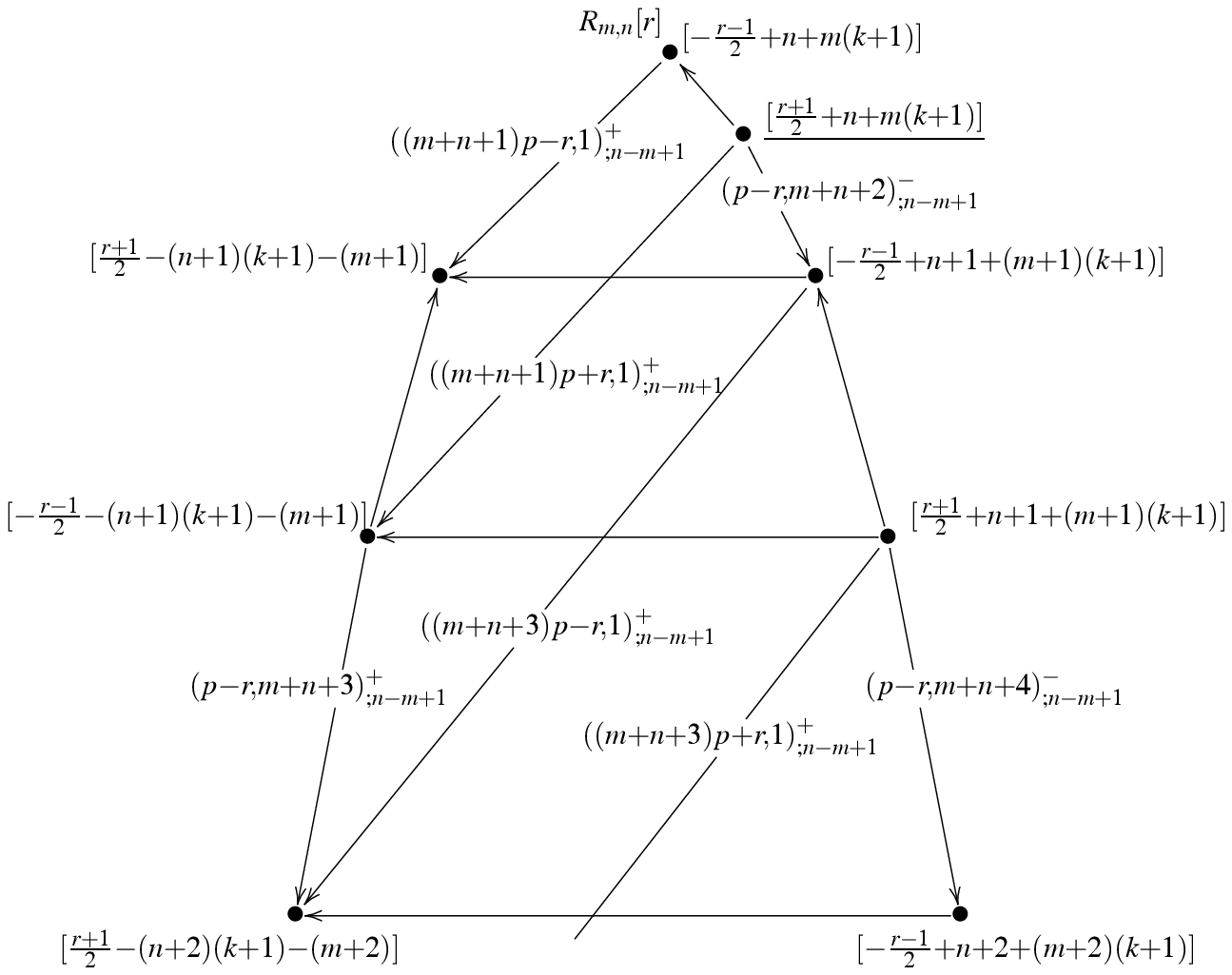}
  \caption{\small Subquotient structure of the left-wing twisted
    Wakimoto module $\Rwak{r}{m,n}$.  For the ease of comparison with
    Fig.~\ref{fig:Lwak}, the picture is ``rotated'' by the spectral
    flow with $\theta=m-n$.  As in Fig.~\ref{fig:Lwak}, the values of
    $h$ are shown in square brackets (one of these is underlined for
    later reference).}
  \label{fig:Rwak}
\end{figure}

Similarly, and with the same conventions, the subquotient structure of
the twisted Wakimoto module $\Rwak{r}{m,n}$ ``generated'' from
$\Rop{r}{m,n}$ is shown in Fig.~\ref{fig:Rwak}.
In the figure, the overall twist is ``undone'' by the same amount as
for the $\Lwak{r}{m,n}$ module; therefore, in describing the module
$\Uwak{r}{m,n}$ with the extremal diagram in~\eqref{eq:Utop}, the two
diagrams in Figs.~\ref{fig:Lwak} and~\ref{fig:Rwak} must be placed
next to each other, in accordance with the grades, which means placing
the top node of $\Rwak{r}{m,n}$ \ $(m+n)p-r+1$ units of charge to the
right of the top node of $\Lwak{r}{m,n}$.  It then follows that
starting with the embedding level of $K^+_2$, each node of
$\Lwak{r}{m,n}$ has a corresponding node of $\Rwak{r}{m,n}$ as the
nearest right neighbor.

The module $\Uwak{r}{m,n}$ is an extension of $\Lwak{r}{m,n}$ and
$\Rwak{r}{m,n}$.  We do not describe all of its structure, which we do
not need, but describe the occurrence of the kernel of the two
screenings below.  For this, we first consider the maps provided by
the screenings.

There are $\hSL2_k$-homomorphisms
\begin{align}\label{eq:Qplus-map}
  \Qplus:\Rwak{r}{m+1,n}\to\Lwak{r}{m,n},\\
  \Qminus:\Lwak{r}{m,n+1}\to\Rwak{r}{m,n},
  \label{eq:Qminus-map}
\end{align}
whose construction can be outlined as follows.  At the level of
extremal states (see~\eqref{eq:Utop}), we have seen that
$\Lop{r}{m+1,n}(z)$ is annihilated by $\Qplus$, but the
nearest-neighbor state
\begin{equation*}
  \Vop{r}{m+1,n}= \thevert{[j-n(k+1)-m]\xiz
    +\frac{j}{k+2}(\psim + \psip) - n\psim - (m+1)\psip}
\end{equation*}
is mapped under $\Qplus$ as
\begin{equation*}
  \Qplus \Vop{r}{m+1,n}(z)=
  \Thevert{[j-n(k+1)-m]\xiz
    +\frac{j}{k+2}(\psim + \psip) - n\psim - m\psip}{z}
  =\Lop{r}{m,n}(z).
\end{equation*}
Further acting with $\Jplus_{n-m-1}$ gives (up to a nonzero factor)
\begin{multline}\label{Qminus-act}
  \Qplus \Rop{r}{m+1,n}(z)=\\
  = (\Jplus_{-1+n-m})^{(m+n+1)(k+2)-r}\,
  \Thevert{[j-n(k+1)-m]\xiz
    +\frac{j}{k+2}(\psim + \psip) - n\psim - m \psip}{z},
\end{multline}
which is just the $\MFFminus{(m+n+1)(k+2)-r,1;n-m}$ singular vector
constructed on the~$\Lop{r}{m,n}$ state.

\begin{figure}[tb]
  \centering
  \includegraphics[bb=1.3in 3.3in 7.4in 10.4in, clip, scale=.8]{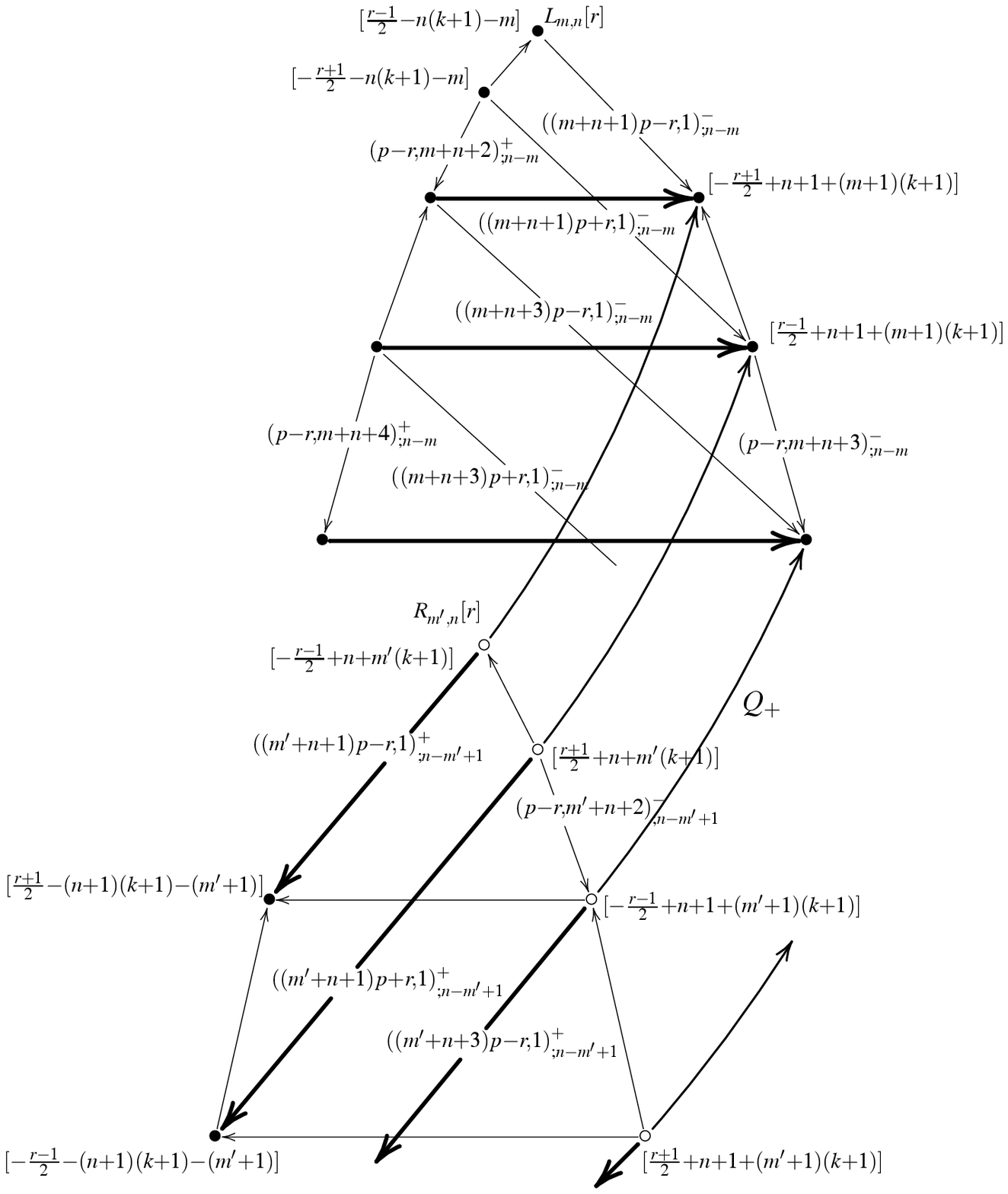}
  \caption{\small The left-wing map
    $\Qplus:\Rwak{r}{m+1,n}\to\Lwak{r}{m,n}$.  Filled dots in the
    $\Rwak{r}{m+1,n}$ module denote subquotients that are in the
    kernel of $\Qplus$.}
  \label{fig:RLmap}
\end{figure}

Figure~\ref{fig:RLmap}
shows further details that make up the definition of
$\Qplus:\Rwak{r}{m+1,n}\to\Lwak{r}{m,n}$.  In the figure, we reproduce
the pictures of the $\Lwak{r}{m,n}$ and $\Rwak{r}{m',n}$ modules, the
latter shown just as in Fig.~\ref{fig:Rwak} for the ease of
comparison, but with $m'$ to be taken equal to $m+1$.  Therefore, the
twist of $\Rwak{r}{m',n}$ is $n-m'-1=n-m-2$, with the result that the
tilted $((m'+n+i)p\pm r,1)^{\plus}$-arrows in the lower part of the
figure (shown boldfaced) should be drawn horizontally in the
conventions applicable to the upper part (we repeat that the module
$\Rwak{r}{m',n}$ is just copied from Fig.~\ref{fig:Rwak}).  But the
map by $\Qplus$ places these tilted $((m'+n+i)p\pm
r,1)^{\plus}$-arrows\pagebreak[3] just over the horizontal arrows
in~$\Lwak{r}{m,n}$ (also boldfaced for this reason), which are
oppositely directed because of the vanishing singular vectors;
therefore, the tilted $\Rwak{r}{m',n}$-arrows point to the kernel
of~$\Qplus$.

\begin{figure}[tb]
  \centering
  \includegraphics[bb=1.3in 2.9in 7.4in 10.4in, clip, scale=.8]{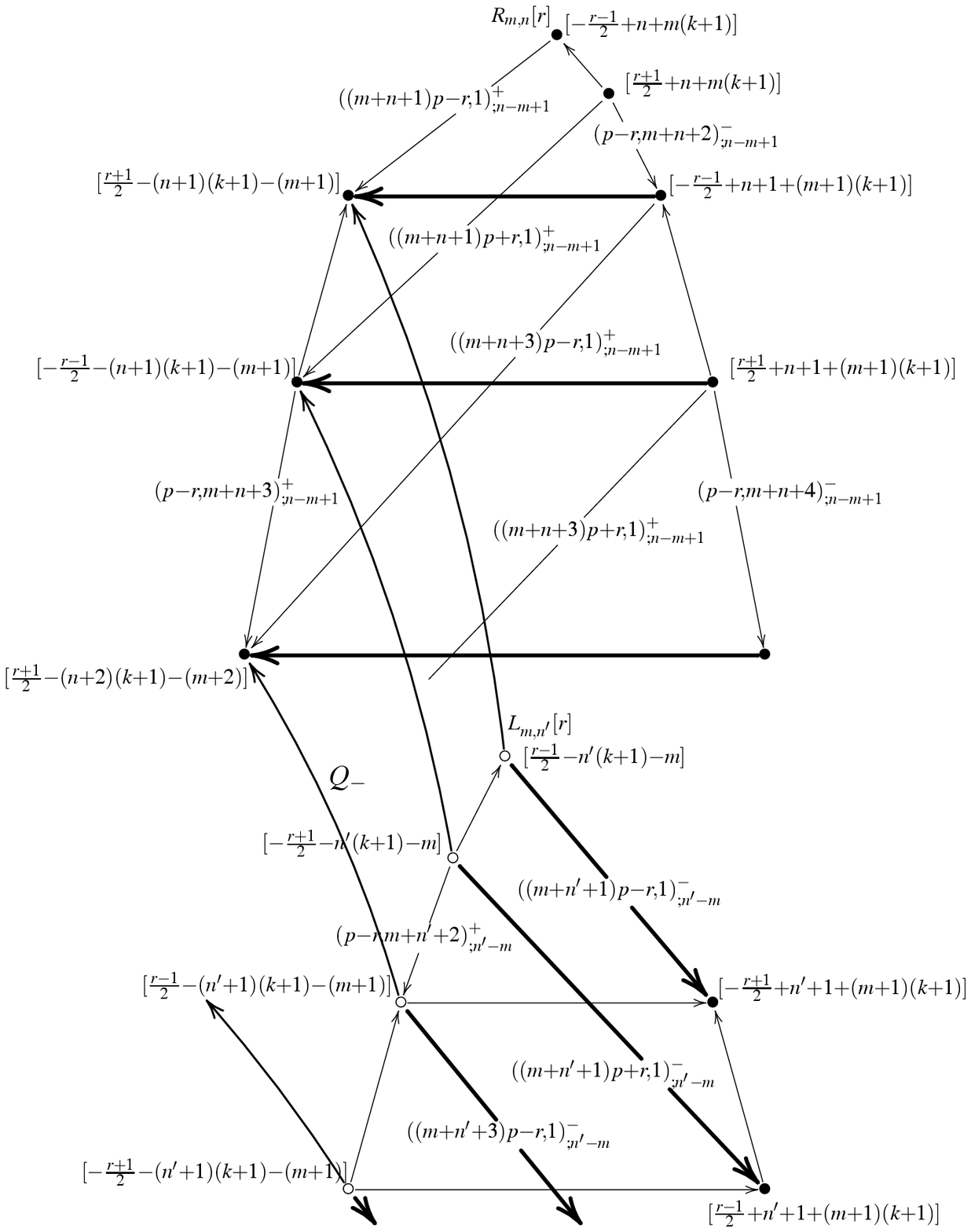}
  \caption{\small The left-wing map
    $\Qminus:\Lwak{r}{m,n+1}\to\Rwak{r}{m,n}$.  Filled dots in
    $\Lwak{r}{m,n+1}$ denote subquotients that are in the kernel of
    $\Qminus$.}
  \label{fig:LRmap}
\end{figure}

The map $\Qminus:\Lwak{r}{m,n+1}\to\Rwak{r}{m,n}$ can be described
similarly (see Fig.~\ref{fig:LRmap}).
We have
\begin{equation*}
  \Qminus\Uop{r}{m,n+1}(z)
  =\Thevert{[n+m(k+1)-j]\xiz
    +\frac{j}{k+2}(\psim + \psip) - n\psim - m \psip}{z}
  =\Rop{r}{m,n}(z)
\end{equation*}
and
\begin{equation}
  \Qminus\Lop{r}{m,n+1}(z)
  =(\Jminus_{m-n-1})^{(m+n+1)(k+2)-r}\Rop{r}{m,n}(z),
\end{equation}
which is just the $\MFFplus{r',1;\theta}$ singular vector with
$r'=(m\,{+}\,n\,{+}\,1)(k\,{+}\,2)\,{-}\,r$ and
$\theta=n\,{-}\,m\,{+}\,1$, constructed on~$\Rop{r}{m,n}$.  The
$\Lwak{r}{m,n'}$ module is shown in Fig.~\ref{fig:LRmap} just as in
Fig.~\ref{fig:Lwak}, with $n'$ to be set equal to $n\,{+}\,1$.  The
tilted $\Lwak{r}{m,n'}$\,{-}\,arrows point to the kernel of~$\Qminus$.

\begin{figure}[tbp]
  \centering
  \includegraphics[bb=2.3in 7.3in 7.4in 10.4in, clip, scale=.8]{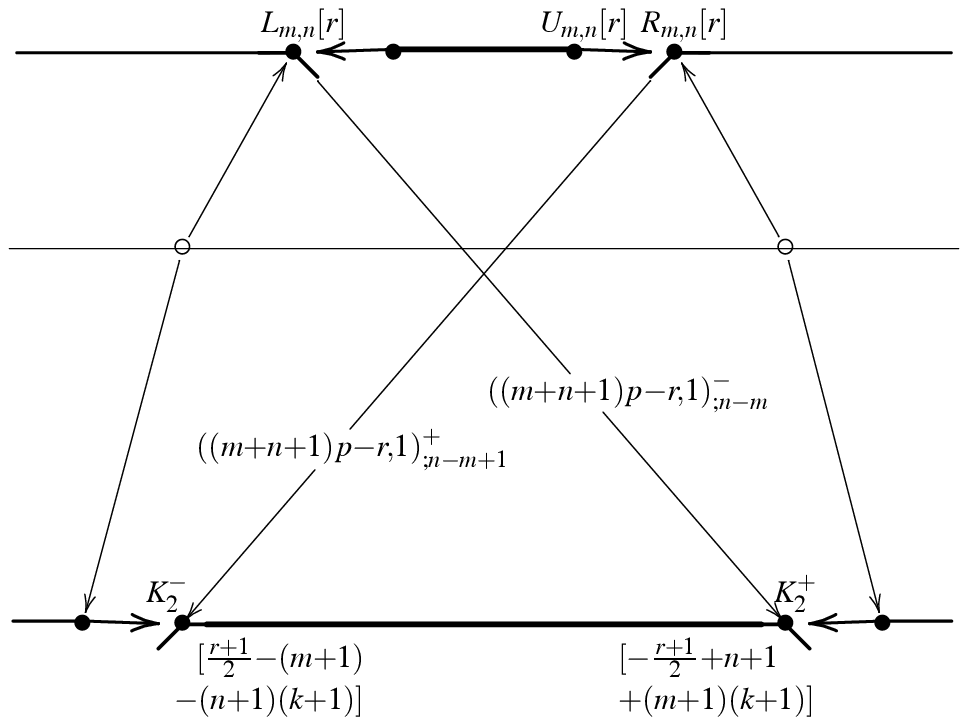}
  \caption{\small In the left-wing module $\Uwak{r}{m,n}$, the first
    embedding level where $\kerker$ is nonzero.}
  \label{fig:Unext}
\end{figure}

In a ``linear combination'' of the notations used in
Eq.~\eqref{eq:Utop} and Figs.~\ref{fig:Verma} and~\ref{fig:TVerma},
the extremal diagram of $\Uwak{r}{m,n}$ and the structure within the
first several embedding levels are as shown in Fig~\ref{fig:Unext}.
(We do \textit{not} fully describe the structure of the first
embedding level.  There occurs a submodule in the kernel of $\Qminus$
and a submodule in the kernel of $\Qplus$, but the intersection of the
kernels is zero.)  As before, expressions in square brackets indicate
the $h$ parameter of the corresponding operators~\eqref{gen-vert}.
The arrows pointing at $K_2^+$ and $K_2^-$ from the respective
nearest-neighbor states indicate that there is a \textit{submodule}
generated from either of the operators at the target nodes of these
arrows,
\begin{multline}\label{K2}
  K_2^+=\polP^+_2\thevert{[-\frac{r+1}{2}+n+1+(m+1)(k+1)]\xiz
    +\frac{j}{k+2}(\psim + \psip) - n \psim - m \psip}\\
  =(\Jplus_{n-m-1})^{(m+n+1)(k+2)-r}
  \thevert{[j-n(k+1)-m]\xiz
    +\frac{j}{k+2}(\psim + \psip) - n\psim - m \psip}
\end{multline}
and
\begin{multline}\label{K21}
  K_2^-=\polP^-_2\thevert{[\frac{r+1}{2}-m-1-(n+1)(k+1)]\xiz
    +\frac{j}{k+2}(\psim + \psip) - n \psim - m \psip}\\
  =(\Jminus_{m-n-1})^{(m+n+1)(k+2)-r}
  \thevert{[n+m(k+1)-j]\xiz
    +\frac{j}{k+2}(\psim + \psip) - n \psim - m \psip},
\end{multline}
where $\polP^\pm_2$ are differential polynomials of degree
$(k+2)(m+n+1)-r$.\pagebreak[3] \ This submodule is in the intersection
of the kernels $\kerker$.  \ Moreover, at each next embedding level,
$\kerker$ is generated by the corresponding operators, which for even
embedding levels are given by
\begin{align}\label{K2i}
  K^+_{2i}&=\polP^+_{2i}\thevert{[-\frac{r+1}{2}+(n+i)+(m+i)(k+1)]\xiz
    +\frac{j}{k+2}(\psim + \psip) - n \psim - m \psip},\\
  K^-_{2i}&=\polP^-_{2i}\thevert{[\frac{r+1}{2}-(m+i)-(n+i)(k+1)]\xiz
    +\frac{j}{k+2}(\psim + \psip) - n \psim - m \psip}.
  \label{K21i}
\end{align}

We actually need the socle $\soc\Uwak{r}{m,n}$ of $\Uwak{r}{m,n}$ and
the kernel
\begin{equation}\label{repK}
  \repK_{m,n}[r]=\kerker\bigr|_{\soc\Uwak{r}{m,n}}
  =\bigoplus_{b\geq1}\repK_{2b}.
\end{equation}
In what follows, we need the characters of the kernels in the diagonal
case $m=n$; it then follows from~\eqref{V-even} that
\begin{multline}\label{LW-even}
  \charSL{\repK_{m,m}[r]}{}(q,z)=
  \sum_{b\geq1}\charSL{\repK_{2b}}{m,m}(q,z)=\\
  {}=
  \mfrac{1}{q^{\frac{1}{8}}\,\vartheta_{1,1}(q,z)}
  \Bigl[\sum_{a\geq m+1}\!\!(a-m)+\sum_{a\leq-m-1}\!\!(-a-m)\Bigr]
  q^{p(\frac{r}{2p}-a)^2}\bigl(z^{-\frac{r+1}{2}+a p}-z^{\frac{r-1}{2}-a p}\bigr).
\end{multline}

\subsubsection{Right wing}\label{sec:RW}
\begin{figure}[tbp]
  \centering  
  \includegraphics[bb=1.3in 7.2in 7.4in 10.4in, clip,
    scale=.9]{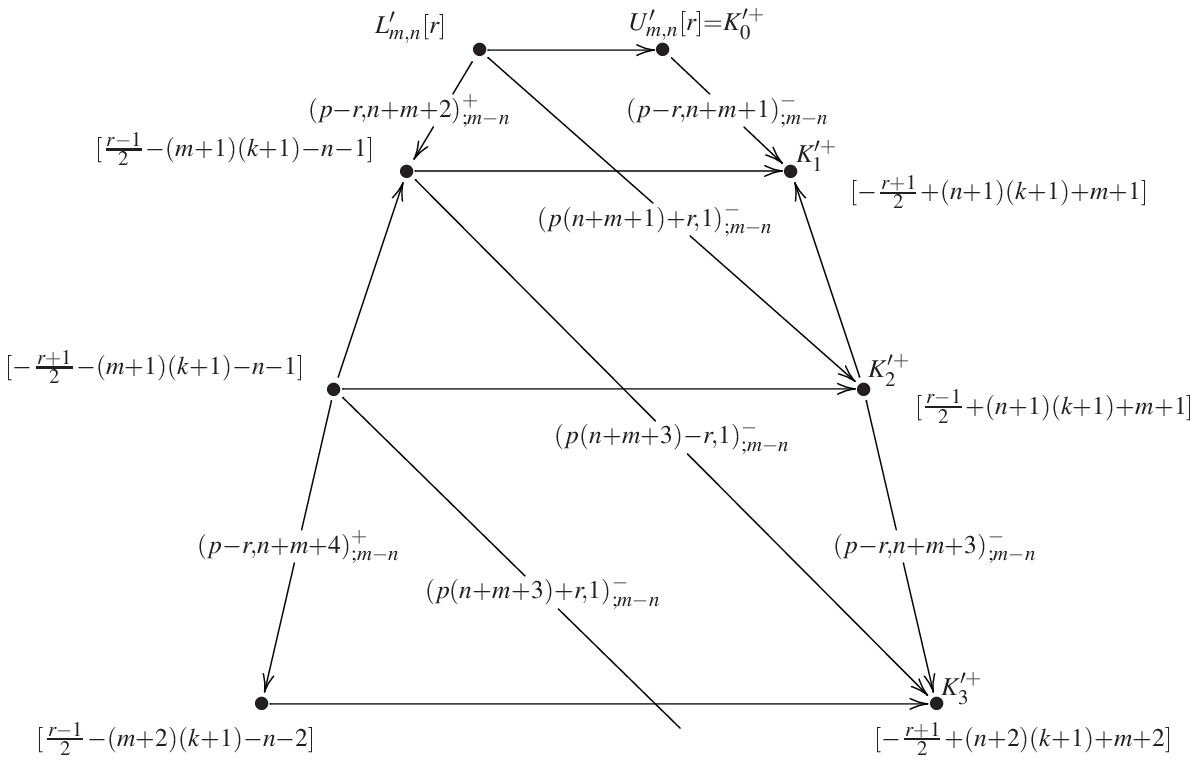}
  \caption{\small Subquotient structure of the right-wing twisted
    Wakimoto module $\pLwak{r}{m,n}$.}
  \label{fig:Lwak-right}
\end{figure}

The structure of the
right-wing modules is essentially dual to that of the left-wing
modules.  We recall the labeling in~\eqref{leftmn}.  For each pair of
nonnegative integers $m$ and $n$, the key operators in the extremal
diagram of the $(m,n)$th module are
\begin{align}
  \pLop{r}{m,n}(z) &= \Thevert{[-j-1-n-m(k+1)]\xiz
    + \frac{j}{k+2}(\psim + \psip)
    + n\psim + m\psip}{z},\\
  \label{pVop}
  \pVop{r}{m,n}(z) &=
  \Thevert{[-j-n-m(k+1)]\xiz+
    \frac{j}{k+2}(\psim + \psip) + n\psim + m \psip}{z}\\
  &{}\doteq\ket{\jplus(p-r,m+n+1);\,m-n+1},\notag\\
  \label{pUop}
  \pUop{r}{m,n}(z) &= \Thevert{[j+m+n(k+1)]\xiz
    + \frac{j}{k+2}(\psim + \psip)
    + n\psim + m\psip}{z}\\
  &{}\doteq\ket{\jplus(r+p(m+n),1);\, m\!-\!n},\notag
  \\
  \pRop{r}{m,n}(z) &= \Thevert{[j+n(k+1)+m+1]\xiz
    + \frac{j}{k+2}(\psim + \psip)
    + n\psim + m\psip}{z},
\end{align}
where we recall that $j=\frac{r-1}{2}$.  The extremal diagram (tilted
in accordance with the twist $m-n$) is ``turned inside out'' compared
with~\eqref{eq:Utop}:
\begin{equation}\label{eq:Utop-right}
  \includegraphics[bb=2in 9in 6.5in 10.3in, clip, 
    scale=.8]{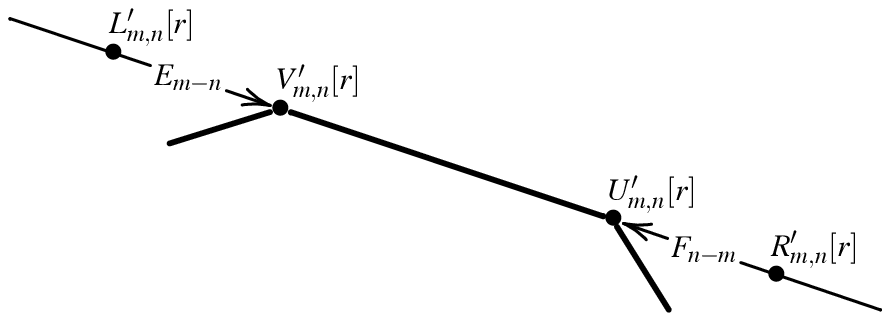}
\end{equation}
The spin of the twisted highest-weight state $\pUop{r}{m,n}$ is
$\jplus(r+p(n+m),1)
$.  The angles are drawn in accordance with the conventions in
Appendix~\bref{app:Verma}, to indicate twisted highest-weight states.
The submodule ``bordered by the two angles'' is in $\kerker$: as is
easy to see, there are regular OPEs
$e^{\psim\ldot\varphi(z)}\,\pUop{r}{m,n}(w)\propto(z-w)^{(m+n)p+r-1}$
and $e^{\psip\ldot\varphi(z)}\,\pUop{r}{m,n}(w)\propto(z-w)^0$, and
hence
\begin{equation*}
  \Qminus\pUop{r}{m,n}(w)=\Qplus\pUop{r}{m,n}(w)=0.
\end{equation*}
Next, we have $\Qplus\pRop{r}{m,n}(w)=\pUop{r}{m+1,n}(w)$ and,
similarly, $\Qminus\pLop{r}{m,n}(w)= \pVop{r}{m,n+1}(w)$, which is the
right-neighbor state of $\pLop{r}{m,n+1}(w)$, as shown
in~\eqref{eq:Utop-right}.  
It is then not difficult to consecutively trace the maps of the
lower-lying subquotients under both $\Qminus$ and $\Qplus$.  About
``half'' the subquotient structure is shown in
Fig.~\ref{fig:Lwak-right}.

It follows that $\kerker$ in $\pUwak{r}{m,n}$ is spanned by the
irreducible subquotients $\repK'_{b}$ generated from the states
labeled~$K'^+_b$, $b=0,1,2,\dots$, in Fig.~\ref{fig:Lwak-right}.
Their characters are readily found as (with the dependence on $m$ and
$n$ indicated as a subscript and the dependence on $r$ suppressed for
notational simplicity)
\begin{alignat}{2}\label{V-even-right}
  \charSL{\repK'_{2b}}{m,n}(q,z) =& \charSL{\repK_{2b+1}}{n,m}(q,z),&
  \quad &b\geq0,\\
  \charSL{\repK'_{2b-1}}{m,n}(q,z) =& \charSL{\repK_{2b}}{n,m}(q,z),&
  \quad &b\geq1
\end{alignat}
(see~\eqref{V-even}--\eqref{V-odd}).  In particular, setting $b=m=n=0$
in~\eqref{V-even-right} gives the character of the integrable
representation $\mI_{r}$ (see~\bref{sec:integrable})
\begin{equation}\label{integrable-char}
  \charSL{\repK'_0}{0,0}(q,z)=\charSL{}{r}(q,z).
\end{equation}

The character of the kernel
\begin{equation}\label{repKp}
  \repK'_{m,n}[r]=\kerker\bigr|_{\soc\pUwak{r}{m,n}}
\end{equation}
for each diagonal right-wing site $(m,m)$ is given by
\begin{multline}\label{RW-odd}
  \charSL{\repK'_{m,m}[r]}{}(q,z)=
  \sum_{b\geq1}\charSL{\repK'_{2b-1}}{m,m}(q,z)=\\
  =\mfrac{1}{q^{\frac{1}{8}}\,\vartheta_{1,1}(q,z)}
  \Bigl[\sum_{a\geq m+1}\!\!(a-m)+\sum_{a\leq-m-1}\!\!(-a-m)\Bigr]
  q^{(\frac{r}{2p}-a)^2}
  \bigl(z^{-\frac{r+1}{2}+ap}-z^{\frac{r-1}{2}-ap}\bigr).
\end{multline}

\subsubsection{The middle} The two-wing creature is obtained by
joining the two wings via the map
\begin{equation*}
  \Qplus\circ \Qminus:\Uwak{r}{1,1}\to\pUwak{r}{0,0}.
\end{equation*}
At the ``corner'' of the right wing, in the module $\pUwak{r}{0,0}$,
the $\pUop{r}{}$ operator~\eqref{pUop} is given by
\begin{equation*}
  \pUop{r}{0,0}(z)=
  e^{\left(j\xiz + \frac{j}{k+2}(\psim + \psip)\!\right)
    \ldot\varphi(z)}.
\end{equation*}
At the corner of the left wing, the corresponding operator is the one
in~\eqref{Uop} with $m=n=1$.  Evidently,
\begin{equation*}
  \Qminus\Uop{r}{1,1}(z)=
  e^{\left(\!(k + 1 - j)\xiz + \frac{j}{k+2}\psim
      + (\frac{j}{k+2} - 1)\psip\!\right) 
    \ldot\varphi(z)}
\end{equation*}
and it then follows from~\eqref{L-type-from-Qminus} that
\begin{equation*}
  (-1)^{p-r}(p-r)!\Qplus
  \Thevert{(k+1-j)\xiz+\frac{j}{k+2}(\psim+\psip)-\psip}{z}
  =(\Jplus_{-1})^{p-r}\,
  \Thevert{j\xiz+\frac{j}{k+2}(\psim+\psip)}{z}.
\end{equation*}
Thus,
\begin{equation*}
  \Qplus \Qminus\Uop{r}{1,1}(z) = \ffrac{1}{(p-r)!}
  \MFFminus{p-r,1;j}\in\pUwak{r}{0,0}(z).
\end{equation*}
The irreducible quotient generated from $\pUwak{r}{0,0}$ is in the
cohomology.  It is not difficult to trace the action of
$\Qplus\Qminus$ on the subquotients in~$\Uwak{r}{1,1}$.

\smallskip

The maps constructed above finally give the butterfly resolution of
integrable representations.

\subsection{Acyclic butterfly complexes}\label{partB}
We next consider the free-field modules whose elements are (the states
associated with) operators descendant from~\eqref{family} for
$2\lambda+1\in\{p+1,\dots,2p-1\}$ (about the second half of the range
in~\eqref{range}); we parameterize the $\lambda$~as
\begin{equation*}
  \lambda=\ffrac{p}{2} + \ffrac{r-1}{2},
\end{equation*}
with $r\in \{1,2,\dots,k+1\}$.  \textit{We keep the notation
  in~\eqref{eq:j}}.

The required modules can then be constructed by replacing $j\mapsto
j+\frac{p}{2}$ in the formulas in~\bref{partA}.  In accordance
with~\eqref{redundancy}, this is equivalent to the shifts $m\mapsto
m-\half$, $n\mapsto n-\half$ for each module in the left wing.
Consequently, we can describe each left-wing module just as in
Eqs.~\eqref{Uop}--\eqref{eq:Utop} and Figs.~\ref{fig:Lwak}
and~\ref{fig:Rwak}, but with
\begin{equation}\label{mn-NS-left}
  m,n\in\{\half,\ffrac{3}{2},\dots\}\quad \text{in the left wing}.
\end{equation}
Taking \textit{both} $m$ and $n$ half-integer leads to no conflict
because only $m+n$ has to be integer (see Figs.~\ref{fig:Lwak}
and~\ref{fig:Rwak}, where all the arrows (singular vectors) depend
only on $m+n$; in particular, the module at the ``corner'' of the left
wing is the one with $m=n=\half$ and hence $m+n=1$).
{}From~\eqref{V-even}, the character of
$\repK_{m,m}[r]=\kerker\bigr|_{\soc\Uwak{r}{m,m}}$ for half-integer
$m$ is given by
\begin{multline}\label{K-left-NS}
  \charW{\repK_{m,m}[r]}{}(q,z)
  =\sum_{b\geq1}\charSL{\repK_{2b}}{m,m}(q,z)\\
  {}=\mfrac{1}{q^{\frac{1}{8}}\,\vartheta_{1,1}(q,z)}\,
  \Bigl[
  \sum_{a\geq\mu+1}\!(a-\mu) + \sum_{a\leq-\mu}\!(1-a-\mu)\Bigr]
  q^{p(\frac{r}{2p}-a+\half)^2}\\*
  {}\times\bigl(z^{-\frac{r+1}{2}+(a-\half)p}
  -z^{\frac{r-1}{2}-(a-\half)p}
  \bigr),
\end{multline}
where $\mu=m+\half$ takes positive integer values in the left wing.

In the right wing, similarly, the modules corresponding to the spin
$\lambda=\frac{p}{2} + \frac{r-1}{2}$ can be described by formulas
in~\bref{sec:RW} with the shift $m\mapsto m+\half$, $n\mapsto
n+\half$, and hence
\begin{equation}\label{mn-NS-right}
  m,n\in\{\half,\ffrac{3}{2},\dots\}
  \quad \text{in the right wing}.
\end{equation}
The kernel $\kerker$ in the socle is given by the sum of
$\repK'_{2b-1}$ for $b\geq1$, and its character is easily expressed as
in~\eqref{K-left-NS}.

As the result of passing to half-integer $m$ and $n$, the resolution
becomes acyclic.\footnote{It is difficult to resist invoking a
  superficial analogy and referring to this case as a Neveu--Schwartz
  one (recall that in the previous case, the cohomology occurred in
  the ``zeroth'' module, which is now absent because of
  half-integer-valued labels).}  It turns out that
$\pUop{r}{\half,\half}$ is now in the image of $\Qminus\circ\Qplus$:
in the module $\Uwak{r}{\half,\half}$, we consider the states at the
level~$r$ relative to the top and at the grades
$-\frac{r+1}{2}-\frac{p}{2}+1$, \dots,
$\frac{r+1}{2}+\frac{p}{2}-1$.  In superimposing the pictures for
$\Lwak{r}{m,m}$ in Fig.~\ref{fig:Lwak} and $\Rwak{r}{m,m}$ in
Fig.~\ref{fig:Rwak} as explained in~\bref{left-wing-A}, these are the
states in between the nodes whose grades are underlined in
Figs.~\ref{fig:Lwak} and Fig.~\ref{fig:Rwak}, for $m=n=\half$.  A
codimension-$1$ submodule in this grade is in the kernel of
$\Qminus\circ\Qplus$, but the one-dimensional quotient is mapped onto
the states between (and including) $\pVop{r}{\half,\half}$ and
$\pUop{r}{\half,\half}$ in~\eqref{eq:Utop-right}.

\subsection{``Steinberg'' modules}\label{sec:steinberg}
\subsubsection{$\lambda=(k+1)/2$}\label{sec:r=p} This case corresponds
to setting $r=p$ in the operators considered in~\bref{partA}.  As a
result, each of the diagrams in Figs.~\ref{fig:Lwak}
and~\ref{fig:Rwak} collapses into a single embedding chain, in
accordance with the degenerations of the MFF singular vectors
discussed in~\bref{degeneration2}.  The details are quite standard,
and we omit them.  The kernel $\kerker$ is spanned by irreducible
subquotients whose highest-weight vectors have the charges
$n+b+m(k+1+b)+\frac{k+1}{2}$, $b\geq0$.  The character of the kernel
in the socle of the left wing is given by
\begin{multline}\label{charK-p}
  \charSL{\repK_{m,n}[p]}{}(q,z)=\\
  =\ffrac{(-1)^{n-m}
  }{
    q^{\frac{1}{8}}\vartheta_{1,1}(q,z)}
  \sum_{b\geq0}\Bigl(
  q^{p(m+b+\half)^2}z^{\frac{p-1}{2}+(m+b)p}
  -q^{p(n+b+\half)^2}z^{-\frac{p+1}{2}-(n+b)p}
  \Bigr).
\end{multline}
In the socle of the right wing, a somewhat different, but easily
reproducible subquotient structure results in the character of
$\kerker$ given by
\begin{multline}\label{charK-p-right}
  \charSL{\repK'_{m,n}[p]}{}(q,z)=\\
  =\ffrac{(-1)^{n-m}
  }{
    q^{\frac{1}{8}}\vartheta_{1,1}(q,z)}
  \sum_{b\geq0}
  \Bigl(
  q^{p(n+b+\half)^2}z^{\frac{p-1}{2}+(n+b)p}
  -q^{p(m+b+\half)^2}z^{-\frac{p+1}{2}-(m+b)p}
  \Bigr).
\end{multline}

It is worth seeing how the middle of the resolution restructures
compared with the case in~\bref{partA} (making the complex acyclic).
In the ``corner'' of the left wing, we have the operator
\begin{equation*}
  \Uop{p}{1,1}(w)=\Thevert{\frac{k+1}{2}\xiz
    +\frac{k+1}{2(k+2)}(\psim+\psip) - \psim - \psip}{w}.
\end{equation*}
It develops a first-order pole in the OPE with
$e^{\psim\ldot\varphi(u)}$ and the resulting operator, moreover, has a
first-order pole with $e^{\psip\ldot\varphi(z)}$; therefore,
\begin{equation*}
  \Qplus\,\Qminus\Uop{p}{1,1}(w)=
  \Thevert{\frac{k+1}{2}\xiz+\frac{k+1}{2(k+2)}(\psim+\psip)}{w}
  =\pUop{p}{1,1}(w).
\end{equation*}

\subsubsection{$\lambda=(2k+3)/2$}\label{sec:r=2p} For $r=2p$, the
structure of $\Uwak{p}{m,n}$ also degenerates; in particular, for
$m=n=1$, (the states corresponding to) the operators
\begin{equation*}
  \Lop{p}{1,1}(z)=\Thevert{-\half\xiz-\frac{1}{2p}(\psim + \psip)}{z}
  \quad\text{and}\quad
  \Rop{p}{1,1}(z)=\Thevert{\half\xiz-\frac{1}{2p}(\psim + \psip)}{z}
\end{equation*}
are ``facing each other,'' i.e., have no extremal states between them
in a picture similar to~\eqref{eq:Utop}.  We omit the details to avoid
further lengthening this already long section.

\section{$W$-algebra, its representations, characters, and modular
  transformations}
\label{sec:W-char}
The aim of this section is to establish the main result stated in the
Introduction.  In~\bref{sec:W-currents}, we first identify the
$W$-algebra generators in the centralizer of the screenings.
In~\bref{sec:W-reps}, we construct $2p$ $W$-algebra representations
$\repY^{\pm}_r$, $1\leq r\leq p$, evaluate their characters
$\charW{\pm}{r}(\tau,\nu)$, and establish their spectral-flow
transformation properties.  The spectral flow closes if $2p$ functions
$\minor^{\pm}_r(\tau,\nu)$, $1\leq r\leq p$, are added.  Modular
transformation properties are studied in~\bref{sec:modular}. \ Certain
combinations of the $\charW{\pm}{r}(\tau,\nu)$,
$\minor^{\pm}_r(\tau,\nu)$ and the integrable characters
$\charSL{}{r}(\tau,\nu)$, $1\leq r\leq p\,{-}\,1$, become parts of
multiplets, i.e., transform in representations of the form
$\oC^n\tensor\pi$, where $\pi$ is some $\SLiiZ$-representation and
$n=2$~and~$3$, where the $\oC^2$ and $\oC^3$ representations are
realized on polynomials in $\tau$ of respective degrees $1$ and $2$
(see~\bref{sec:C2-def} and~\bref{sec:C3-def}).  In addition, a certain
triangular structure emerges, with terms of the form $\nu$ times
``lower'' characters occurring in modular transformations of the
``higher'' characters.  The precise result is in
\hbox{Lemmas~\bref{lemma:mod1} and~\bref{lemma:mod2}}.

\subsection{The $W^{\pm}(z)$ currents}\label{sec:W-currents}
The $W$-algebra representing the symmetry of the model is defined as
the maximum local algebra acting in the kernel.  As discussed in the
Introduction, we somewhat restrict this definition by taking only the
generators that map between $\hSL2_k$ modules of the same twist.

\subsubsection{Locality and the vacuum representation} We first select
operators that are local with respect to all operators in the kernel.
The kernel $\kerker$ is spanned by operators of the general form
\begin{equation}\label{modes-exp}
  (\mathrm{modes})\cdot
  e^{(a\xiz+\frac{r-1}{2p}(\psim+\psip)+m\psim + n\psip)\ldot\varphi(z)}
\end{equation}
with integer~$a$, integer $r$, and \textit{simultaneously} integer or
half-integer $m$ and $n$, where $(\mathrm{modes})$ are differential
polynomials in the $\hSL2_k$ currents.  In the OPE
\begin{equation*}
  e^{a\xiz\ldot\varphi(z)}\,e^{(b\xiz + \frac{r-1}{2p}(\psim+\psip) +
    m\psim + n\psip)\ldot\varphi(w)}\propto(z-w)^{a(m-n)},
\end{equation*}
the exponent $a(m-n)$ is therefore always integer; noninteger
exponents thus occurs only in the OPEs
\begin{equation*}
  e^{\frac{r-1}{2p}(\psim+\psip)\ldot\varphi(z)}\,
  e^{\frac{r'-1}{2p}(\psim+\psip)\ldot\varphi(w)}
\propto(z-w)^{\frac{(r-1)(r'-1)}{2p}}
\end{equation*}
and in the OPEs involving $e^{(m \psim + n \psip)\ldot\varphi}$ in the
case of half-integer $m$ and~$n$.  It follows that the vertices as
in~\eqref{modes-exp} with $r=1$ and integer~$m$ and~$n$ produce
integer-valued exponents\,---\,no nonlocalities\,---\,in the OPEs with
all of the vertices encountered in the kernel (in checking the OPE
with $e^{(m' \psim + n' \psip)\ldot\varphi}$, it is essential that
$m'$ and $n'$ can only be half-integer simultaneously).  We therefore
identify the vacuum representation of the $W$-algebra with the kernel
$\kerker$ in the socle of the butterfly resolution of the $r=1$
integrable representation.  That is, the vacuum representation of the
$W$-algebra is given by
\begin{equation*}
  \bigoplus_{m\geq1}\repK_{m,m}[1]\oplus
  \bigoplus_{m\geq0}\repK'_{m,m}[1],
\end{equation*}
where $\repK_{m,n}[r]$ are defined in~\eqref{repK} and
$\repK'_{m,n}[r]$ in~\eqref{repKp}.

\subsubsection{$W^{\pm}(z)$ currents}
The fields $\Wright(z)$ and $\Wleft(z)$ generating the $W$-algebra are
associated with certain singular vectors as follows.

\begin{itemize}
\item $\Wright(z)$ corresponds to the singular vector
  $\MFFplus{p\,{-}\,1,3;1}$ constructed on the vertex
  $\pVop{1}{1,1}=e^{(-p\xiz + \psim + \psip)\ldot\varphi}$ (see
  \eqref{pVop} and~\eqref{eq:Utop-right}; from~\eqref{pVop}, this
  vertex represents the twisted highest-weight state
  $\ket{\jplus(p\,{-}\,1,3);1}$).\footnote{The $k=0$ example given
    below is already sufficiently generic, and may help visualize the
    positions of the various states.}  Therefore, $\Wright(z)$ is a
  ($\theta\!=\!1$)-twisted highest-weight operator of dimension
  $4p\,{-}\,2$ and charge $-2p\!+\!1$.  (Its \textit{spin} is
  $-2p\!+\!1+\frac{k}{2}=\jplus(1,4)=\jminus(4p\,{-}\,1,1)$.)  \ More
  explicitly,
  \begin{align}
    \label{eq:Wright}
    \Wright(z)&=
    (\Jminus_{-1})^{3 p - 1}(\Jplus_{0})^{2 p - 1}(\Jminus_{-1})^{p - 1}
    (\Jplus_{0})^{-1}(\Jminus_{-1})^{-p - 1}
    \Thevert{-p\xiz + \psim + \psip}{z}.
  \end{align}
  
\item $\Wleft(z)$ corresponds to the singular vector
  $\MFFminus{3p\,{-}\,1, 1}$ constructed on the vertex
  $\Lop{1}{1,1}=\thevert{-p\xiz -\psim -\psip}$ (see~\eqref{Lop}).
  This singular vector the $((m+n+1)p-r,1)^{\minus}_{;n-m}$ arrow in
  Fig.~\ref{fig:Lwak}, where we now set $m=n=1$ and $r=1$.  It follows
  that $\Wleft(z)$ is a highest-weight operator of the same dimension
  $4p\,{-}\,2$ and of the charge
  $\jplus(4p\,{-}\,1,1)=\jminus(1,4)=2p\,{-}\,1$:
  \begin{align}
    \label{eq:Wleft}
    \Wleft(z)&=(\Jplus_{-1})^{3 p - 1}
    \Thevert{-p\xiz  -\psim -\psip}{z}
    =\bigl(\dd^{3p-1}e^{\psip\ldot\varphi(z)}\bigr)
    \Thevert{(2p-1)\xiz - \psim - 2 \psip}{z}
  \end{align}
  (where we used~\eqref{Qminus-act} to evaluate a power of
  $\Jplus_{-1}$).
\end{itemize}

Once again, the meaning of~\eqref{eq:Wright} is that $\Wright(z)$ is
the operator whose corresponding state is given by the appropriate
singular vector (expressed as in the MFF formulas) evaluated on the
state corresponding to the vertex $\Thevert{-p\xiz + \psim +
  \psip}{z}$.

The OPE of the currents starts as
\begin{equation*}
  \Wleft(z)\,\Wright(w)=\ffrac{\mathscr{O}_{p-1}(w)}{(z-w)^{7p-3}}+\dots,
\end{equation*}
where $\mathscr{O}_{p-1}$ is the charge-$0$ dimension-$(p\,{-}\,1)$
operator at the first embedding level (the level of $K'^+_1$) in
Fig.~\ref{fig:Lwak-right} with $m=n=0$ and $r=1$: up to a nonzero
factor, therefore,
\begin{equation*}
  \mathscr{O}_{p-1}=\Jminus_{0}^{p - 1} \MFFminus{p - 1,1}
  =\Jminus_{0}^{p - 1} \Jplus_{-1}^{p - 1}\pUop{0,0}{1}
\end{equation*}
(where $\pUop{0,0}{1}(w)=\one$ is the unit operator). \ For $k=0$ and
$1$,
in particular,
\begin{align*}
  \mathscr{O}_{1}(z) &= 2\Jnaught(z)
  = \dd\varphim(z) - \dd\varphip(z),\\
  \mathscr{O}_{2}(z) &= -4 \Jplus\Jminus(z) + 8 \Jnaught\Jnaught(z)
  + 4 \dd\Jnaught(z)\\
  &= \dd\varphip\dd\varphip(z)
  - 4 \dd\varphip\dd\varphim(z) 
  + \dd\varphim\dd\varphim(z)
  + \dd^2\varphip(z) + \dd^2\varphim(z)
\end{align*}
up to nonzero factors.

\subsubsection{Example}\label{ex:k=0}\setcounter{figure}{1}%
For $k=0$, some details of the vacuum representation are shown in
Fig.~\thefigure. \
\afterpage{%
  \includegraphics[bb=.4in 7.8in 8.8in 10.3in, clip, angle=90]{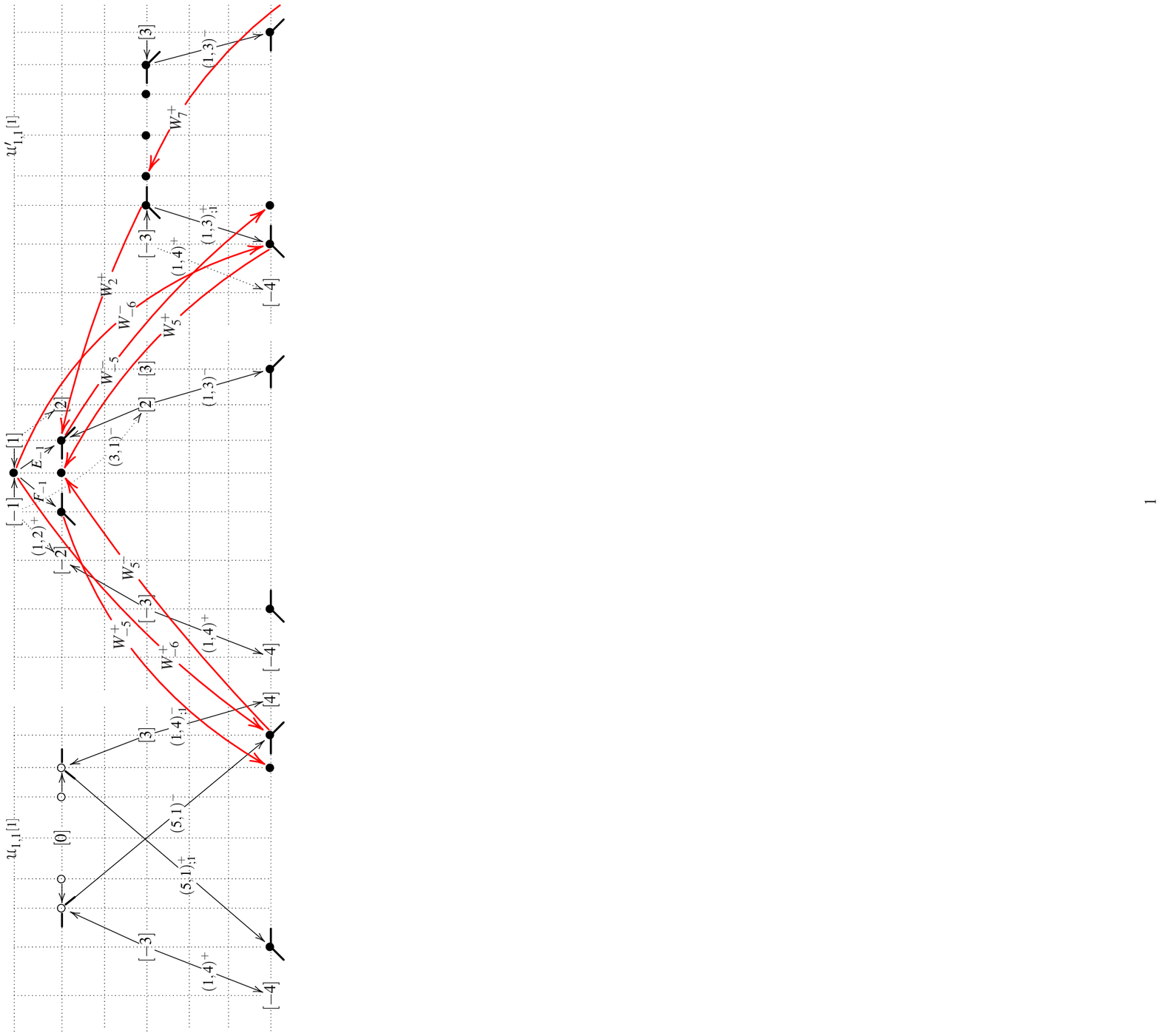}
  \qquad \rotatebox[origin=lB]{90}{\parbox{.88\textheight}{%
      \textsc{Figure}~\thefigure. \ \small Some details of the
      $W$-algebra vacuum representation at $k=0$.  Three copies of a
      two-dimensional lattice indicate the
      $(\mathrm{charge},\,\mathrm{dimension})$ bigrade.  The picture
      shows three modules: $\Uwak{1}{1,1}$ \textup{(}from the left
      wing of the butterfly\textup{)} and $\pUwak{1}{0,0}$ and
      $\pUwak{1}{1,1}$ \textup{(}from the right wing\textup{)}.  Some
      sites are shown with their \textit{charges} in square brackets.
      Boldfaced sites are in $\kerker$, but not all of them are in the
      socle: as the maps show, the extremal states of the right-hand
      module $\pUwak{1}{1,1}$ in the grades $-2,\dots,2$ \textup{(}the
      ``lid''\textup{)}, although in $\kerker$, are not in the socle,
      and hence not in the vacuum representation of the $W$-algebra
      (among the descendants of these states, only those in the
      submodule with extremal states in the grades $(-3,6)$, $\dots$,
      $(3,6)$ are in the vacuum representation).  The states
      corresponding to the $W^{\pm}(z)$ generators are at the grades
      $(\pm(2p\,{-}\,1),4p\,{-}\,2)$ \textup{(}in the current case
      where $k=0$, at $(3,6)$ for $\Wleft$ and $(-3,6)$ for
      $\Wright$\textup{)}; the $W^{\pm}_{-4p+2}=W^{\pm}_{-6}$ arrows
      map into them from the vacuum. ``Angles'' denote highest-weight
      state or twisted highest-weight states.  Tilted downward
      $(r',s')^{\pm}$-arrows show some of the nonvanishing singular
      vectors in the corresponding modules $\Lwak{1}{1,1}$,
      $\Rwak{1}{1,1}$, and $\pLwak{1}{m,m}$, $\pRwak{1}{m,m}$
      ($m=0,1$), as in Figs.~\ref{fig:Lwak}, \ref{fig:Rwak},
      and~\ref{fig:Lwak-right}.  The twist is additionally indicated
      with ${}_{;1}$.  Tilted upward arrows are the reverse of arrows
      that would lead to vanishing singular vectors.  In the middle
      module, the action of $\Jplus_{-1}$ and $\Jminus_{-1}$ on the
      vacuum \textup{(}the top middle state\textup{)} is also
      shown.
    }}%
  \addtocounter{figure}{1}\mbox{}\\
  \mbox{}\\
  \mbox{}}
As regards explicit expressions for the generators, with the negative
powers involved in $\Wright(z)$ understood in the standard MFF setting
in~\eqref{properties}, we find that the five factors
in~\eqref{eq:Wright} evaluate as
\begin{multline*}
  \Jminus_{-1}^{3 p - 1} \Jplus_{0}^{2 p - 1}  \Jminus_{-1}^{p - 1}  
  \Jplus_{0}^{-1}  \Jminus_{-1}^{-p - 1}=
  \Jminus_{-1}^{5} \Jplus_{0}^{3}  \Jminus_{-1}  
  \Jplus_{0}^{-1}  \Jminus_{-1}^{-3}\\
  =
  120 \Jminus_{-3} + 60 \Jnaught_{-2}  \Jminus_{-1} + 
  120 \Jnaught_{-1}  \Jminus_{-2} + 60 \Jnaught_{-1}^{2}  \Jminus_{-1} - 
  12 \Jplus_{-1}  \Jminus_{-1}^{2}
  + \Jplus_{0}^{2}  \Jminus_{-1}^{3}\\
  {}- 
  30 \Jplus_{0}  \Jminus_{-2}  \Jminus_{-1} - 
  18 \Jplus_{0}  \Jnaught_{-1}  \Jminus_{-1}^{2}.
\end{multline*}
The corresponding free-field expression is
\begin{multline*}
  \Wright(z)=
  \bigl(
  -6 \dd\varphim \dd^2\varphip(z) - 
  18 \dd^2\varphim \dd\varphim(z) - 
  12 \dd^2\varphim \dd\varphip(z)
  + 3 \dd\varphim \dd\varphim \dd\varphim(z) \\
  {}+ 
  6 \dd\varphim \dd\varphim \dd\varphip(z)
  + 3 \dd\varphim \dd\varphip \dd\varphip(z)
  + 12 \dd^3\varphim(z)
  \bigr)\Thevert{-3\xiz+\psim+\psip}{z}.
\end{multline*} 
{}From~\eqref{eq:Wleft}, we also have
\begin{multline*}
  \Wleft(z)=
  \bigl(
  10 \dd^3\varphip \dd^2\varphip(z)
  + 5 \dd^4\varphip \dd\varphip(z)
  + 15 \dd^2\varphip \dd^2\varphip \dd\varphip(z)\\
  {}+ 10 \dd^3\varphip \dd\varphip \dd\varphip(z)
  + 10 \dd^2\varphip \dd\varphip \dd\varphip \dd\varphip(z)
  + \dd\varphip \dd\varphip \dd\varphip \dd\varphip \dd\varphip(z)\\
  {}+ \dd^5\varphip(z) 
    \bigr)\Thevert{3\xiz-\psim-\psip}{z}.
\end{multline*}

\subsection{``Narrow'' $W$-algebra representations $\repY^{\pm}_r$ and
  their characters}\label{sec:W-reps}
First, the integrable $\hSL2_k$ representations $\mI_{r}$, $1\leq
r\leq p\,{-}\,1$, are $W$-algebra representations.  Next, the
resolutions in Sec.~\bref{sec:resolutions} allow constructing $2p$ \
$W$-algebra representations, denoted by $\repY^{\pm}_r$, $1\leq r\leq
p$, in what follows.  Their underlying vector spaces are the sums of
irreducible $\hSL2_k$ subquotients in the kernel $\kerker$ evaluated
in the socle of each module along the horizontal symmetry line of the
butterfly in~\eqref{eq:bfly2}.  That is, $\repY^+_r$ is the sum of
$\repK_{m,m}[r]$ in~\eqref{repK} and $\repK'_{m,m}[r]$
in~\eqref{repKp}:
\begin{align*}
  \repY^+_r &= \bigoplus_{m\geq1}\repK_{m,m}[r] \oplus
  \mathop{{\bigoplus}^{\ast}}\limits_{m\geq0}\repK'_{m,m}[r]\\
  \intertext{where the asterisk at the direct sum is a ``lazy
    notation'' to indicate that at $m=0$, the subquotient given by the
    integrable representation $\mI_{r}$ (see~\eqref{integrable-char})
    is to be omitted.  Similarly, $\repY^-_r$ is given by an analogous
    construction in the ``Neveu--Schwartz sector,'' with summations
    going over half-integer values:}  
  \repY^-_r &=
  \smash[b]{\bigoplus_{m\geq\half}}\repK_{m,m}[r] \oplus
  \smash[b]{\bigoplus_{m\geq\half}}\repK'_{m,m}[r].
\end{align*}

We next find the characters
\begin{equation*}
  \charW{\pm}{r}(q,z)=\charW{\repY^\pm_r}{}(q,z).
\end{equation*}
In what follows, we write $\theta_r$ for $\theta_{r,p}$ and similarly
for theta-function derivatives, and sometimes omit the theta-function
argument when it is just $(q,z)$ or, equivalently, $(\tau,\nu)$ (see
Appendix~\bref{app:theta} for the theta-function conventions).

\begin{lemma}\label{lemma:the-chars}
  The $W$-algebra characters $\charW{\pm}{r}(q,z)$ are given by
  \textup{(}see~\eqref{Omega} for $\Omega(q,z)$\textup{)}
  \begin{align*}
    \charW{+}{r}(q,z) &=\mfrac{1}{\Omega(q,z)}\biggl(
    \ffrac{r^2}{4p^2}\bigl(\theta_{-r}-\theta_{r}\bigr)
    +\ffrac{r}{p^2}\bigl(\theta'_{-r}+\theta'_{r}\bigr)
    +\ffrac{1}{p^2}\bigl(\theta''_{-r}-\theta''_{r}\bigr)
    \!\!\biggr),\\
    \charW{-}{r}(q,z) &=\mfrac{1}{\Omega(q,z)}\biggl(\!\!
    \Bigl(\ffrac{r^2}{4p^2} -
    \ffrac{1}{4}\Bigr)\bigl(\theta_{p-r}-\theta_{p+r}\bigr)
    {}+\ffrac{r}{p^2}\bigl(\theta'_{p-r}+\theta'_{p+r}\bigr)
    +\ffrac{1}{p^2}\bigl(\theta''_{p-r}-\theta''_{p+r}\bigr)
    \!\!\biggr)
  \end{align*}
  for $1\leq r\leq p\,{-}\,1$ \textup{(}with the theta-function
  arguments $(q,z)$ omitted in the right-hand sides\textup{)},
  and\pagebreak[3]
  \begin{equation*}
    \charW{+}{p}(q,z)
    = \frac{2\theta'_{p}(q,z)
    }{p\,\Omega(q,z)},
    \qquad
    \charW{-}{p}(q,z)
    =\frac{2\theta'_{0}(q,z)
    }{p\,\Omega(q,z)}.
  \end{equation*}
\end{lemma}
This readily follows by direct calculation: for $1\leq r\leq
p\,{-}\,1$, the sum of characters in~\eqref{LW-even}
and~\eqref{RW-odd} yields
\begin{multline*}
  \charW{+}{r}(q,z)
  =\sum_{m\geq1}\charSL{\repK_{m,m}[r]}{}(q,z) +
  \mathop{{\sum}^{\ast}}\limits_{m\geq0}\charSL{\repK'_{m,m}[r]}{}(q,z)\\
  {}=\mfrac{1}{q^{\frac{1}{8}}\,\vartheta_{1,1}(q,z)} \sum_{a\in\oZ}
  a^2q^{p(\frac{r}{2p}+a)^2}\bigl(z^{-\frac{r+1}{2}-ap}
  -z^{\frac{r-1}{2}+ap}\bigr),
\end{multline*}
where the asterisk affects the $m=0$ term just as above.  For
$\repY^-_r$, no term is excluded at $m=0$, and summation over
half-integer values of $m$ gives
\begin{multline*}
  \charW{-}{r}(q,z)
  =\sum_{m\geq\half}\charSL{\repK_{m,m}[r]}{}(q,z)
  +
  \sum_{m\geq\half}\charSL{\repK'_{m,m}[r]}{}(q,z)
  =2\sum_{m\geq\half}\charSL{\repK'_{m,m}[r]}{}(q,z)
  =\\
  {}=\mfrac{1}{q^{\frac{1}{8}}\,\vartheta_{1,1}(q,z)}
  \sum_{a\in\oZ+\half}
  (a^2 - \ffrac{1}{4})q^{p(\frac{r}{2p}+a)^2}\bigl(z^{-\frac{r+1}{2}-ap}
  -z^{\frac{r-1}{2}+ap}\bigr).
\end{multline*}
In terms of theta-functions, this gives the above formulas.  For
$r=p$, we find from~\eqref{charK-p} and \eqref{charK-p-right} that
\begin{equation*}
  \charW{+}{p}(q,z)=
  \sum_{m\geq1}\charSL{\repK_{m,m}[p]}{}(q,z)
  +\sum_{m\geq0}\charSL{\repK'_{m,m}[p]}{}(q,z)
  {}=\frac{2\theta'_{p}(q,z)}{p\,\Omega(q,z)},
\end{equation*}
and a similar calculation leads to~$\charW{-}{p}(q,z)$.

\subsubsection{Spectral flow transformation properties}
We next study the spectral-flow orbit of the above characters
$\charW{\pm}{r}(q,z)$.  It follows
from~\eqref{theta-sf}--\eqref{theta''-sf} that their spectral flow
transformation properties are given by
\begin{align*}
  \charW{+}{r;1}(q,z)&=-\charW{-}{r}(q,z)
  -
  \minor^-_r(q,z) - \half\charSL{}{p-r}(q,z),\\
  \charW{-}{r;1}(q,z)&=-\charW{+}{r}(q,z)
  -
  \minor^+_r(q,z)
  \\[-4pt]
  \intertext{for $1\leq r\leq p\!-\!1$, and}
  \charW{+}{p;1}(q,z)&=-\charW{-}{p}(q,z)
  -
  \minor^-_p(q,z),\\
  \charW{-}{p;1}(q,z)&=-\charW{+}{p}(q,z)
  -
  \minor^+_p(q,z),
\end{align*}
where $\charSL{}{r}(q,z)$ are the integrable representation characters
and
\begin{align}
  \minor^+_r(q,z)
  &=\mfrac{1}{\Omega(q,z)}\Bigl(
  \ffrac{r}{2p}\bigl(\theta_{r}(q,z)+\theta_{-r}(q,z)\bigr)
  -
  \ffrac{1}{p}\bigl(\theta'_{r}(q,z)-\theta'_{-r}(q,z)\bigr)
  \!\Bigr),
  \notag\\[-.4\baselineskip]
  \mbox{}\label{omega-general}\\[-.5\baselineskip]
  \minor^-_r(q,z)
  &=\mfrac{1}{\Omega(q,z)}\Bigl(
  \ffrac{r}{2p}\bigl(\theta_{p-r}(q,z)+\theta_{r-p}(q,z)\bigr)
  -
  \ffrac{1}{p}\bigl(\theta'_{r-p}(q,z)-\theta'_{p-r}(q,z)\bigr)
  \!\Bigr)\notag
  \\
  \intertext{for $ 1\leq r\leq p\!-\!1$, and}
  \label{omega-boundary}
  \minor^+_p(q,z)&=
  \mfrac{\theta_{p}(q,z)}{\Omega(q,z)},\qquad
  \minor^-_p(q,z)=
  \mfrac{\theta_{0}(q,z)}{\Omega(q,z)}.
\end{align}

Defining the spectral-flow transformation rule for $\minor^{\pm}_r$ as
for characters (see~\eqref{spectral-sl2-general}), we further
calculate the transformation laws
\begin{align*}
  \minor^+_{r;1}(q,z)&=-\minor^-_r(q,z)
  -
  \half\charSL{}{p-r}(q,z),\\
  \minor^-_{r;1}(q,z)&=-\minor^+_r(q,z)
  +
  \half\charSL{}{r}(q,z)
\end{align*}
for $1\leq r\leq p\!-\!1$, while
$\minor^{\pm}_{p;1}(q,z)=\minor^{\pm}_{p}(q,z)$.  We do not detail the
representation-theory interpretation of the $\minor^{\pm}_r$ in this
paper.

\subsection{Modular transformation properties}\label{sec:modular} We
now evaluate the modular transformation properties of the $5p\,{-}\,1$
functions given by the above $W$-algebra characters $\charW{\pm}{r}$,
$1\leq r\leq p$, the integrable representation characters
$\charSL{}{r}$, $1\leq r\leq p\,{-}\,1$, and the $\minor^{\pm}_r$,
$1\leq r\leq p$.

\subsubsection{The ``minimal'' $\SLiiZ$-representation~$\Rmin$}
We first recall the well-known transformation formulas
\begin{align}
  \label{min-T-action}
  \charSL{}{r}(\tau+1,\nu)&=
  \lambda_{r,p}\charSL{}{r}(\tau,\nu),
  \qquad\lambda_{r,p}=e^{i\pi(\frac{r^2}{2p}-\frac{1}{4})},\\
  \label{min-S-action}
  \charSL{}{r}(-\ffrac{1}{\tau},\ffrac{\nu}{\tau})
  &=\sqrt{\ffrac{2}{p}}\, e^{i\pi
    k\frac{\nu^2}{2\tau}}\sum_{s=1}^{p-1}
  \sin\!\ffrac{\pi r s}{p}\,\charSL{}{s}(\tau,\nu),
\end{align}
which just state that the integrable representation characters
$\charSL{}{r}$ span a $(p\,{-}\,1)$-dimensional
$\SLiiZ$-represen\-tation $\Rmin$.

\begin{rem}
  Strictly speaking, to define $\Rmin$, we have to ``eliminate'' the
  $e^{i\pi k\frac{\nu^2}{2\tau}}$ factor in~\eqref{min-S-action}; this
  then gives the $\SLiiZ$-representation uniquely defined by the
  $T$-transformation as in~\eqref{min-T-action} and the
  $S$-transformation
  \begin{equation}\label{min-rep}
    S\,\charSL{}{r}
    =\sqrt{\ffrac{2}{p}}\,
    \sum_{s=1}^{p-1}
    \sin\!\ffrac{\pi r s}{p}\,\charSL{}{s}.
  \end{equation}
  The relation between~\eqref{min-S-action} and~\eqref{min-rep} can be
  understood in terms of an automorphy factor.  The argument (with
  some details omitted, see~\cite{Mumford} and
  also~\cite[Sec.~4.1]{[FHST]} is based on the fact that
  $j(\gamma;\tau,\nu)$ defined for $\gamma\,{=}\,\smatrix{a}{b}{c}{d}$
  as
  \begin{equation}\label{automorphy}
    j(\gamma;\tau,\nu)=
    \zeta_{c,d}^{-1}\,(c\tau{+}d)^{-\half}_{}\,
    e^{-i\pi\frac{c\nu^2}{c\tau + d}}_{},
  \end{equation}
  where $\zeta_{c,d}$ is a certain eighth root of
  unity~\cite{Mumford},
  satisfies the cocycle condition $j(\gamma\gamma';\tau,\nu)=
  j(\gamma';\tau,\nu)\, j(\gamma;\gamma'\tau,\gamma'\nu)$,
  $j(\one;\tau,\nu) = 1$.

  In what follows, we write the modular transformations as they follow
  from calculations for the characters, with the understanding that
  the $e^{i\pi k\frac{\nu^2}{2\tau}}$ factors are to be omitted when
  we speak of finite-dimensional $\SLiiZ$ representations.
  \textit{Matrix} automorphy factors are also used to unravel the
  structure of representations derived below.
\end{rem}

\begin{lemma}\label{lemma:mod1}
  The functions
  \begin{align}
    \minorplus_0(\tau,\nu)&=\minor^-_p(\tau,\nu),\notag\\
    \label{minorplus-funcs}
    \minorplus_r(\tau,\nu)&=\minor^+_r(\tau,\nu)+\minor^-_{p-r}(\tau,\nu),
    \quad 1\leq r\leq p\,{-}\,1,\\
    \minorplus_p(\tau,\nu)&=\minor^+_p(\tau,\nu)\notag
  \end{align}
  span a $(p\!+\!1)$-dimensional $\SLiiZ$-represen\-tation $\Rpi$:
  \begin{equation}\label{minorplus-S}
    \begin{aligned}
      \minorplus_r(\tau+1,\nu)
      &=\lambda_{r,p}\minorplus_r(\tau,\nu),\\
      \minorplus_r(-\ffrac{1}{\tau},\ffrac{\nu}{\tau})
      &=i\sqrt{\ffrac{2}{p}}\, e^{i\pi k\frac{\nu^2}{2\tau}} \biggl(
      \ffrac{1}{2}\minorplus_0(\tau,\nu)
      +\ffrac{(-1)^r}{2}\minorplus_p(\tau,\nu)
      +\sum_{s=1}^{p-1}\cos\!\ffrac{\pi r
        s}{p}\,\minorplus_s(\tau,\nu) \!\!\biggr)
    \end{aligned}
  \end{equation}
  for $0\leq r\leq p$.
  The functions
  \begin{equation}
    \label{skewminor-funcs}
    \begin{split}
      \skewminor_r(\tau,\nu) &= (p-r)\minor^+_r(\tau,\nu) -
      r\minor^-_{p-r}(\tau,\nu),\\
      \tauskewminor_r(\tau,\nu) &= \tau\skewminor_r(\tau,\nu),
    \end{split}
    \qquad 1\leq r\leq p\!-\!1,
  \end{equation}
  transform as
  \begin{align}\label{skewminor-T}
    \skewminor_r(\tau+1,\nu)&=\lambda_{r,p}
    \skewminor_r(\tau,\nu),
    \quad
    \tauskewminor_r(\tau+1,\nu)=\lambda_{r,p}
    \bigl(\tauskewminor_r(\tau,\nu) +\skewminor_r(\tau,\nu)\bigr),\\
    \label{skewminor-S}
    \skewminor_r(-\ffrac{1}{\tau},\ffrac{\nu}{\tau})
    &=\sqrt{\ffrac{2}{p}}\, e^{i\pi k\frac{\nu^2}{2\tau}}
    \sum_{s=1}^{p-1}\sin\!\ffrac{\pi r s}{p}
    \Bigl(\tauskewminor_s(\tau,\nu)
    - \ffrac{p\nu}{2}\charSL{}{s}(\tau,\nu)\Bigr),\\
    \tauskewminor_r(-\ffrac{1}{\tau},\ffrac{\nu}{\tau})
    &=\sqrt{\ffrac{2}{p}}\, e^{i\pi k\frac{\nu^2}{2\tau}}
    \sum_{s=1}^{p-1}\sin\!\ffrac{\pi r s}{p}
    \Bigl(-\skewminor_s(\tau,\nu)
    + \ffrac{p\nu}{2\tau}\charSL{}{s}(\tau,\nu)\Bigr).
  \end{align}  
\end{lemma}

This is shown by straightforward calculation based
on~\eqref{eq:theta-S}--\eqref{eq:theta''-S} for the
$S$-transfor\-mations.  For~$T$, the formulas are obvious.

For notational simplicity, we no longer use a special notation for
functions like $\tau\skewminor_r$ in~\eqref{skewminor-funcs}.  It must
be clear from~\bref{sec:C2-def} how the occurrence of $\tau$ give rise
to $\oC^2$ tensor factors in $\SLiiZ$-representations.

\subsubsection{$\SLiiZ$ representation structure: a deformed
  $\oC^2\tensor\Rmin$}\label{sec:2-orbit}
The admixture of $\nu$ times integrable representation characters
in~\eqref{skewminor-S} fits into the representation structure
described in~\bref{nu-C2}, with a direct sum of representations
deformed by a matrix automorphy factor.  The functions $\skewminor_r$,
$\tau\skewminor_r$, and $\charSL{}{r}$ are combined into a column
{\small$\begin{pmatrix}
    f(\tau)\skewminor\\
    \charSL{}{}
  \end{pmatrix}$}, where $f(\tau)$ is a polynomial of degree
${\leq}\,1$ and we omit the indices, letting $\skewminor$ and
$\charSL{}{}$ denote a vector in $\oC^{p-1}$ each.  The column of the
above form (read from bottom up) can therefore be considered an
element of $\oC^{p-1}\oplus\oC^2{\tensor}\,\oC^{p-1}$.  The
$\SLiiZ$-action defined by~\eqref{skewminor-T}--\eqref{skewminor-S}
(in the version where this is a \textit{right} action) differs from
that on $\oC^{p-1}\oplus\oC^2{\tensor}\,\oC^{p-1}
=\Rmin\oplus\oC^2{\tensor}\,\Rmin$ by a matrix automorphy factor: the
action is given by
\begin{equation}\label{2-orbit}
  \begin{pmatrix}
    f(\tau)\skewminor\\
    \charSL{}{}
  \end{pmatrix}\!\actedby\gamma
  =
  \begin{pmatrix}
    (c\tau+d)f(\gamma\tau)\skewminor\actedby\gamma
    + \beta\nu c f(\gamma\tau)\charSL{}{}\actedby\gamma\\
    \charSL{}{}\actedby\gamma
  \end{pmatrix},\qquad
  \beta=
  -
  \ffrac{p}{2},
\end{equation}
where  $\gamma= \mbox{\footnotesize$\displaystyle
  \begin{pmatrix}
  a&b\\
  c&d
\end{pmatrix}$}$ in the right-hand side acts on each $\oC^{p-1}$ as on
the integrable representation characters and $\gamma\tau$ is defined
in~\bref{sec:C1}.  At $\beta=0$ (or, formally, $\nu=0$), obviously,
the matrix automorphy factor becomes the identity matrix and the
transformation law in~\eqref{2-orbit} becomes that on
$\Rmin\oplus\oC^2{\tensor}\,\Rmin$.

In the next lemma, we encounter a $\tau^2\skewmajor_r$ and hence a
$\oC^3$ tensor factor (cf.~\bref{sec:C3-def}).

\begin{lemma}\label{lemma:mod2}
  The functions
  \begin{align*}
    \rho_0(\tau,\nu)&=\charW{-}{p}(\tau,\nu),\\
    \rho_r(\tau,\nu)&=\charW{+}{r}(\tau,\nu) + \charW{-}{p-r}(\tau,\nu)
    + \ffrac{r}{2p}\charSL{}{r}(\tau,\nu),\quad
    1\leq r\leq p\!-\!1,\\
    \rho_p(\tau,\nu)&=\charW{+}{p}(\tau,\nu)
  \end{align*}
  transform as
  \begin{align*}
    \rho_r(\tau+1,\nu)&=\lambda_{r,p}\rho_r(\tau,\nu),\\
    \rho_r(-\ffrac{1}{\tau},\ffrac{\nu}{\tau})
    &=i\sqrt{\ffrac{2}{p}}\,
    e^{i\pi k\frac{\nu^2}{2\tau}}
    \biggl(\ffrac{1}{2}(\tau\rho_0(\tau,\nu)
    +
    \nu\minorplus_0(\tau,\nu))
    +\ffrac{(-1)^r}{2}(\tau\rho_p(\tau,\nu)
    +
    \nu\minorplus_p(\tau,\nu))\\
    &\qquad\qquad\qquad{}+\sum_{s=1}^{p-1}
    \cos\!\ffrac{\pi r s}{p}\,
    \bigl(\tau\rho_s(\tau,\nu)
    +
    \nu \minorplus_s(\tau,\nu)\bigr)\!\!\biggr).
  \end{align*}
  
  The functions
  \begin{equation*}
    \skewmajor_r(\tau,\nu)
    =(p-r)\charW{+}{r}(\tau,\nu) - r\charW{-}{p-r}(\tau,\nu)
    -\Bigl(\ffrac{r^2}{4p}
    +\ffrac{1}{8i\pi\tau}
    \Bigr)\charSL{}{r}(\tau,\nu),
    \quad 1\leq r\leq p\!-\!1,
  \end{equation*}
  transform as
  \begin{align*}
    \skewmajor_r(\tau+1,\nu)&=\lambda_{r,p}\skewmajor_r(\tau,\nu),\\
    \skewmajor_r(-\ffrac{1}{\tau},\ffrac{\nu}{\tau})
    &=\sqrt{\ffrac{2}{p}}\,
    e^{i\pi k\frac{\nu^2}{2\tau}}
    \sum_{s=1}^{p-1}\sin\!\ffrac{\pi r s}{p}\,
    \Bigl(\tau^2\skewmajor_s(\tau,\nu)
    +
    \nu\tau\skewminor_s(\tau,\nu)
    - \ffrac{p\nu^2}{4}
    \charSL{}{s}(\tau,\nu)
    \Bigr).
  \end{align*}
\end{lemma}
This also follows by a direct calculation based
on~\eqref{eq:theta-S}--\eqref{eq:theta''-S} for~$S$.  For~$T$, apart
from the same eigenvalues $\lambda_{r,p}$, the transformations amount
to substituting $\tau\mapsto\tau+1$ in polynomials of degree not
greater than~$2$, which is not a difficult calculation.

\subsubsection{$\SLiiZ$-representation structure: deformed
  $\oC^2\tensor\Rpi$ and $\oC^3\tensor\Rmin$}\label{sec:3-orbit}
It follows from~\bref{lemma:mod2} and~\bref{nu-C2} that the $\rho$
(with suppressed indices, i.e., viewed as a vector in $\oC^{p+1}$)
transform under $\SLiiZ$ (again in the right-action version)
as\pagebreak[3]
\begin{equation}\label{2-orbit-next}
  \begin{pmatrix}
    f(\tau)\rho\\
    \minorplus
  \end{pmatrix}\!\actedby\gamma
  =
  \begin{pmatrix}
    (c\tau+d)f(\gamma\tau)\rho\actedby\gamma
    + \alpha\nu c f(\gamma\tau)\minorplus\actedby\gamma\\
    \minorplus\actedby\gamma
  \end{pmatrix},\qquad
  \alpha= 1,
\end{equation}
where $f(\tau)$ is a polynomial of degree ${\leq}\,1$ and in the
right-hand side the $\SLiiZ$ action on each $\oC^{p+1}$ is as on
$\Rpi$, see~\eqref{minorplus-S}.  This representation is also a
deformation of $\Rpi\oplus\oC^2{\tensor}\,\Rpi$ via a matrix
automorphy factor.

Similarly, it follows from~\bref{lemma:mod2} and~\bref{alpha-beta-C3}
that the $\SLiiZ$-action on the $\skewmajor$ (now viewed as a vector
from $\oC^{p-1}$ endowed with the $\SLiiZ$ representation isomorphic to
$\Rmin$) is a ``composition'' of the finite-dimensional
representations and an ``even larger'' matrix automorphy factor: the
transformations derived in the last lemma are equivalent to the
$\SLiiZ$ action given by
\begin{multline}\label{3-orbit}
  \begin{pmatrix}
    f(\tau)\skewmajor\\
    \skewminor\\
    \charSL{}{}
  \end{pmatrix}\!\actedby\gamma\\
  {}=
  \begin{pmatrix}
    (c\tau+d)^2f(\gamma\tau)\skewmajor\actedby\gamma
    + \alpha\nu\,c(c\tau+d)f(\gamma\tau)\skewminor\actedby\gamma
    + \frac{\alpha\beta}{2}\nu^2\,c^2
    f(\gamma\tau)\charSL{}{}\actedby\gamma
    \\
    \skewminor\actedby\gamma\\
    \charSL{}{}\actedby\gamma
  \end{pmatrix},
\end{multline}
where $f(\tau)$ is a polynomial of degree ${\leq}\,2$, in the
right-hand side $\SLiiZ$ acts on each $\oC^{p-1}$ as on $\Rmin$, and
\begin{equation*}
  \alpha= 1,\qquad \beta= -\frac{p}{2}
\end{equation*}
as above.  At zero values of $\alpha$ and $\beta$ we recover a direct
sum of finite-dimensional representations, the ``$\skewmajor$'' one
being $\oC^3\!\tensor\Rmin$.


\subsubsection{Example: $k=0$} For $k=0$ (and $9p\,{-}\,3=15$), the
$W$-algebra generators are given in~\bref{ex:k=0}.  We here have a
single integrable representation character $\charSL{}{1}=1$.  The
other $8+6=14$ generalized characters are a triplet $(\minorplus_0,
\minorplus_1, \minorplus_2)$, a ``$\tau$''-doublet $(\skewminor_1,
\tau\skewminor_1)$, a $\oC^2$ (due to $\tau$) tensored with another
triplet $(\rho_0, \rho_1,$\linebreak[0]$\rho_2)$, with $\nu\pi_r$
occurring in its $S$-transform, and, finally, a ``$\tau$''-triplet
$(\varphi_1, \tau\varphi_1,$\linebreak[0]$\tau^2\varphi_1)$, with both
$\skewminor_1$ and $\charSL{}{1}$ occurring in its $S$-transform.

We also note that the $c=0$ logarithmic model of $\hSL2_0$ is somewhat
``smaller in size'' than the celebrated $(3,2)$ logarithmic model with
the central charge $c=0$~\cite{[Cardy]}.  The logarithmic $(p=3,q=2)$
model involves $\half(p\,{-}\,1)(q\,{-}\,1)+2 p q=13$ irreducible
$W$-algebra representations, while the space of torus amplitudes has
(based on the modular-group argument) dimension
$\half(3p\,{-}\,1)(3q\,{-}\,1)$\linebreak[0]${}=20$~\cite{[FGST3]}.
For $\hSL2_0$, these two numbers seem to be $5p\,{-}\,1=9$ and
$9p\,{-}\,3=15$ respectively.

\section{Hamiltonian reduction to the $W$-algebra of the $(p,1)$
  model}
\label{sec:HRR}
In this section, we apply the Hamiltonian reduction functor to the
$W$-algebra of the logarithmic $\hSL2_k$ model.\pagebreak[3] We show
that the $W$-generators~\eqref{eq:Wright} and~\eqref{eq:Wleft} reduce
to generators of the triplet $W$-algebra~\cite{[K-first],[GK2],[GK3]}
of the $(p,1)$ logarithmic model, which were defined in~\cite{[FHST]}
in terms of a Virasoro screening.

It is well-known that the result of Hamiltonian reduction of the
$\hSL2_k$ algebra itself is the Virasoro algebra
\begin{equation}\label{eq:TT}
  T(z)\,T(w)=\ffrac{d/2}{(z\!-\!w)^4} + \ffrac{2T(w)}{(z\!-\!w)^2} +
  \ffrac{\dd T(w)}{z\!-\!w}
\end{equation}
with central charge
\begin{equation}\label{eq:d(k)}
  d = 13 -\ffrac{6}{k\!+\!2} - 6(k\!+\!2).
\end{equation}

\begin{Thm}\addcontentsline{toc}{subsection}{\thesubsection.  \ \
    Reduction to the triplet $(p,1)$ algebra}\label{thm:HamRed}
  Hamiltonian reduction of the $W$-algebra generated by the
  currents~\eqref{eq:Wright} and~\eqref{eq:Wleft} is the triplet
  $W$-algebra of the $(p\!=\!k\!+\!2,1)$ logarithmic model.
\end{Thm}

\subsection{Construction of the ``inverse reduction''} To find how the
$W$-algebra representations are reduced, it is useful to recall how
the $\hSL2$ currents are \textit{reconstructed} from the reduction
result~$T(z)$.

\begin{lemma}[\cite{[S-inv]}]\label{lemma:invertHR}
  Let $\phi$ and $\varphi$ be two scalar fields with the operator
  products
  \begin{equation*}
    \dd\varphi(z)\,\dd\varphi(w)=\ffrac{1}{(z\!-\!w)^2},\quad
    \dd\phi(z)\,\dd\phi(w)=\ffrac{-1}{(z\!-\!w)^2}
  \end{equation*}
  and let $T(z)$ satisfy~\eqref{eq:TT} with $d$ given
  by~\eqref{eq:d(k)}.  For $k\neq0$, the currents
  \begin{align}
    \Jplus(z) ={}& e^{\sqrt{\frac{2}{k}}(\varphi(z) - \phi(z))},
    \label{Bplus}
    \\
    \Jnaught(z) ={}& \sqrt{\ffrac{k}{2}}\,\dd\varphi(z),
    \label{Bnaught}
    \\
    \Jminus(z) ={}& \Bigl(\!  (k\!+\!2) T(z) - \ffrac{k}{2}\,\dd\phi
    \dd\phi(z) - \sqrt{\ffrac{k}{2}}\,(k\!+\!1) \dd^2\phi(z)\!\Bigr)
    e^{-\sqrt{\frac{2}{k}}(\varphi(z) - \phi(z))}
    \label{Bminus}
  \end{align}
  then satisfy the $\hSL2$ OPEs in~\eqref{sl2-OPE}.
\end{lemma}

We emphasize that no free-field representation is required of~$T(z)$.

This construction of $\Jplus(z)$, $\Jnaught(z)$, and $\Jminus(z)$ can
be considered an ``inversion'' of the Hamiltonian reduction starting
with its result, the energy-momentum tensor~$T(z)$.  The reduction of
any expression built out of the $\hSL2$ currents thus amounts to
simply expressing the currents in terms of $T(z)$, $\varphi(z)$, and
$\phi(z)$ (and, depending on one's taste, eventually setting the two
free fields equal to zero).  In particular, the Sugawara
energy-momentum tensor evaluates in accordance with this procedure~as
\begin{equation*}
  T_{\mathrm{Sug}}(z)
  = T(z) + \half\dd\varphi\dd\varphi(z) -
  \half\dd\phi\dd\phi(z) - \sqrt{\ffrac{k}{2}}\,\dd^2\phi(z).
\end{equation*}

\subsubsection{Remark} The above formulas do not allow the limit
$k\to0$.  But this can be considered an artifact of the choice of
scalar fields in~\eqref{Bplus}--\eqref{Bminus}.  Changing them
as\pagebreak[3]
\begin{equation*}
  \varphi=\ffrac{1}{\sqrt{2}}\bigl(\ffrac{2}{\sqrt{k}}\,\bar X
  + \ffrac{\sqrt{k}}{2}\,X\bigr),
  \quad
  \phi=\ffrac{1}{\sqrt{2}}\bigl(\ffrac{2}{\sqrt{k}}\,\bar X
  - \ffrac{\sqrt{k}}{2}\,X\bigr)
\end{equation*}
where the new fields $X$ and $\bar X$ have the OPEs
\begin{equation*}
  \dd X(z)\,\dd\bar X(w)=\ffrac{1}{(z-w)^2},
  \quad
  \dd X(z)\,\dd X(w)=\mathrm{reg},
  \quad
  \dd\bar X(z)\,\dd\bar X(w)=\mathrm{reg},
\end{equation*}
maps~\eqref{Bplus}--\eqref{Bminus} into a form where $k$ can be set
equal to zero, with the result
\begin{align*}
  \Jplus(z) &= e^{X(z)},\\
  \Jnaught(z) &= \dd\bar X(z),\\
  \Jminus(z) &= (2 T(z) - \dd\bar X \dd\bar X(z) - \dd^2\bar X(z))
  e^{-X(z)}
\end{align*}
for the $\hSL2_0$ currents.  We restrict ourself to the observation
that this simple change of variables suffices to resolve the apparent
$k=0$ problem, and proceed with~\eqref{Bplus}--\eqref{Bminus}, which
we prefer for essentially esthetical reasons.

\subsubsection{Reducing the representations}\label{sec:HRreps}
The simple recipe to ``invert'' the reduction extends to
representations.  A spin-$\lambda$ (nontwisted) $\hSL2$ highest-weight
state$/$operator $U_\lambda(z)$\linebreak[0]${}\doteq\ket{\lambda}$ is
obtained by ``dressing'' the Virasoro primary $\mathscr{V}_\delta(z)$
of dimension\footnote{Hereinafter, all Virasoro primaries and their
  dimensions are with respect to $T(z)$
  in~\eqref{eq:TT}--\eqref{eq:d(k)}.}
\begin{equation}\label{eq:delta}
  \delta=\ffrac{\lambda(\lambda+1)}{k+2}-\lambda
\end{equation}
as
\begin{equation*}
  U_\lambda(z)
  =\mathscr{V}_\delta(z)e^{\lambda\sqrt{\frac{2}{k}}(\varphi(z)-\phi(z))}.
\end{equation*}
Taking $\lambda=\jplus(r,s)$ (see~\bref{MFFthm}) corresponds to the
Virasoro dimension
\begin{equation*}
  \delta_{r,s}=
  \ffrac{\jplus(r, s)(\jplus(r, s) + 1)}{k + 2} - \jplus(r, s)
  =\ffrac{r^2 - 1}{4(k + 2)} + (k + 2)\ffrac{s^2 - 1}{4}
  + \ffrac{1 - r s}{2}.
\end{equation*}
For example, the spin-$\half$ highest-weight state is thus constructed
as $e^{\half\sqrt{\frac{2}{k}}(\varphi(z) -
  \phi(z))}\,\mathscr{V}_{[21]}(z)$, where
$\mathscr{V}_{[21]}(z)\equiv \mathscr{V}_{\delta_{2,1}}(z)$ is the
``21'' vertex operator for the Virasoro algebra of $T(z)$.  Moreover,
the differential equation $(k+2)\dd^2 \mathscr{V}_{[21]}(z) - T(z)
\mathscr{V}_{[21]}(z)=0$ satisfied by $\mathscr{V}_{[21]}(z)$ then
implies the singular vector vanishing
$\Jminus_{0}\Jminus_{0}(e^{\half\sqrt{\frac{2}{k}}(\varphi(z) -
  \phi(z))}\,\mathscr{V}_{[21]}(z))=0$.

\subsubsection{Expressing the $W^{\pm}(z)$ currents}
Hamiltonian reduction of the fields generating the $W$-algebra,
Eqs.~\eqref{eq:Wright} and~\eqref{eq:Wleft}, is particularly simple
for $\Wleft(z)$ because of the simple formula for the $\MFFminus{r',
  1}$ singular vectors, which, moreover, immediately evaluate
explicitly in realization~\eqref{Bplus}.  In accordance with the
construction of nontwisted highest-weight states in~\bref{sec:HRreps},
we have
\begin{equation*}
  \Lop{r}{m,m}(z)\Bigr|_{\eqref{Bplus}-\eqref{Bminus}}
  =
  \mathscr{V}_{[r,2m+1]}(z)
  e^{(j - (k + 2) m)\sqrt{\frac{2}{k}}(\varphi(z)-\phi(z))}
\end{equation*}
(recall notation~\eqref{eq:j});\pagebreak[3] therefore, in evaluating
$\Wleft(z)=(\Jplus_{-1})^{3 p - 1}\Lop{1}{1,1}(z)$
(see~\eqref{eq:Wleft}), we have $\Lop{1}{1,1}(z)$ expressed through
$\mathscr{V}_{[1,3]}(z)=\mathscr{V}_{2k+3}(z)$; next, it follows
from~\eqref{Bplus} that the action of $\Jplus_{-1}$ affects only the
free-field sector, and therefore
\begin{equation}\label{Wleft-reduced}
  \Wleft(z)\Bigr|_{\eqref{Bplus}-\eqref{Bminus}}
  =\mathscr{V}_{2k+3}(z)\,e^{(2k+3)\sqrt{\frac{2}{k}}(\varphi(z)-\phi(z))}.
\end{equation}
This identifies the Hamiltonian reduction of $\Wleft(z)$ as the
dimension-$(2p-1)$ Virasoro primary.

The Hamiltonian reduction of $\Wright(z)$ is evaluated explicitly to
within the ``explicitness'' of the MFF construction for singular
vectors.  With the $n=m$ operators~\eqref{pUop} represented in the
inverse Hamiltonian reduction setting as
\begin{equation*}
  \pUop{r}{m,m}(z)\Bigr|_{\eqref{Bplus}-\eqref{Bminus}}=
  \mathscr{V}_{[r+2 m p,1]}(z)
  e^{(j+(k+2)m)\sqrt{\frac{2}{k}}(\varphi(z)-\phi(z))},
\end{equation*}
we evaluate $\Wright(z)$ as a free-field vertex times a nonvanishing
Virasoro singular vector built on the dimension-$1$ primary
$\mathscr{V}_{[2p+1,1]}(z)=\mathscr{V}_{1}(z)$:
\begin{equation}\label{Wright-reduced}
  \Wright(z)\Bigr|_{\eqref{Bplus}-\eqref{Bminus}}
  =\bigl(D_{2k+2}(T)\mathscr{V}_{1}(z)\bigr)
  e^{-(2k+3)\sqrt{\frac{2}{k}}(\varphi(z)-\phi(z))},
\end{equation}
where $D_{2k+2}(T)\mathscr{V}_{1}(z)
=\alpha\dd^{2k+2}\mathscr{V}_{1}(z)+\dots+\gamma
T(z)^{k+1}\mathscr{V}_{1}(z)$ is a normal-ordered differential
polynomial in $T(z)$ and $\mathscr{V}_{1}(z)$, linear in
$\mathscr{V}_{1}(z)$, of the total degree $2k+2$, if we set $\deg
T(z)=2$, $\deg \mathscr{V}_{1}(z)=1$, and $\deg\dd=1$.  Clearly,
$\oint \mathscr{V}_{1}$ is one of the two Virasoro screenings.

We thus see from~\eqref{Wleft-reduced} and~\eqref{Wright-reduced} that
the two currents
\begin{equation}\label{w-reduced}
  \begin{split}
    w^+(z)&=\mathscr{V}_{2p-1}(z),\\
    w^-(z)&=D_{2p-2}(T)\mathscr{V}_{1}(z)
  \end{split}
\end{equation}
are the result of the Hamiltonian reduction of $\Wleft(z)$ and
$\Wright(z)$.  Moreover, their construction in terms of Virasoro
generators and vertices shows that they are the
dimension-$(2p\,{-}\,1)$ currents generating the $W$-algebra of the
$(p,1)$ logarithmic conformal field theory model, which were
constructed in~\cite{[FHST]} using a Virasoro screening operator.

\subsubsection{}\textit{If} we further use the bosonization
\begin{equation*}
  T(z)=\half\dd f(z)\dd f(z)
  +\ffrac{p-1}{\sqrt{2p}}\,\dd^2f(z)
\end{equation*}
for the energy-momentum tensor and
\begin{equation}\label{V-bosonized}
  \mathscr{V}_{[r,s]}(z)=e^{\sqrt{\frac{2}{p}}\jplus(r, s)f(z)}
\end{equation}
for the vertices in terms of a free field with the OPE $\dd f(z)\,\dd
f(w)=\ffrac{1}{(z-w)^2}$, then the  vertex operators 
in~\eqref{w-reduced} become
$\mathscr{V}_{[1,3]}(z)\equiv
\mathscr{V}_{2p-1}(z)=e^{-\sqrt{2p}f(z)}$ and
$\mathscr{V}_{[2p+1,1]}(z)\equiv
\mathscr{V}_{1}(z)=e^{\sqrt{2p}f(z)}$, as in the free-field
construction in~\cite{[FHST]}, with the screening given by~$\oint
e^{\sqrt{2p}f}$.

\subsubsection{Example}
For $k=1$, the differential polynomial in~\eqref{Wright-reduced} is
given by
\begin{multline}\label{p=3}
  D_4\mathscr{V}_{1}(z)=9\bigl(-60\,T(z)\dd^2\mathscr{V}_{1}(z) - 6\dd
  T(z)\, \dd \mathscr{V}_{1}(z)\\
  {}-18 \dd^2T(z)\,\mathscr{V}_{1}(z) + 64 T(z) T(z)\mathscr{V}_{1}(z)
  + 9\dd^4\mathscr{V}_{1}(z)\bigr)
\end{multline}
(with nested normal ordering from right to left, as usual).  In terms
of the bosonization in~\eqref{V-bosonized}, we then find
\begin{equation*}
  w^+(z)=e^{-\sqrt{6}f(z)}
\end{equation*}
and a straightforward evaluation of~\eqref{p=3} gives
\begin{multline*}
  w^-(z)
  =\bigl(3120\,\dd^2 f(z)\dd^2 f(z) - 
  1440\,\dd^3 f(z)\dd f(z) - 
  960\sqrt{6}\,\dd^2 f(z)\dd f(z)\dd f(z)\\*
  {}+ 1440\,\dd f(z)\dd f(z) \dd f(z)\dd f(z) + 
  40\sqrt{6}\,\dd^4 f(z)\bigr)e^{\sqrt{6}f(z)},
\end{multline*}
which is $-80$ times the $(p\!=\!3,1)$-model operator $\Wleft(z)$
in~\cite[Example 2.2.1]{[FHST]}.\footnote{The $\pm$ conventions seem
  to be particularly difficult to match.}

\subsection{``Reduction'' of the characters}\label{Hrchar}
We recall from~\cite{[MP]} that taking residues of integrable or
admissible $\hSL2$ characters at $z=q^n$ gives either zero or
Virasoro-representation characters (times a two-boson character), in
agreement with the Hamiltonian reduction (under which \textit{some}
irreducible $\hSL2$ representations map into the trivial Virasoro
representation).  This can be considered a ``Hamiltonian reduction''
at the level of characters.  For example, the integrable $\hSL2_k$
characters $\charSL{}{r}(q,z)$ are holomorphic functions of~$z$, and
hence have zero residues.

This observation extends to the logarithmic$/${}$W$-algebra realm as
follows.  We take the residues at $z=1$.  As just noted, the
integrable representation characters have zero residue at this point
in particular.  Next, for the $W$-algebra characters
in~\bref{lemma:the-chars}, we also have
\begin{equation*}
  \res_{z=1}\charW{\pm}{r}(q,z)=0,\qquad 1\leq r\leq p
\end{equation*}
(although these characters do have poles elsewhere, as is easy to
see).  On the other hand, the $\minor^{\pm}_r(q,z)$
in~\eqref{omega-general}--\eqref{omega-boundary} have nonvanishing
residues at $z=1$, yielding the $2p$ characters of the $(p,1)$-model
triplet $W$-algebra in~\cite{[FHST]}, times the factor $1/\eta(q)^2$
that accounts for the character of two free
bosons:
\begin{equation}\label{residues}
  \begin{split}
    \res_{z=1}\minor^{+}_r(q,z)&=\mfrac{1}{\eta(q)^2}\,
    \mfrac{
      r\theta_r(q)
      -
      2\theta'_r(q)}{p\,\eta(q)},\\
    \res_{z=1}\minor^{-}_r(q,z)&=\mfrac{1}{\eta(q)^2}\,
    \mfrac{
      r\theta_{p-r}(q)
      +
      2\theta'_{p-r}(q)}{p\,\eta(q)},
  \end{split}
\end{equation}
where the theta-constants in the right-hand sides are the
corresponding theta-functions at $z=1$.  The resulting characters are
indeed those in~\cite{[FHST]} (identification requires readjusting the
conventions for the theta-derivative and for the $\pm$ labeling).

Another interpretation of the above residue formulas can be given in
terms of \textit{specialization} of the $W$-algebra characters to
$z=1$.\footnote{I thank I.~Tipunin for the suggestion.}
That is (switching to the $\nu$-language), setting $\nu=0$ in the
modular transformation properties in~\bref{lemma:mod1}
and~\bref{lemma:mod2}, we have the well-defined specialized characters
\begin{alignat*}{2}
  \charSL{}{r}(\tau)&=\lim_{\nu\to0}\charSL{}{r}(\tau,\nu),&\quad
  &1\leq r\leq p-1,\\*
  \charW{\pm}{r}(\tau)&=\lim_{\nu\to0}\charW{\pm}{r}(\tau,\nu),&\quad
  &1\leq r\leq p.\\*
  \intertext{Taking the limit in the definitions in~\bref{lemma:mod2},
    we obtain the corresponding $\rho_r(\tau)$, $1\leq r\leq p+1$, and
    $\skewmajor_r(\tau)$, $1\leq r\leq p-1$. \ In addition, we define}
  \widehat\minor^{\pm}_r(\tau)
  &=\lim_{\nu\to0}\bigl(\nu\minor^{\pm}_r(\tau,\nu)\bigr), &\quad
  &1\leq r\leq p,
\end{alignat*}
which are just the residues in~\eqref{residues} times inessential
constant factors.  In accordance with the definitions
in~\bref{lemma:mod1}, this then gives the corresponding
$\widehat\minorplus_r(\tau)$, $1\leq r\leq p+1$, and
$\widehat\skewminor_r(\tau)$, $1\leq r\leq p-1$.

Remarkably, these definitions allow taking $\nu\to0$ in the modular
transformation formulas in~\bref{lemma:mod1} and~\bref{lemma:mod2}
(because the $\skewminor_r$ enter the right-hand sides of the
transformations in~\bref{lemma:mod2} only in the combination
$\nu\skewminor_r(\tau,\nu)$).  But the $S$-transform formulas for the
thus ``specialized'' characters acquire negative powers of $\tau$, for
example
\begin{align*}
  \widehat\minorplus_r(-\ffrac{1}{\tau}) &=i\sqrt{\ffrac{2}{p}}\,
  \biggl(\ffrac{1}{2\tau}\,\widehat\minorplus_0(\tau)
  +\ffrac{(-1)^r}{2\tau}\,\widehat\minorplus_p(\tau)
  +\ffrac{1}{\tau}
  \sum_{s=1}^{p-1}\cos\!\ffrac{\pi r s}{p}\,\widehat\minorplus_s(\tau)
  \!\!\biggr),
  \\
  \intertext{as well as implicit negative powers of $\tau$ in}
  \skewmajor_r(-\ffrac{1}{\tau}) &=\sqrt{\ffrac{2}{p}}\,
  \sum_{s=1}^{p-1}\sin\!\ffrac{\pi r s}{p}\,
  \Bigl(\tau^2\skewmajor_s(\tau)
  +
  \tau\widehat\skewminor_s(\tau)\!\Bigr),
\end{align*}
because $S$-transforming $\tau^2\skewmajor_r(\tau)$ gives rise to
$\tau^{-1}\widehat\skewminor_s(\tau)$ in the right-hand side.

Rewriting the modular transformation formulas in terms of the
$(p,1)$-model characters
$\widetilde\skewminor_s(\tau)=\eta(\tau)^2\widehat\skewminor_s(\tau)$
and
$\widetilde\minorplus_s(\tau)=\eta(\tau)^2\widehat\minorplus_s(\tau)$,
we, in particular, reproduce the $\SLiiZ$ representation
$\Rpi\oplus\oC^2\!\tensor\Rmin$ on these characters in~\cite{[FHST]}:
\begin{align*}
  \widetilde\minorplus_r(-\ffrac{1}{\tau}) &=i\sqrt{\ffrac{2}{p}}\,
  \biggl(\ffrac{1}{2}\,\widetilde\minorplus_0(\tau)
  +\ffrac{(-1)^r}{2}\,\widetilde\minorplus_p(\tau) +
  \sum_{s=1}^{p-1}\cos\!\ffrac{\pi r
    s}{p}\,\widetilde\minorplus_s(\tau) \!\!\biggr),
  \\
  \widetilde\skewminor_r(-\ffrac{1}{\tau}) &=\sqrt{\ffrac{2}{p}}\,
  \sum_{s=1}^{p-1}\sin\!\ffrac{\pi r s}{p}\,\tau
  \widetilde\skewminor_s(\tau).
\end{align*}
Transformations of the remaining specialized characters involve both
the $(p,1)$ characters and the free-boson characters $\eta(\tau)^{-2}$
in the right-hand sides:\pagebreak[3]
\begin{align*}
  \rho_r(-\ffrac{1}{\tau}) &=i\sqrt{\ffrac{2}{p}}\,
  \biggl(\ffrac{1}{2}\Bigl(\tau\rho_0(\tau) +
  \ffrac{\widetilde\minorplus_0(\tau)}{\eta(\tau)^2}\Bigr)
  +\ffrac{(-1)^r}{2}\Bigl(\tau\rho_p(\tau) +
  \ffrac{\widetilde\minorplus_p(\tau)}{\eta(\tau)^2}\Bigr)\\
  &\qquad\qquad\qquad\qquad\qquad\qquad{}
  +\sum_{s=1}^{p-1} \cos\!\ffrac{\pi r s}{p}\, \Bigl(\tau\rho_s(\tau)
  +
  \ffrac{\widetilde\minorplus_s(\tau)}{\eta(\tau)^2}\Bigr)\!\!\biggr),
  \\
  \skewmajor_r(-\ffrac{1}{\tau}) &=\sqrt{\ffrac{2}{p}}\,
  \sum_{s=1}^{p-1}\sin\!\ffrac{\pi r s}{p}\,
  \Bigl(\tau^2\skewmajor_s(\tau) +
  \ffrac{\tau\widetilde\skewminor_s(\tau)}{\eta(\tau)^2}\Bigr).
\end{align*}

\section{Further prospects}\label{sec:todo}
This section is essentially a todo list or, depending on one's
standpoint, a list of things that the author has failed to accomplish.
It may nevertheless be helpful in setting logarithmic $\hSL2$ models
in some new perspectives.

\subsection{\textit{Three} fermionic screenings} It would be
exceptionally interesting to see how the present construction
generalizes to the case with three fermionic screenings, which is
indeed exceptional because the $W$-algebra that commutes with three
fermionic screenings is a Hamiltonian reduction of the exceptional
affine Lie superalgebra $\widehat{D}(2|1;\alpha)$.  It is at the same
time (see~\cite{[FS-D]}) the algebra of the conformal field theory of
the coset
\begin{equation*}
  \mfrac{\hSL2_{k_1}\oplus\hSL2_{k_2}}{\hSL2_{k_1 + k_2}},
\end{equation*}
which makes it particularly interesting in applications.

\subsection{Kazhdan--Lusztig correspondence, the dual quantum group,
  and fusion}\label{sec:KLcorr}
We recall that the property of a chiral algebra and a quantum group
associated with the screenings to be each other's centralizers
underlies the Kazhdan--Lusztig correspondence~\cite{[KL]} between
representation categories of the chiral algebra and the quantum
group.\enlargethispage{\baselineskip}

Logarithmic conformal field theory models have nice quantum-group
counterparts, which capture at least part of the structure of
logarithmic models and are therefore quite useful in investigating
them~\cite{[FGST],[FGST2],[FGST-q]}.  On the quantum-group side, the
central role is played by two objects, the \textit{center} and the
Grothendieck ring.  In the known examples, the center carries the same
modular group representation as is realized on generalized characters
of the logarithmic model.\footnote{A decomposition of the
  $\SLiiZ$-representation on a quantum group center involving $\oC^n$
  tensor factors (actually, $\oC^2$) was first, to our knowledge,
  observed in~\cite{[Kerler]}.}  Elements of the quantum-group center
also translate into boundary states in boundary conformal field
theories~\cite{[GT]}.  The Kazhdan--Lusztig-dual quantum group in the
$\hSL2_k$ context\,---\,a quantum supergroup
$\overline{\mathscr{U}}_{\q}s\ell(2|1)$ at a root of unity\,---\,may
actually correspond to the logarithmic $\hSL2_k/u(1)$ theory, because
the number of screenings is then equal to the number of free fields.
Indications of a complicated structure of its center can already be
found in~\cite{[AAB]}.  For the center to carry a modular group action
at all, the quantum group must be ribbon and factorizable
(cf.~\cite{[FGST],[FGST-q]}); finding these structures, or at least
some analogues thereof, for a quantum \textit{super}group is a
separate, quite interesting task.  Hamiltonian reduction of
logarithmic conformal field theories in Sec.~\bref{sec:HRR} must also
have a counterpart for the respective Kazhdan--Lusztig-dual quantum
groups, or at least for their centers and the modular group actions on
them.

The previous experience with Kazhdan--Lusztig-dual quantum groups
shows that their Grothendieck rings give (a ``$K_0$-version'' of) the
fusion for the corresponding $W$-algebras
\cite{[FHST],[FGST],[FGST3],[EF]}.  Although this ``experimental law''
might require modification in the current case of the dual pair given
by the above $W$-algebra and $\overline{\mathscr{U}}_{\q} s\ell(2|1)$
at a root of unity, the Grothendieck ring of the dual quantum group is
certainly related to the $W$-algebra representation theory, being at
the same time an object that is much easier to evaluate.\footnote{We
  recently became aware of the results in~\cite{[Erd]}, which go far
  beyond the Grothendieck ring for the quantum groups closely related
  to those dual to the $(p,1)$ logarithmic models: tensor products of
  the indecomposable quantum group representations are evaluated
  there.  We thank K.~Erdmann for the communication.}

Another, rather speculative, inference from the (so far hypothetical)
Kazhdan--Lusztig correspondence with a quantum~$s\ell(2|1)$ is that
because~$s\ell(2|1)$ has typical (``wide'') and atypical (``narrow'')
representations, a similar picture, inasmuch as it survives
specializing to a root of unity and imposing constraints in the
quantum group, is to be expected for the $W$-algebra representations.
The $\repY^{\pm}_r$ representations constructed above are then
certainly the ``narrow'' ones.  The problem of
``typical$/$wide$/$massive'' representations and of the role they may
play is left for the future (their $\hSL2$ ``building blocks'' may
have the extremal diagrams shaped like those of \hbox{admissible or
  even relaxed representations,
  cf.~\cite{[LMRS]}}).

\subsection{More general models: ``triplet'' $W$-algebras instead of
  $\smash{\protect\hSL2}$}
For integer $k$, there must exist logarithmic models generalizing the
logarithmic $\hSL2_k$ differently than $\hSL2\to\hSL{n}$, which first
suggests itself.  The bosonization of $\hSL2_k$ in~\bref{bosonize} is
the $n=2$ case of a general pattern of algebras $\WW{n}(k)$, $n\geq1$,
constructed very similarly~\cite{[W2n]} (the $n\!=\!1$ case is merely
the $\beta\gamma$\ system and the $n\!=\!3$ case is the
Bershadsky--Polyakov algebra~\cite{[Pol],[Ber]}).  For $n>2$, the set
of $n+1$ vectors in $\oC^{n+1}$ generalizing the data $\xiz$, $\psim$,
$\psip$ in~\bref{sec:bosonization} is $\xiz$, $a_{n-2}$, \dots, $a_1$,
$\psim$, $\psip$ with the Gram matrix generalizing~\eqref{Gram2} as
\begin{equation*}
  \addtolength{\arraycolsep}{-2pt}
  \mbox{\small$\displaystyle
    \begin{pmatrix}
      0\rlap{\kern-70pt$\smash{\begin{array}[t]{r}
            \xiz\\{a}_{n-2}\\{a}_{n-3}\\ \\ \\ {a}_1\\
            \psim\\\psip
          \end{array}
          \rule[-109pt]{.5pt}{117pt}}$}&0&\hdotsfor{3}&0&1&-1\\
      0&2(k\,{+}\,n)
      &-k\,{-}\,n  &0   &\hdotsfor{3}&0 
      \\
      0&-k\,{-}\,n          &2(k\,{+}\,n)&-k\,{-}\,n&0&\hdotsfor{2}&0 
      \\
      \hdotsfor{8}\\
      \hdotsfor{8}\\
      0&0&\hdotsfor{1}&0     &-k\,{-}\,n&2(k\,{+}\,n)&-k\,{-}\,n &0 
      \\
      1&0&\hdotsfor{2}&0     &-k\,{-}\,n&1     &k\,{+}\,n\,{-}\,1 
      \\
      -1&0&\hdotsfor{3}&0&    k\,{+}\,n\,{-}\,1&1     
    \end{pmatrix}$}\kern-40pt
\end{equation*}
(the determinant is given by $-n(k\,{+}\,n)^{n-1}$).  These vectors
and an $(n\!+\!1)$-tuple of scalar fields~${\varphi}$ are used to
construct the screenings (with the dot denoting the Euclidean scalar
product in~$\oC^{n+1}$)
\begin{equation*}
  E_i=\oint e^{{a}_i\ldot{\varphi}},~i=1,\dots,n-2,
  \qquad
  \Qminus=\oint e^{\psim\ldot{\varphi}},\quad
  \Qplus=\oint e^{\psip\ldot{\varphi}}
\end{equation*}
representing the nilpotent subalgebra of $\Univ_q\SSL{n}{1}$.  The
centralizer of the screenings is a $W$-algebra $\WW{n}(k)$ generated
by two currents $E(z)$ and $F(z)$ with the OPEs
\begin{gather*}
  E(z)\,F(w)=\mfrac{\lambda_{n-1}(n,k)
  }{
    (z\,{-}\,w)^n}
  {}\,{+}\,\mfrac{
    n
    \lambda_{n-2}(n,k)
    \,H(w)}{(z\,{-}\,w)^{n-1}}
  +\mbox{\Large$\dots$},
\end{gather*}
where $\lambda_m(n,k)=
\prod_{i=1}^{m}\bigl(i(k\,{+}\,n\,{-}\,1)\,{-}\,1\bigr)$,
and, further,
\begin{gather*}
  H(z)\,E(w)=\mfrac{E(w)}{z\,{-}\,w},
  \quad
  H(z)\,F(w)=-\mfrac{F(w)}{z\,{-}\,w},
  \quad
  H(z)\,H(w)=
  \mfrac{\frac{n\,{-}\,1}{n}\,k\,{+}\,n\,{-}\,2
  }{(z\,{-}\,w)^2}.
\end{gather*}  
Quite an explicit construction of $E(z)$ and $F(z)$ is
available~\cite{[W2n]}.

The entries of the above Gram matrix are integer for integer $k$ and
the fermionic screenings commute for $k+n\geq 1$; ``integrable''
representations of $\WW{n}(k)$, mentioned in~\cite{[W2n]} (also
see~\cite{[FJM]}), are then a good starting point for the construction
of logarithmic models.

\subsection{Other logarithmic $\smash{\protect\hSL2_k}$?}
\label{sec:other-logs}
There are two aspects of ``other'' logarithmic $\hSL2_k$-models with
nonnegative integer~$k$.  First, there are various possibilities of
constructing ``essentially larger'' models, e.g., by taking the kernel
of only one screening.  Second, with just two screenings, an a priori
different logarithmic extension of $\hSL2_k$ conformal models with
integer $k\geq-1$ is possible based on the ``nonsymmetric''
bosonization of $\hSL2_k$ (with a fermionic and a bosonic screening).
It would be not entirely trivial if the results actually coincide with
those in this paper.

One more possibility is to ``bosonize'' $\hSL2_k$ just by the
construction in~\eqref{Bnaught} in terms of two free fields and an
energy-momentum tensor $T(z)$ that is \textit{not} represented through
a free field.  The screening$/$kernel machinery then involves the
screening~$S$ constructed in terms of the fields
in~\bref{lemma:invertHR} as~\cite{[FFHST]}
\begin{equation}\label{I-S}
  S = \oint e^{\sqrt{\frac{k}{2}}\phi}\,\mathscr{V}_{[12]},
\end{equation}
where $\mathscr{V}_{[12]}(z)$ is the ``12'' vertex operator for
$T(z)$, of dimension $\delta_{12}=\frac{3k}{4} + 1$.  The proof that
$S$ is a screening uses the differential equation $\dd^2
\mathscr{V}_{[12]}(z) - (k\!+\!2)\,T(z)\mathscr{V}_{[12]}(z){} = 0$
satisfied by~$\mathscr{V}_{[12]}(z)$.  It is readily seen that $S$ is
a \textit{fermionic} screening.  Moreover, like the Virasoro vertex
operator $\mathscr{V}_{[12]}$, it has two components that map
differently between Virasoro modules.  The construction of the
logarithmic $\hSL2_k$ model may then be repeated with this screening
action.

\subsection{Rational $k$} Constructing a logarithmic $\hSL2_k$ theory
for rational $k$ appears to be a more complicated (and certainly
bulkier) problem.\pagebreak[3] The bosonization
in~\eqref{bosonization} is still applicable for rational~$k$, and the
two fermionic screenings given by~\eqref{the-screenings} exist, but
the corresponding vertex operators $e^{\psi_{\pm}\ldot\varphi(z)}$ are
no longer local with respect to each other.  The extension to
rational~$k$ may turn out to be easier to achieve in the
``nonsymmetric'' bosonization of $\hSL2$, with one bosonic and one
fermionic screenings, but this task seems to be rather involved
anyway.  The ultimate logarithmic theory is then to include the
elegant constructions in~\cite{[G-alg],[LMRS]}.

\subsection{Projective modules} 
Taking the kernel of screenings and identifying the relevant
$W$-algebra and its irreducible representations is only the first step
in constructing (the chiral sector of) a logarithmic conformal field
theory model because \textit{extensions} of these representations must
be taken in order to obtain modules where the relevant generators
(e.g., $L_0$) act nonsemisimply.  The extensions are to be taken ``up
to the limit,'' which means constructing projective covers of
irreducible $W$-algebra modules.  The full space of states in a given
chiral sector is then the sum of all nonisomorphic indecomposable
projective modules.  This is just another major difference from the
semisimple$/$rational case, where the chiral space of states is
exhausted by irreducible representations.  At chosen fractional values
of $k$, constructions of some indecomposable $\hSL2$-modules were
given in~\cite{[G-alg],[LMRS]}. \ A complication to be encountered
with the $W$-algebra in this paper may be expected from the fact that
the modular group representation is infinite-dimensional before the
matrix automorphy factors are isolated, which may suggest some
pathologies in the $W$-algebra projective modules.

\subsubsection*{Acknowledgments} I am indebted to my ``logarithmic''
collaborators over the last several years, B.~Feigin, J.~Fuchs,
A.~Gainutdinov, S.~Hwang, and I.~Tipunin, without whom this paper
would not have been written and who were with me in spirit, if not in
person, while it was actually being written.  Special thanks go to
I.~Tipunin for sharing his insights with me.  I thank J.~Fuchs and
A.~Dosovitsky for valuable comments.  S.~Parkhomenko has shown to me
that applying spectral flow to the butterfly makes it fly.  This work
was (almost) finished at the Institute for Mathematical Sciences,
London, and was certainly influenced by a stimulating atmosphere at
the ``Themes in the interface of representation theory and physics''
conference; I am grateful to C.~Hull and P.~Martin for the warm
hospitality in London.  I am also grateful to A.~Isaev and P.~Pyatov
for the hospitality at the conference ``Classical and quantum
integrable systems'' in Dubna, where the paper was (ultimately)
finished.  \hbox{This paper was supported in part by the RFBR Grant
  07-01-00523}.

\appendix

\section{Embedding structure of Verma
  $\smash{\protect\widehat{s\ell}(2)_k}$-modules}\label{app:Verma} We
recall the embedding structure of some $\hSL2_k$ Verma modules with
positive integer $k+2$.  It is of course well known, and is given here
for convenience of reference; the reader may also find the picture of
a twisted Verma module useful in deciphering the figures in the main
body of the paper.

We use the notation \raisebox{6pt}{\
  \footnotesize$\xymatrix@=10pt{\verma}$} for the highest-weight state
of a Verma module, to make it reminiscent of the Verma module extremal
diagram: in accordance with the annihilation conditions
in~\eqref{twistedhw} for $\theta=0$, the $\hSL2$ operators act on the
highest-weight vector as
\begin{equation}\label{eq:vermahw}
  \includegraphics[bb=1.3in 9.1in 7.4in 10.4in, clip, scale=.65]{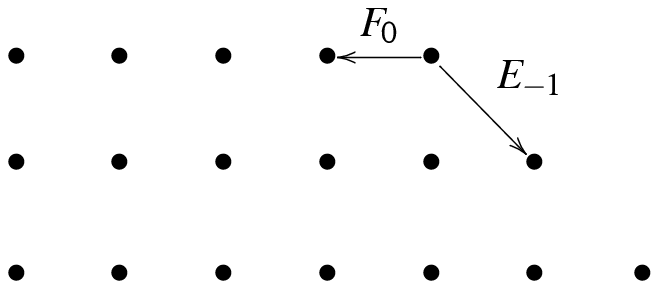}
\end{equation}
with the occupied states, shown with dots, extending from west to
south-east, but not east or north.  This notational convention is
naturally extended to $(\theta=1)$-twisted Verma modules, whose
extremal diagram has the form
\begin{equation}\label{eq:cvermahw}
  \includegraphics[bb=1.3in 9.1in 7.4in 10.4in, clip,
    scale=.65]{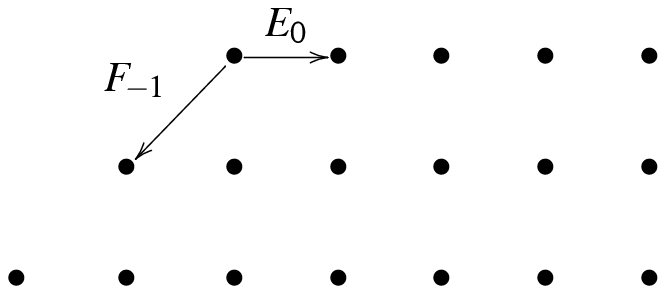}
\end{equation}
and we therefore let \raisebox{6pt}{\
  \footnotesize$\xymatrix@=10pt{\iverma}$} denote the twisted
highest-weight vector with twist~$1$.

\begin{figure}[tb]
  \centering
  \includegraphics[bb=1.3in 4.7in 7.6in 10.3in, clip,scale=.8]{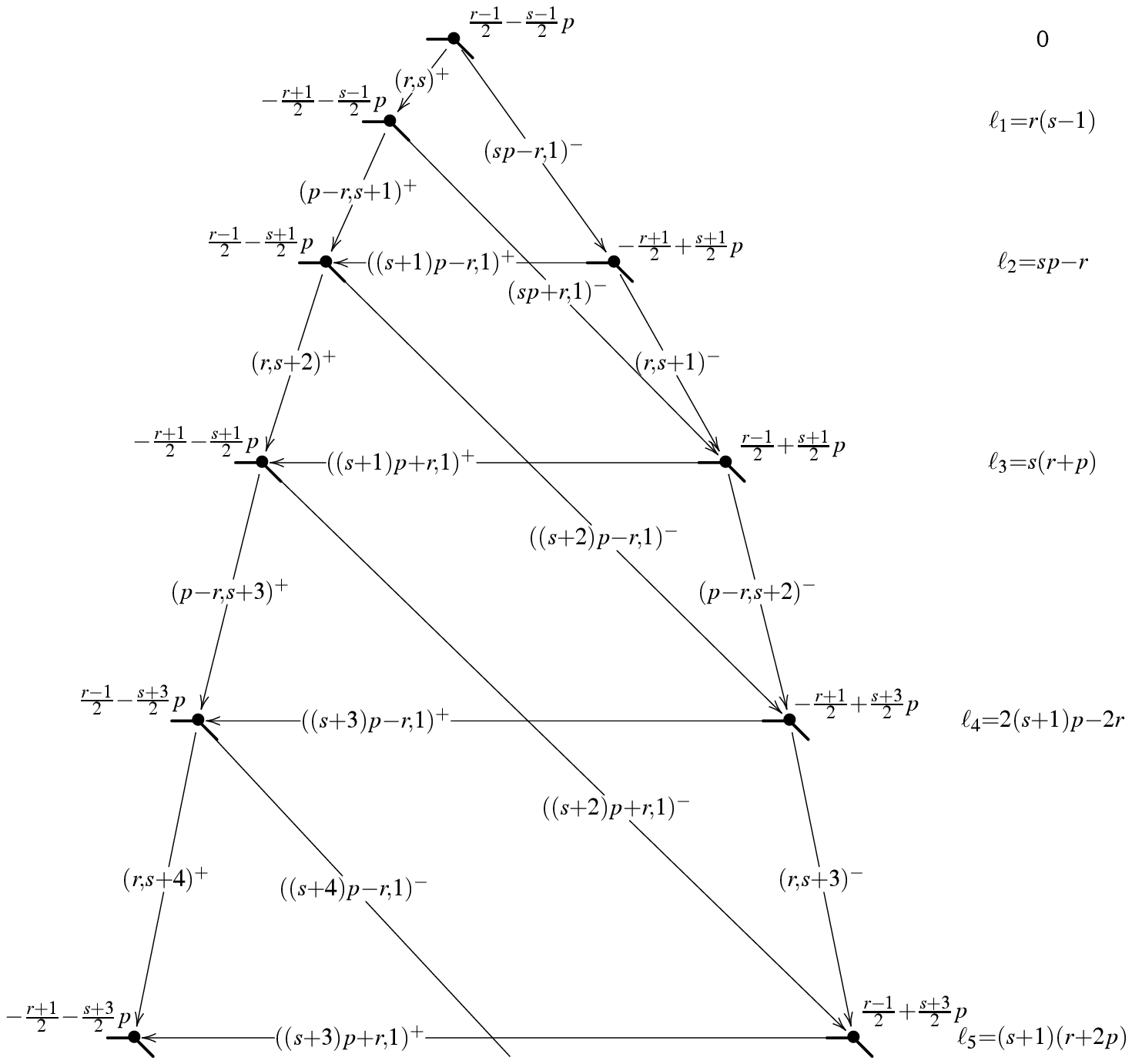}
  \caption{\small Embedding structure of the Verma module
    $\Verma_{\jplus(r,s)}$.}
  \label{fig:Verma}
\end{figure}

Let $p=k+2\in\{2,3,\dots\}$ and
$\lambda=\jplus(r,s)=\frac{r-1}{2}-p\frac{s-1}{2}$.  Submodules in the
Verma module $\Verma_{\lambda}$ are then arranged as in
Fig.~\ref{fig:Verma}.
Arrows labeled $(r',s')^\pm$ denote $\MFFpm{r',s'}$ singular vectors
(see~\bref{MFFthm}).  The labels at the Verma submodules (which are
represented by dots) show the spin of the respective highest-weight
vector.  The right column gives the relative level with respect to the
highest-weight vector of the module:
\begin{equation}\label{Verma-dimensions}
  \ell_i=
  \begin{cases}
    \ffrac{i}{2}\bigl((s + \ffrac{i}{2} - 1) p - r\bigr),
    &\text{ even } i,\\[6pt]
    \bigl(s + \ffrac{i\!-\!3}{2}\bigr)
    \bigl(r + \ffrac{i\!-\!1}{2}\,p\bigr),
  &\text{ odd } i.
  \end{cases}
\end{equation}

\begin{figure}[tb]
  \centering
  \includegraphics[bb=1.3in 5.1in 7.6in 10.3in, clip,
    scale=.8]{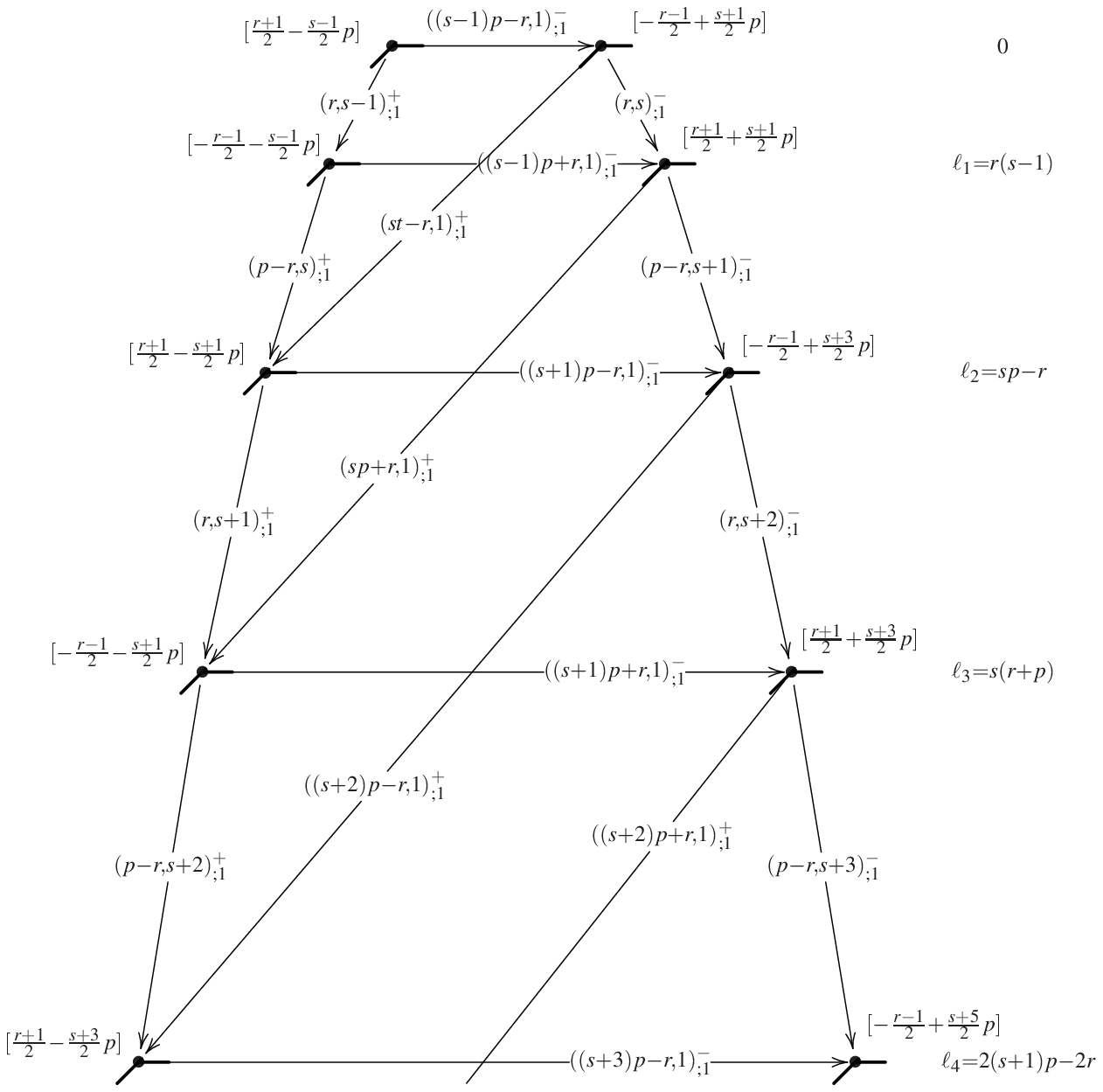}
    \caption{\small Embedding structure of the $(\theta=1)$-twisted
      Verma module $\Verma_{\jplus(r,s);1}$.  Labels in square
      brackets indicate \textit{charges} of the corresponding twisted
      highest-weight vectors.}
  \label{fig:TVerma}
\end{figure}

It may be instructive to consider the picture in Fig.~\ref{fig:Verma}
transformed by the spectral flow with $\theta=1$.  More precisely, we
take the twisted Verma module $\Verma_{\lambda+1+\frac{k}{2};1}$,
whose highest-weight vector is in the grade $(\lambda+1,\ell)$, where
$(\lambda,\ell)$ is the grade of the highest-weight vector of
$\Verma_{\lambda}$ (see~\eqref{twistedhw}).  For
$\lambda=\frac{r-1}{2}-p\frac{s-1}{2}$ as above, the embedding diagram
of $\Verma_{\lambda+1+\frac{k}{2};1}$ is shown in
Fig.~\ref{fig:TVerma},
where an arrow labeled $(a,b)^\pm_{;1}$ denotes the $\MFFpm{a,b;1}$
singular vector.  The labels at the twisted highest-weight vectors
\textit{indicate their charge, not spin} (i.e., the eigenvalue of
$\Jnaught_0$, not $\lambda$ in~\eqref{twistedhw}).

We leave it to the reader to place the two diagrams, of
$\Verma_{\lambda}$ and $\Verma_{\lambda+\frac{p}{2};1}$, into the
corresponding grades in the same picture and see how an extension of
one module by the other can then be constructed (the key is the
matching lengths of the horizontal arrows in the two diagrams).

We next recall the characters of some irreducible subquotients
occurring in Fig.~\ref{fig:Verma}.  Let $\mN_i$ denote the
\textit{right}-hand irreducible subquotient at the $i$th embedding
level (i.e., on the level $\ell_i$ relative to the top of the
diagram).  {}From the BGG resolution, its character follows as
\begin{multline} \label{Verma-char-even}
  \charSL{\mN_i}{}(q,z)=
  \frac{
    q^{\frac{r^2}{4p}-\frac{s-1}{2}r+\frac{(s-1)^2}{4}p}
  }{
    q^{\frac{1}{8}}\vartheta_{1,1}(q,z)}\\
  {}\times
  \Bigl(\sum_{a\geq0} + \!\!\sum_{a\leq -s-i-1}\Bigr)
  q^{\frac{i+2a}{2}\left((s-1+\frac{i+2a}{2})p-r\right)}
  \Bigl(z^{-\frac{r+1}{2}+\frac{s+i+2a-1}{2}p}
  -z^{\frac{r-1}{2}-\frac{s+i+2a-1}{2}p}
  \Bigr)
\end{multline}
for even $i$ and
\begin{multline} \label{Verma-char-odd}
  \charSL{\mN_i}{}(q,z)=
  \frac{
    q^{\frac{r^2}{4p}-\frac{s-1}{2}r+\frac{(s-1)^2}{4}p}
  }{
    q^{\frac{1}{8}}\vartheta_{1,1}(q,z)}\\
  {}\times
  \Bigl(\sum_{a\geq1} + \!\!\sum_{a\leq 1-s-i}\Bigr)
  q^{\frac{i+1+2a}{2}\left((s+\frac{i+2a-1}{2})p - r\right)}
  \Bigl(z^{\frac{r-1}{2}-\frac{s+i+2a}{2}p} -
  z^{-\frac{r+1}{2}+\frac{s+i+2a}{2}p}
  \Bigr)
\end{multline}
for odd $i$.

\section{Theta-function conventions}\label{app:theta}
\renewcommand{\theequation}{B.\arabic{equation}}%
The higher-level theta-functions are defined as
\begin{gather}\label{thetaKac}
  \theta_{r, p}(q,z)=\sum_{\iota\in\oZ+\frac{r}{2 p}}
  q^{p \iota^2}z^{p \iota}.
\end{gather}
We set
\begin{equation}
  \theta'_{r, p}(q,z)=z\ffrac{\dd}{\dd z}\,\theta_{r, p}(q,z),
  \quad
  \theta''_{r, p}(q,z)=\Bigl(z\ffrac{\dd}{\dd z}\Bigr)^2
  \theta_{r, p}(q,z).
\end{equation}
We also use the classic theta-functions
\begin{gather}\label{vartheta}
  \vartheta_{1,1}(q, z)
  =  \sum_{m\in\oZ}q^{\half(m^2 - m)} (-z)^{-m}
  = \prod_{m\geq0}\!(1\!-\!z^{-1} q^m)
  \prod_{m\geq1}\!(1\!-\!z q^m)\prod_{m\geq1}\!(1\!-\!q^m),\\
  \label{plain-theta}
  \vartheta(q,z)
  =\sum_{m\in\oZ}q^{\frac{m^2}{2}}z^m
\end{gather}
related to~\eqref{thetaKac} as
\begin{equation*}
  \theta_{r, p}(q,z)
  =z^{\frac{r}{2}}\,q^{\frac{r^2}{4 p}}\,
  \vartheta(q^{2 p},z^{p} q^r).
\end{equation*}

The quasiperiodicity properties of theta-functions are expressed as
\begin{align}\label{theta-sf}
  \theta_{r}(q,z q^n) &=
  q^{-p\frac{n^2}{4}}z^{-p\frac{n}{2}}\,
  \theta_{r+p n}(q,z),\\
  \intertext{with $\theta_{r + p n, p}(q,z)=\theta_{r}(q,z)$
    for \textit{even} $n$, and}
  \vartheta_{1,1}(q,z\,q^n) &=
  (-1)^n\,q^{-\half(n^2+n)}\,
  z^{-n}\vartheta_{1,1}(q,z),
  \qquad n\in\oZ.
\end{align}
It then follows that
\begin{align}\label{theta'-sf}
  \theta'_{r}(q,zq^n)&=q^{-p\frac{n^2}{4}}z^{-p\frac{n}{2}}
  \Bigl(\theta'_{r + pn, p}(q,z)
  - \ffrac{pn}{2}\theta_{r + pn}(q,z)\Bigr),\\
  \label{theta''-sf}
  \theta''_{r}(q,zq^n)&=q^{-p\frac{n^2}{4}}z^{-p\frac{n}{2}}
  \Bigl(\theta''_{r + pn, p}(q,z)
  - p n\theta'_{r + pn}(q,z)
  + \ffrac{p^2n^2}{4}\theta_{r + pn}(q,z)
  \Bigr).
\end{align}

We resort to the standard abuse by writing $f(\tau,\nu)$ for
$f(e^{2i\pi\tau},e^{2i\pi\nu})$; it is tacitly assumed that
$q=e^{2i\pi\tau}$ (with $\tau$ in the upper complex half-plane) and
$z=e^{2i\pi\nu}$.  

The modular $T$-transform of the theta-function is expressed as
\begin{align}
  \theta_{r, p}(\tau+1, \nu)
  &=e^{i\pi\frac{r^2}{2p}}\,\theta_{r, p}(\tau, \nu)\\
  \intertext{and the $S$-transform as}
  \label{eq:theta-S}
  \theta_{r, p}(-\ffrac{1}{\tau}, \ffrac{\nu}{\tau})
  &=e^{i\pi\frac{p\nu^2}{2\tau}}
  \sqrt{\ffrac{-i\tau}{2 p}}\sum_{s=0}^{2p-1}\!
  e^{-i\pi\frac{r s}{p}}\theta_{s}(\tau,\nu).\\
  \intertext{Therefore,}
  \label{eq:theta'-S}
  \theta'_{r, p}(-\ffrac{1}{\tau}, \ffrac{\nu}{\tau})
  &=e^{i\pi\frac{p\nu^2}{2\tau}} \sqrt{\ffrac{-i\tau}{2
      p}}\sum_{s=0}^{2p-1}\!  e^{-i\pi\frac{r s}{p}} \Bigl(
  \tau\theta'_{s}(\tau,\nu) + \ffrac{p\nu}{2}\,\theta_{s}(\tau,\nu)
  \Bigr)\\
  \intertext{(the price paid for abusing notation is that
    $\theta'_{r}(\tau,\nu)=\frac{1}{2i\pi}\frac{\dd}{\dd\nu}
    \theta_{r}(\tau,\nu)$) and}
  \label{eq:theta''-S}
  \theta''_{r, p}(-\ffrac{1}{\tau}, \ffrac{\nu}{\tau})
  &=e^{i\pi\frac{p\nu^2}{2\tau}}
  \sqrt{\ffrac{-i\tau}{2 p}}\sum_{s=0}^{2p-1}\!
  e^{-i\pi\frac{r s}{p}}
  \Bigl(
  \tau^2\theta''_{s}(\tau,\nu)
  + p\nu\tau\theta'_{s}(\tau,\nu)\\*
  &\qquad\qquad\qquad\qquad\qquad\qquad\qquad\quad
  {}+\bigl(\ffrac{p^2\nu^2}{4}+\ffrac{p\tau}{4i\pi}\bigr)
  \theta_{s}(\tau,\nu)
  \!\Bigr).\notag
\end{align}

We also note the formula
\begin{align*}
  \Omega(-\ffrac{1}{\tau},\ffrac{\nu}{\tau})
  &=-i\sqrt{-i\tau}\,e^{i\pi\frac{\nu^2}{2\tau}}\Omega(\tau,\nu)
\end{align*}
for the function
$\Omega(q,z)=q^{\frac{1}{8}}z^{\half}\vartheta_{1,1}(q,z)$.

The eta function
\begin{gather}\label{eta}
  \eta(q)  
  =q^{\frac{1}{24}}
  \smash{\prod\limits_{m=1}^\infty(1-q^{m})}
\end{gather}
transforms as 
\begin{gather}\label{modular-eta}
  \eta(\tau+1)=\smash{e^{\frac{i\pi}{12}}\eta(\tau),\qquad
    \eta(-\ffrac{1}{\tau})=
    \sqrt{-i\tau}\,\eta(\tau)}.
\end{gather}

In calculating residues in~\bref{Hrchar}, we also need the formula
\begin{equation*}
  \ffrac{\dd\vartheta_{1,1}(q,z)}{\dd z}\Bigr|_{z=q^n}=
  (-1)^{n}\,q^{-\frac{1}{8}}\eta(q)^3\,q^{-\frac{n^2}{2} -
    \frac{3n}{2}},
  \qquad n\in\oZ.
\end{equation*}

\section{Some elementary tricks with $\SLiiZ$ representations}
\label{app:SL2Z}
\renewcommand{\theequation}{C.\arabic{equation}}

\subsection{%
  $\smash{\protect\SLiiZ\curvearrowright\upperH\,{\times}\,\oC}$}
\label{sec:C1}
We first recall the standard $\SLiiZ$-action
on~$\upperH\,{\times}\,\oC$ (where $\upperH$ is the upper half-plane),
\begin{equation*}
  \gamma=
  \begin{pmatrix}
    a\;&b\\
    c\;&d
  \end{pmatrix}
  \!{:}\quad (\tau,\nu)\,\mapsto\,
  \Bigl(\ffrac{a\tau+b}{c\tau+d}, \ffrac{\nu}{c\tau+d}\Bigr).
\end{equation*}
The space $\mathcal{F}$ of suitable (e.g., meromorphic or just
fractional-linear in $\tau$) functions on $\upperH\,{\times}\,\oC$ is
then endowed with an $\SLiiZ$-action.

\subsection{$\oC^2$}\label{sec:C2}
\subsubsection{}\label{sec:C2-def}
The defining two-dimensional representation\,---\,\textit{the}
doublet\,---\,of $\SLiiZ$ is the representation where
\begin{equation*}
  S=
  \begin{pmatrix}
    0&-1\\
    1 & 0
  \end{pmatrix}
  \quad\text{and}\quad
  T=
  \begin{pmatrix}
    1&1\\
    0&1
  \end{pmatrix}
\end{equation*}
act on $\oC^2$ just by these matrices.  We choose a basis $\uone$,
$\utau$ in $\oC^2$ such that $S\utau=-\uone$, $S\uone=\utau$,
$T\utau=\utau+\uone$, and $T\uone=\uone$.

If $\pi$ is any finite-dimensional $\SLiiZ$-representation, with
$S_\pi=\Sact$ and $T_\pi=\Tact$ acting on vectors denoted by $\omega$,
then $\oC^2\!\tensor\pi$ is spanned by $\uone\omega$ and
$\utau\omega$, with the action
\begin{equation*}
  S(\utau\omega)=-\Sact\omega,\quad
  S(\uone\omega)=\utau\,\Sact\omega,\quad
  T(\utau\omega)=\utau\,\Tact\omega + \uone\,\Tact\omega,\quad
  T(\uone\omega)=\uone\,\Tact\omega.
\end{equation*}
This $\oC^2\!\tensor\pi$ representation can be realized using the
$\SLiiZ$ action on $\mathcal{F}$ as follows: we identify $\uone=1$ and
$\utau=\tau$, view $\omega$ as a function of $\tau$, which allows
considering $f(\tau)\omega$ with $f\in\mathcal{F}$, and redefine the
action of $S$ and $T$ as
\begin{equation}\label{ST-redef}
  S(f(\tau)\omega)=\tau\,f(-\ffrac{1}{\tau})
  \,\Sact\omega,\quad
  T(f(\tau)\omega)=f(\tau+1)\,\Tact\omega
\end{equation}
(in other words, $f$ is prescribed to transform with weight~$1$).
Indeed, these formulas immediately imply, e.g., that $S(\tau\omega)
=-\frac{1}{\tau}\,\tau\,\Sact\omega=-\Sact\omega$.\footnote{By
  $f(-\frac{1}{\tau})$, we everywhere mean
  $f(\tau)\Bigr|_{\tau\to-\frac{1}{\tau}}$, and similarly for
  $f(\tau+1)$. For example, the $\tau\to-\frac{1}{\tau}$ operation
  sends $f(\frac{\tau}{\tau+1})$ into $f(-\frac{1}{\tau-1})$.}
Wishing to deal with the more standard $c\tau+d$ (rather than
$-c\tau+a$ for the inverse matrix), we have to consider the
\textit{right} action, which is determined by~\eqref{ST-redef} as
\begin{equation*}
  \gamma:f(\tau)\omega
  \mapsto
  (c\tau+d)f(\gamma\tau)\,\omega\actedby\gamma,
  \qquad\gamma=
  \mbox{\footnotesize$\begin{pmatrix}
    a&b\\
    c&d
  \end{pmatrix}$}\in\SLiiZ.
\end{equation*}
Clearly, extending $f(\tau)$ beyond polynomials of degree $1$ yields
an infinite-di\-mensional $\SLiiZ$ representation on the space of
suitable functions times~$\pi$, containing $\oC^2\tensor\pi$ as a
subrepresentation.

\subsubsection{}\label{nu-C2}
If $\pi'$ is another $\SLiiZ$-representation of the same dimension
as~$\pi$, with $S_{\pi'}=\Sact'$ and $T_{\pi'}=\Tact'$ acting on
vectors denoted by $\chi$, then the direct sum
$\pi'\oplus\oC^2\!\tensor\pi$ admits a family of deformations achieved
by introducing a (matrix) automorphy factor as follows.  We start from
the direct sum $\pi'\oplus\oC^2\!\tensor\pi$ realized as
\begin{equation}\label{2-col-ori}
  \begin{pmatrix}
    f(\tau)\omega\\
    \chi
  \end{pmatrix}
\end{equation}
(where $\omega$ is from the representation space of $\pi$ and $\chi$
is from $\pi'$), with $f(\tau)$ being a first- or zeroth-degree
polynomial, in accordance with the above realization of the doublet.
The entire representation space is then spanned by
{\small$\begin{pmatrix}
    f(\tau)\omega + \beta\nu g(\tau)\chi\\
    \chi
  \end{pmatrix}$}.  The terms containing $\nu$ in the top row are
always linear in $\nu$ and proportional to a chosen parameter $\beta$,
because of their origin that becomes clear momentarily.

The $\SLiiZ$ action is uniquely defined~by the $T$ and $S$ actions
\begin{align*}
  T
  \begin{pmatrix}
    f(\tau)\omega + \beta\nu g(\tau)\chi\\
    \chi
  \end{pmatrix}
  &=
  \begin{pmatrix}
    f(\tau+1)\,\Tact\omega
    + \beta\nu g(\tau+1)\,\Tact'\chi\\
    \Tact'\chi
  \end{pmatrix},
  \\
  S
  \begin{pmatrix}
    f(\tau)\omega + \beta\nu g(\tau)\chi\\
    \chi
  \end{pmatrix}
  &=
  \begin{pmatrix}
    \tau\,f(-\frac{1}{\tau})\,\Sact\omega
    + \beta\frac{\nu}{\tau}\,g(-\frac{1}{\tau})\Sact'\chi
    + \beta\nu f(-\frac{1}{\tau})\,\Sact'\chi\\
    \Sact'\chi
  \end{pmatrix}.
\end{align*}
The third term in the top row of the $S$-transformation formula is the
origin of terms proportional to $\beta\nu$.  The prescription for the
$S$-transformation rule is to act with and without an extra $\tau$
factor on terms without~$\nu$ and with~it in the top row.  That is,
$f$ in $f(\tau)\omega$ is assigned weight~$1$ as before, whereas any
$g$ in $\nu g(\tau)\chi$ is considered to have weight~$0$.

It is easy to see that the $\SLiiZ$-orbit (again in the right-action
version) of elements~\eqref{2-col-ori} can then be written as
in~\eqref{2-orbit}.

\subsection{$\oC^3$}\label{sec:C3}
\subsubsection{}\label{sec:C3-def}
Taking the symmetrized square of $\oC^2$ gives the $\oC^3$
representation;\pagebreak[3] following~\bref{sec:C2-def}, we can
represent its basis as $\uone\tensor\uone$,
$\utau\tensor\uone+\uone\tensor\utau$, and $\utau\tensor\utau$.  A
somewhat shorter notation for the same basis is $\uone$, $\uiitau$,
and $\utauii$, with the $\SLiiZ$-action immediately found as
\begin{gather*}
  S\uone=\utauii,\quad S\,\uiitau=-\uiitau,\quad S\utauii=\uone,\\
  T\uone=\uone,\quad T\,\uiitau=\uiitau + 2\cdot\uone,\quad
  T\utauii=\utauii + \uiitau + \uone.
\end{gather*}
This form suggests a ``functional realization'' of $\oC^3\!\tensor\pi$
(for a finite-dimensional $\SLiiZ$ representation $\pi$ determined by
$\varphi\mapsto\Sact\varphi$ and $\varphi\mapsto\Tact\varphi$) as the
representation spanned by $\varphi$, $2 \tau\varphi$, and
$\tau^2\varphi$, with the $\SLiiZ$-action defined by
\begin{gather*}
  S(f(\tau)\varphi)=\tau^2 f(-\ffrac{1}{\tau})\,\Sact\varphi,
  \qquad
  T(f(\tau)\varphi)=f(\tau+1)\,\Tact\varphi.
\end{gather*}
In other words, $f$ is assigned weight~$2$ and $\SLiiZ$ acts (in the
right-action version) as
\begin{equation*}
  \gamma:f(\tau)\varphi\mapsto
  (c\tau+d)^2 f(\gamma\tau)\varphi\actedby\gamma.
\end{equation*}
Extending $f(\tau)$ beyond degree-$2$ polynomials gives an
infinite-dimensional $\SLiiZ$ representation in which
$\oC^3\tensor\pi$ is a subrepresentation.

\subsubsection{}
\label{alpha-beta-C3}
With $\oC^3\!\tensor\pi$ realized as in~\bref{sec:C3-def}, we
construct a deformation of $\pi''\oplus\pi'\oplus\oC^3\!\tensor\pi$
for three representations $\pi''$, $\pi'$, and $\pi$ of the same
dimension.  We write
\begin{equation}\label{3-col-ori}
  \begin{pmatrix}
    f(\tau)\varphi\\
    \omega\\
    \chi
  \end{pmatrix}
\end{equation}
for an arbitrary element of $\pi''\oplus\pi'\oplus\oC^3\!\tensor\pi$,
where $f(\tau)$ a polynomial of degree at most two.  The ``new''
$\SLiiZ$-action on such elements gives rise to and is defined on
elements of the form
\begin{equation*}
  \begin{pmatrix}
    f(\tau)\varphi + \nu g(\tau) \omega + \nu^2 h(\tau)\chi\\
    \omega\\
    \chi
  \end{pmatrix},
\end{equation*}
where the $\nu$-dependent terms are in fact proportional to the
parameters $\alpha$ and $\beta$ as becomes clear from the $S$-action
formula below.  

The $T$ generator acts as (we omit the primes distinguishing the
actions in $\pi$, $\pi'$, and $\pi''$)
\begin{small}
\begin{equation*}
  \begin{pmatrix}
    f(\tau)\varphi + \nu g(\tau) \omega + \nu^2 h(\tau)\chi\\
    \omega\\
    \chi
  \end{pmatrix}
  \stackrel{T}{\longmapsto}
  \begin{pmatrix}
    f(\tau+1)\Tact\varphi
    + \nu g(\tau+1)\Tact\omega
    + \nu^2 h(\tau+1)\Tact\chi\\
    \Tact\omega\\
    \Tact\chi
  \end{pmatrix}
\end{equation*}
{\normalsize and $S$ as}
\begin{multline*}
  \begin{pmatrix}
    f(\tau)\varphi + \nu g(\tau) \omega + \nu^2 h(\tau)\chi\\
    \omega\\
    \chi
  \end{pmatrix}
  \stackrel{S}{\longmapsto}\\
  \begin{pmatrix}
    \tau^2f(-\frac{1}{\tau})\Sact\varphi
    + \nu g(-\frac{1}{\tau})\Sact\omega
    + \frac{\nu^2}{\tau^2} h(-\frac{1}{\tau})\Sact\chi
    +\alpha\nu\tau f(-\frac{1}{\tau})\Sact\omega
    +\nu^2\left(\frac{\alpha\beta}{2}f(-\frac{1}{\tau})
      +\frac{\beta}{\tau}g(-\frac{1}{\tau})\right)\Sact\chi\\
    \Sact\omega\\
    \Sact\chi
  \end{pmatrix}.
\end{multline*}%
\end{small}%
The $S$-transformation rule in the top row can be described as
follows.  The $\nu$-independent term transforms as a triplet (with a
factor of~$\tau^2$ as in~\bref{sec:C3-def}); the term linear in $\nu$
transforms as ``$\nu$ times a doublet,'' i.e., $S:\nu\,{\cdot}\,
g(\tau)\omega \mapsto{}$\linebreak[3]$\frac{\nu}{\tau}\,{\cdot}\,\tau
g(-\frac{1}{\tau})\Sact\omega$ (with the extra $\tau$ factor as
in~\bref{sec:C2-def}); the $\nu^2$-term involves no additional
$\tau$-dependent factors: it transforms in accordance with the action
on functions on $\upperH\,{\times}\,\oC$; in addition, the pair
{\small$\begin{pmatrix}
    f(\tau)\varphi\\
    \omega
\end{pmatrix}$} gives rise to an extra term $\alpha\nu\tau
f(-\frac{1}{\tau})\Sact\omega$, the pair {\small$\begin{pmatrix}
  \nu g(\tau)\omega\\
  \chi
\end{pmatrix}$} to an extra term
$\nu^2\frac{\beta}{\tau}g(-\frac{1}{\tau})\Sact\chi$, and the pair
{\small$\begin{pmatrix}
  f(\tau)\varphi\\
  \chi
\end{pmatrix}$} to
$\frac{\alpha\beta}{2}\nu^2f(-\frac{1}{\tau})\Sact\chi$.

It is not difficult to see that the (right-action) $\SLiiZ$-orbit of
elements~\eqref{3-col-ori} can be written as in~\eqref{3-orbit}.

\newcommand{\reftitle}[1]{\textit{#1}}

\end{document}